\documentclass[twocolumn,twocolappendix]{aastex63}
\usepackage{natbib,aas_macros,amsmath}
\usepackage{multirow,color}
\usepackage{natbib}

\def\tcb{\textcolor{black}}

\newcommand{\oiii}{[O\,{\sc iii}]}

\newcommand{\cii}{[C\,{\sc ii}]}
\newcommand{\ci}{[C\,{\sc i}]}

\newcommand{\hst}{{\it HST}}

\submitjournal{}
\accepted{June 18, 2024}
\received{March 07, 2023}
\revised{June 05, 2024}

\shorttitle{ALCS: Deep 1.2-mm Number Counts and IR LFs at $z\simeq1-8$}
\shortauthors{Fujimoto et al.}

\begin{document}

\title{
ALMA Lensing Cluster Survey: \\
Deep 1.2 mm Number Counts and Infrared Luminosity Functions at \boldmath $z\simeq1-8$ 
}
`
\correspondingauthor{Seiji Fujimoto}
\email{fujimoto@utexas.edu}

\author[0000-0001-7201-5066]{Seiji Fujimoto}
\altaffiliation{Hubble Fellow}
\affiliation{
Department of Astronomy, The University of Texas at Austin, Austin, TX, USA
}
\affiliation{
Cosmic Dawn Center (DAWN), Denmark
}
\affiliation{
Niels Bohr Institute, University of Copenhagen, Jagtvej 128, DK2200 Copenhagen N, Denmark
}

%%%%%%%%%%%%%%%%%%%%%%%%%%%%%%%%%%%%%%%%%%%%%%%%%%%%%%%%%%%%%%%%%%%%%%%%%%%%
%%% Authors who provided significant contributions over the years %%%

\author[0000-0002-4052-2394]{Kotaro Kohno}
\affiliation{
Institute of Astronomy, Graduate School of Science, The University of Tokyo, 2-21-1 Osawa, Mitaka, Tokyo 181-0015, Japan
}
\affiliation{
Research Center for the Early Universe, Graduate School of Science, The University of Tokyo, 7-3-1 Hongo, Bunkyo-ku, Tokyo 113-0033, Japan
}

\author[0000-0002-1049-6658]{Masami Ouchi}
\affiliation{National Astronomical Observatory of Japan, 2-21-1 Osawa, Mitaka, Tokyo 181-8588, Japan}
\affiliation{Institute for Cosmic Ray Research, The University of Tokyo, 5-1-5 Kashiwanoha, Kashiwa, Chiba 277-8582, Japan}
\affiliation{Kavli Institute for the Physics and Mathematics of the Universe (WPI), University of Tokyo, Kashiwa, Chiba 277-8583, Japan}

\author[0000-0003-3484-399X]{Masamune Oguri}
\affiliation{Center for Frontier Science, Chiba University, 1-33 Yayoi-cho, Inage-ku, Chiba 263-8522, Japan}
\affiliation{Department of Physics, Graduate School of Science, Chiba University, 1-33 Yayoi-Cho, Inage-Ku, Chiba 263-8522, Japan}

\author[0000-0002-5588-9156]{Vasily Kokorev}
\affiliation{
Kapteyn Astronomical Institute, University of Groningen, P.O. Box 800, 9700AV Groningen, The Netherlands
}
\affiliation{
Cosmic Dawn Center (DAWN), Denmark
}
\affiliation{
Niels Bohr Institute, University of Copenhagen, Jagtvej 128, DK2200 Copenhagen N, Denmark
}

\author[0000-0003-2680-005X]{Gabriel Brammer}
\affiliation{
Cosmic Dawn Center (DAWN), Denmark
}
\affiliation{
Niels Bohr Institute, University of Copenhagen, Jagtvej 128, DK2200 Copenhagen N, Denmark
}

\author[0000-0002-4622-6617]{Fengwu Sun}
\affiliation{Steward Observatory, University of Arizona, 933 N. Cherry Avenue, Tucson, 85721, USA}

\author[0000-0003-3926-1411]{Jorge Gonz\'alez-L\'opez}
\affiliation{
N\'ucleo de Astronom\'ia de la Facultad de Ingenier\'ia y Ciencias, Universidad Diego Portales, Av. Ejército Libertador 441, Santiago, Chile
}
\affiliation{
Las Campanas Observatory, Carnegie Institution of Washington, Casilla 601, La Serena, Chile
}

%%%%%%%%%%%%%%%%%%%%%%%%%%%%%%%%%%%%%%%%%%%%%%%%%%%%%%%%%%%%%%%%%%%%%%%%%%%%
%%% Authors who help with the data analysis and key scientific arguments in the paper, including mass model constructions, follow-up observations #### (alphabetically)

\author[0000-0002-8686-8737]{Franz E. Bauer}
\affiliation{Instituto de Astrof{\'{\i}}sica, Facultad de F{\'{i}}sica, Pontificia Universidad Cat{\'{o}}lica de Chile, Campus San Joaquín, Av. Vicuña Mackenna 4860, Macul Santiago, Chile, 7820436} 
\affiliation{Centro de Astroingenier{\'{\i}}a, Facultad de F{\'{i}}sica, Pontificia Universidad Cat{\'{o}}lica de Chile, Campus San Joaquín, Av. Vicuña Mackenna 4860, Macul Santiago, Chile, 7820436} 
\affiliation{Millennium Institute of Astrophysics, Nuncio Monse{\~{n}}or S{\'{o}}tero Sanz 100, Of 104, Providencia, Santiago, Chile} 

\author[0000-0001-6052-3274]{Gabriel B. Caminha}
\affiliation{
Kapteyn Astronomical Institute, University of Groningen, Postbus 800, 9700 AV Groningen, The Netherlands
}

\author[0000-0001-6469-8725]{Bunyo Hatsukade}
\affiliation{National Astronomical Observatory of Japan, 2-21-1 Osawa, Mitaka, Tokyo 181-8588, Japan}
\affiliation{Graduate Institute for Advanced Studies, SOKENDAI, Osawa, Mitaka, Tokyo 181-8588, Japan}
\affiliation{
Institute of Astronomy, Graduate School of Science, The University of Tokyo, 2-21-1 Osawa, Mitaka, Tokyo 181-0015, Japan
}

\author[0000-0001-5492-1049]{Johan Richard}
\affiliation{
Univ Lyon, Univ Lyon1, Ens de Lyon, CNRS, Centre de Recherche Astrophysique de Lyon UMR5574, F-69230, Saint-Genis-Laval,France
}

\author[0000-0003-3037-257X]{Ian Smail}
\affiliation{Centre for Extragalactic Astronomy, Department of Physics, Durham University, Durham DH1 3LE, UK}

\author[0000-0002-0498-5041]{Akiyoshi Tsujita}
\affiliation{
Institute of Astronomy, Graduate School of Science, The University of Tokyo, 2-21-1 Osawa, Mitaka, Tokyo 181-0015, Japan
}

\author[0000-0001-7821-6715]{Yoshihiro Ueda}
\affiliation{Department of Astronomy, Kyoto University, Kyoto 606-8502, Japan}

\author[0000-0001-6653-779X]{Ryosuke Uematsu}
\affiliation{Department of Astronomy, Kyoto University, Kyoto 606-8502, Japan}

\author[0000-0002-0350-4488]{Adi Zitrin}
\affiliation{
Physics Department, Ben-Gurion University of the Negev, P.O. Box 653, Be’er-sheva 8410501, Israel
}

%%%%%%%%%%%%%%%%%%%%%%%%%%%%%%%%%%%%%%%%%%%%%%%%%%%%%%%%%%%%%%%%%%%%%%%%%%%%%%
%%%%% ALCS program architects %%%%% (alphabetically)
\author[0000-0001-7410-7669]{Dan Coe}
\affiliation{Space Telescope Science Institute (STScI), 
3700 San Martin Drive, Baltimore, MD 21218, USA}
\affiliation{Association of Universities for Research in Astronomy (AURA), Inc.
for the European Space Agency (ESA)}
\affiliation{Center for Astrophysical Sciences, Department of Physics and Astronomy, The Johns Hopkins University, 3400 N Charles St. Baltimore, MD 21218, USA}

\author[0000-0002-4616-4989]{Jean-Paul Kneib}
\affiliation{
Laboratoire d\'Astrophysique, Ecole Polytechnique F\'ed\'erale de Lausanne, Observatoire de Sauverny, CH-1290 Versoix, Switzerland
}

\author[0000-0002-9365-7989]{Marc Postman}
\affiliation{Space Telescope Science Institute (STScI), 
3700 San Martin Drive, 
Baltimore, MD 21218, USA}

\author[0000-0002-7196-4822]{Keiichi Umetsu}
\affiliation{Institute of Astronomy and Astrophysics, Academia Sinica (ASIAA), 
AS/NTU Astronomy-Mathematics Building, 
No. 1, Sec. 4, Roosevelt Rd., Taipei 10617, Taiwan}

%%%%%%%%%%%%%%%%%%%%%%%%%%%%%%%%%%%%%%%%%%%%%%%%%%%%%%%%%%%%%%%%%%%%%%%%%%%%
% Authors who help with the comparisons with the simulations 

\author[0000-0003-3021-8564]{Claudia del P. Lagos}
\affiliation{International Centre for Radio Astronomy Research (ICRAR), M468, University of Western Australia, 35 Stirling Hwy, Crawley, \\WA 6009, Australia}
\affiliation{ARC Centre of Excellence for All Sky Astrophysics in 3 Dimensions (ASTRO 3D)}
\affiliation{
Cosmic Dawn Center (DAWN), Denmark
}

\author[0000-0003-1151-4659]{Gerg\"{o} Popping}
\affiliation{European Southern Observatory, Karl-Schwarzschild-Str. 2, D-85748, Garching, Germany}

%%%%%%%%%%%%%%%%%%%%%%%%%%%%%%%%%%%%%%%%%%%%%%%%%%%%%%%%%%%%%%%%%%%%%%%%%%%%
%%% Authors who contribute to enrich and strengthen the scientific arguments in the paper #### (alphabetically)

\author[0000-0003-3139-2724]{Yiping Ao}
\affiliation{
Purple Mountain Observatory and Key Laboratory for Radio Astronomy, Chinese Academy of Sciences, Nanjing, China
}

\author[0000-0002-7908-9284]{Larry Bradley}
\affiliation{Space Telescope Science Institute}

\author[0000-0001-8183-1460]{Karina Caputi}
\affiliation{
Kapteyn Astronomical Institute, University of Groningen, P.O. Box 800, 9700AV Groningen, The Netherlands
}
\affiliation{
Cosmic Dawn Center (DAWN), Denmark
}

\author[0000-0003-0348-2917]{Miroslava Dessauges-Zavadsky}
\affiliation{D\'epartement d\'Astronomie, Universit\'e de Gen\`eve, Chemin Pegasi 51, 1290 Versoix, Switzerland}

\author[0000-0003-1344-9475]{Eiichi Egami}
\affiliation{
Steward Observatory, University of Arizona, 933 N. Cherry Ave, Tucson, AZ 85721, USA
}

\author[0000-0002-8726-7685]{Daniel Espada}
\affiliation{Departamento de F\'{i}sica Te\'{o}rica y del Cosmos, Campus de Fuentenueva, Edificio Mecenas, Universidad de Granada, E-18071, Granada, Spain}
\affiliation{Instituto Carlos I de F\'{i}sica Te\'{o}rica y Computacional, Facultad de Ciencias, E-18071, Granada, Spain}

\author[0000-0001-5118-1313]{R.\,J.~Ivison}
\affiliation{European Southern Observatory, Karl-Schwarzschild-Strasse~2, D-85748 Garching, Germany}

\author[0000-0003-1974-8732]{Mathilde Jauzac}
\affiliation{Centre for Extragalactic Astronomy, Durham University, South Road, Durham DH1 3LE, UK}
\affiliation{Institute for Computational Cosmology, Durham University, South Road, Durham DH1 3LE, UK}

\author[0000-0002-7821-8873]{Kirsten K. Knudsen}
\tcb{\affiliation{Department of Space, Earth and Environment, Chalmers University of Technology, SE-412 96 Gothenburg, Sweden}}

\author[0000-0002-6610-2048]{Anton M. Koekemoer}
\affiliation{Space Telescope Science Institute, 3700 San Martin Dr., Baltimore, MD 21218, USA} 

\author[0000-0002-4872-2294]{Georgios E. Magdis}
\affiliation{
Cosmic Dawn Center (DAWN), Denmark
}
\affiliation{DTU-Space, Technical University of Denmark, Elektrovej 327, 2800, Kgs. Lyngby, Denmark}
\affiliation{Niels Bohr Institute, University of Copenhagen, Jagtvej 128, 2200, Copenhagen N, Denmark}

\author[0000-0003-3266-2001]{Guillaume Mahler}
\affiliation{Centre for Extragalactic Astronomy, Durham University, South Road, Durham DH1 3LE, UK}
\affiliation{Institute for Computational Cosmology, Durham University, South Road, Durham DH1 3LE, UK}

\author[0000-0002-8722-516X]{A.~M.~Mu\~noz~Arancibia}
\affiliation{Millennium Institute of Astrophysics, Nuncio Monse\~nor S\'otero Sanz 100, Providencia, Santiago, Chile}
\affiliation{Center for Mathematical Modeling (CMM), Universidad de Chile, Beauchef 851, Santiago, Chile}

\author[0000-0002-7028-5588]{Timothy Rawle}
\affiliation{
European Space Agency (ESA), ESA Office, Space Telescope Science Institute, 3700 San Martin Drive, Baltimore, MD 21218, USA
}

\author[0000-0002-2597-2231]{Kazuhiro Shimasaku}
\affiliation{Department of Astronomy, School of Science, The
University of Tokyo, 7-3-1 Hongo, Bunkyo-ku, Tokyo 113-0033, Japan}
\affiliation{Research Center for the Early Universe, The University of
Tokyo, 7-3-1 Hongo, Bunkyo-ku, Tokyo 113-0033, Japan}

\author[0000-0003-3631-7176]{Sune Toft}
\affiliation{
Cosmic Dawn Center (DAWN), Denmark
}
\affiliation{
Niels Bohr Institute, University of Copenhagen, Jagtvej 128, DK2200 Copenhagen N, Denmark
}

\author[0000-0003-1937-0573]{Hideki Umehata}
\affiliation{Institute for Advanced Research, Nagoya University, Furocho, Chikusa, Nagoya 464-8602, Japan}
\affiliation{Department of Physics, Graduate School of Science, Nagoya University, Furocho, Chikusa, Nagoya 464-8602, Japan}
\affiliation{Cahill Center for Astronomy and Astrophysics, California Institute of Technology, 1200 E California Blvd, MC 249-17, Pasadena, CA
91125, USA}

\author[0000-0001-6477-4011]{Francesco Valentino}
\affiliation{European Southern Observatory, Karl-Schwarzschild-Str. 2, D-85748, Garching, Germany}
\affiliation{
Cosmic Dawn Center (DAWN), Denmark
}
\affiliation{Niels Bohr Institute, University of Copenhagen, Jagtvej 128, 2200, Copenhagen N, Denmark}

\author[0000-0002-2504-2421]{Tao Wang}
\affiliation{School of Astronomy and Space Science, Nanjing University, Nanjing 210093, China}
\affiliation{Key Laboratory of Modern Astronomy and Astrophysics (Nanjing University), Ministry of Education, Nanjing 210093, China}

\author[0000-0003-2588-1265]{Wei-Hao Wang}
\affiliation{Institute of Astronomy and Astrophysics, Academia Sinica, Taipei 10617, Taiwan}

\def\apj{ApJ}%
         % Astrophysical Journal
\def\apjl{ApJ}%
         % Astrophysical Journal, Letters
\def\apjs{ApJS}% 
         % Astrophysical Journal, Supplement

\def\rme{\rm e}
\def\rmstar{\rm star}
\def\rmFIR{\rm FIR}
\def\itHubble{\it Hubble}
\def\rmyr{\rm yr}

%---------------------------------------------------------------------
\begin{abstract}
We present a statistical study of 180 dust continuum sources identified in 33 massive cluster fields by the ALMA Lensing Cluster Survey (ALCS) over a total of 133 arcmin$^{2}$ area, homogeneously observed at 1.2~mm. 
ALCS enables us to detect extremely faint mm sources by lensing magnification, including near-infrared (NIR) dark objects showing no counterparts in existing {\it Hubble Space Telescope} and {\it Spitzer} images. The dust continuum sources belong to a blind sample ($N=141$) with S/N $\gtrsim$ 5.0 (a purity of $>$ 0.99) or a secondary sample ($N=39$) with S/N=4.0--5.0 screened by priors. 
With the blind sample, we securely derive 1.2-mm number counts down to $\sim7$~$\mu$Jy, and find that the total integrated 1.2mm flux is 20.7$^{+8.5}_{-6.5}$~Jy~deg$^{-2}$, resolving $\simeq$ 80\% of the cosmic infrared background light. The resolved fraction varies by a factor of 0.6--1.1 due to the completeness correction depending on the spatial size of the mm emission. 
We also derive infrared (IR) luminosity functions (LFs) at $z=0.6$--7.5 with the $1/V_{\rm max}$ method, finding the redshift evolution of IR LFs characterized by positive luminosity and negative density evolution.
The total (=UV+IR) cosmic star-formation rate density (SFRD) at $z>4$ is estimated to be $161^{+25}_{-21}$\% of 
\tcb{the \cite{madau2014} measurements mostly based on rest-frame UV surveys.} 
Although our general understanding of the cosmic SFRD is unlikely to change beyond a factor of 2, these results add to the weight of evidence for an additional ($\approx 60$\%) SFRD component contributed by the faint-mm population, including NIR dark objects.
\end{abstract}

%---------------------------------------------------------------------
\keywords{ galaxies: formation --- galaxies: evolution --- galaxies: high-redshift }
%---------------------------------------------------------------------

\section{Introduction}
\label{sec:intro}

Extragalactic background light (EBL) is electromagnetic radiation that fills the Universe, arising from several distinct physical processes and energetically dominated by cosmic microwave (CMB), optical (COB), and infrared (CIB) radiation. 
The CMB and COB are explained by the leftover radiation from the Big Bang and the stellar continuum of galaxies, respectively (e.g., \citealt{totani2001,planck2014}; cf. \citealt{lauer2022}). 
On the other hand, the origin of the CIB has not yet been fully accounted for as yet, despite its importance implied by the fact that the total energy of the CIB has been known to be comparable to that of the COB since its initial discovery with the {\it Cosmic Background Explorer } (\textit{COBE}) satellite \citep{puget1996,fixsen1998,hauser1998,hauser2001,dole2006}.   
Therefore, concluding whether the CIB is explained by known physical mechanisms and objects in the universe is an essential open question in modern astrophysics.

Some fraction of the stellar continuum in galaxies is absorbed by dust, re-emitted as thermal infrared (IR) emission, and thought to contribute to the CIB. 
Thus, to understand the origin of the CIB, the first approach is detecting individual IR-emitting galaxies and evaluating their contribution to the CIB. 
In these studies, one of the most advantageous wavelength regimes is the submm/mm, owing to the negative $k$-correction. 
Rare IR-bright, high-redshift dusty star-forming galaxies were first identified a few decades ago at submm/mm wavelengths, referred to as submm galaxies (SMGs) due to their brightness at the submm/mm wavelengths ($S_{\rm 1mm}\gtrsim1$ mJy; e.g., \citealt{hughes1998,blain2002,casey2014}). 
However, bright SMGs only contribute $\sim10$--20\% to the CIB, based on previous blank field observations with single-dish telescopes \citep[e.g.,][]{perera2008,hatsukade2011,scott2012}, 
indicating that the bulk of the CIB is produced by populations distinct from the SMGs. 
\tcb{
While several previous studies using the stacking analyses with {\it Herschel} and SCUBA2 data have pushed the detection limit beyond the SMGs and hinted that faint dusty galaxies may dominantly account for the CIB at submm wavelengths \citep[e.g.,][]{bethermin2012b, viero2013, zavala2017}, direct individual detection is essential to avoid potential systematics inherent in the stacking approach. Moreover, recent studies suggest the presence of remarkably extended diffuse dust emission beyond individual galaxy scales \citep[e.g.,][]{xiao2023}, which is challenging to distinguish given the poor spatial resolution of these instruments ($\sim$10--30$''$). Consequently, deep submm/mm observations with high angular resolution are crucial for directly verifying the origin of the CIB, whether it arises from compact emission of individual galaxies or diffuse emission around galaxies, for example. 
}

Atacama Large Millimeter/submillimeter Array (ALMA) observations allow us to explore a faint submm/mm regime ($S_{\rm 1mm}<1$ mJy) without uncertainties from source confusion and blending, owing to ALMA's high sensitivity and angular resolution. 
Faint submm/mm sources have been investigated in a large variety of ALMA observations: 
via dedicated blind surveys of blank \citep[e.g.,][]{aravena2016,wwang2016, dunlop2017,umehata2017,hatsukade2018,franco2018,gonzalez2020,zavala2021,adscheid2024} and lensing cluster fields \citep[e.g.,][]{fujimoto2016,arancibia2019,munoz-arancibia2022}; 
and as serendipitous sources in single pointing observations \citep[e.g.,][]{hatsukade2013,ono2014,carniani2015,fujimoto2016,zavala2018,bethermin2020} and calibrator fields \citep[e.g.,][]{oteo2016,klitsch2020}. 
These studies show that faint submm/mm sources newly identified with ALMA strongly outnumber the SMGs. For instance, the deepest observations, aided by strong gravitational lensing, report that $\sim$70$-$100\% of CIB is resolved down to $\sim$0.01 mJy at $\approx$1~mm \citep[e.g.,][]{fujimoto2016,arancibia2019}. 
On the other hand, 
the deepest ALMA blank field surveys so far in Hubble Ultra Deep Field (HUDF) 
estimate a 1.2~mm total flux density integrated down to zero flux by extrapolation to be $6.8\pm0.4$ Jy~deg$^{-2}$ \citep{gonzalez2020}, 
which corresponds to $\sim40\%$ of the CIB at 1.2~mm. 
Because the effective survey areas in these previous deepest ALMA studies are less than a few arcmin$^{2}$
at the faintest 1-mm flux density regime of 0.01--0.1 mJy, the difference might result from cosmic variance \citep[e.g.,][]{trenti2008} and concluding whether the CIB is fully resolved or not has been still challenging even with ALMA. 
\tcb{
Naturally, this also advocates the difficulty in constraining the IR luminosity function (LF) and subsequently dust-obscured cosmic star-formation rate density, especially at high redshifts ($z\gtrsim4$; e.g., \citealt{novak2017, liu2018, gruppioni2020}). 
}

To enlarge the survey area at the faintest regime and overcome cosmic variance, 
making full use of gravitational lensing is one of the most effective approaches. 
Assuming a magnification factor of $\mu$ and a given survey area in the observed frame of $A$, the effective survey area and the data depth after the lens correction decrease by a factor of $\mu$ from the observed frame. 
We thus need $\mu$ times observation time to recover the given survey area of $A$ (= $A/\mu \times \mu$). 
On the other hand, if we achieve the $\mu$ times deep data at the area of $A$ without the lensing support, we generally need $\mu^{2}$ times integration time for the observations.  
Therefore, the gravitational lensing support saves in observation time by a factor of $\sim\mu$ to achieve observations at a given survey area and data depth. 
\tcb{Following the successful outcome from the SCUBA-2 lensing cluster survey \citep[e.g.,][]{cowie2022},} a complete ALMA survey towards massive galaxy clusters will allow us to detect large numbers of faint submm/mm sources even out to the epoch of reionization \citep{watson2015,laporte2017a,inami2022} in the most effective way and offers a unique opportunity to assess the origin of the CIB. 

%%%%%%%%%%%%%%%%%%%%%%%
\begin{figure*}[t!]
\includegraphics[trim=0cm 0cm 0cm 0cm, clip, angle=0,width=1.0\textwidth]{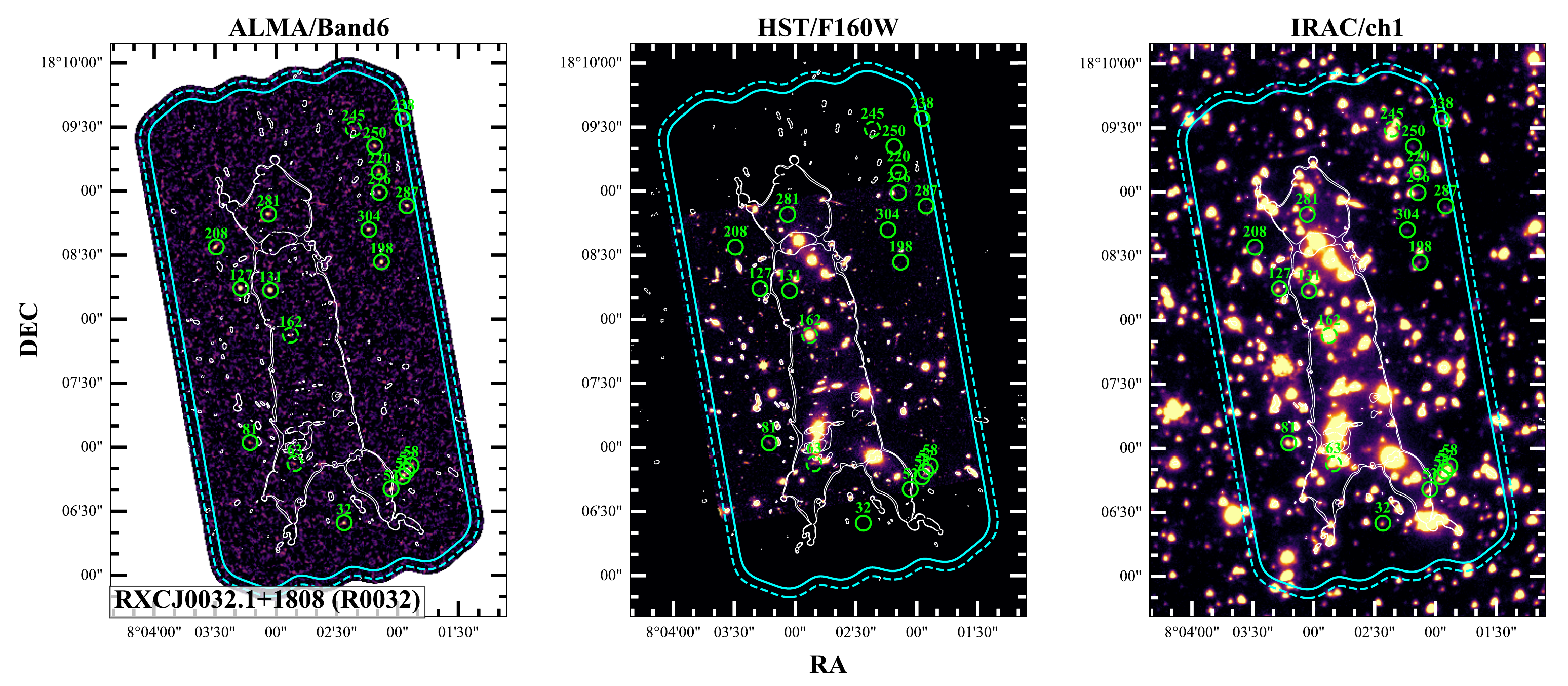}
 \caption{
Footprints of ALMA Band 6 (left), HST/F160W (middle), and IRAC/ch1 (right) in one of the ALCS fields.  
The dashed and solid cyan lines show the relative sensitivity response to the deepest 30\% and 50\% of the mosaic, respectively. 
The white lines denote the $\mu=200$ magnification curve at $z=2$, estimated from our fiducial lens model of this cluster. 
The labeled solid and dashed green circles represent the ALMA continuum catalog sources with SNR$_{\rm nat}$ $\geq 5$ and $= 4-5$, respectively. 
\label{fig:fig1}}
\end{figure*}
%%%%%%%%%%%%%%%%%%%%%%% 

In this paper, we present statistics of 180 faint 1.2-mm sources identified from the large imaging and spectral campaign of 33 massive galaxy clusters in the ALMA Lensing Cluster Survey (ALCS) to assess the origins of the CIB. 
Making full use of the rich ancillary data sets, including {\it HST} and {\it Spitzer} IRAC images and photometry,
we also study the IR luminosity function (LF) at $z=1-8$ and the cosmic star-formation rate (SFR) density (SFRD) based on the largest statistics so far for the faint submm/mm population with ALMA.   
The structure of this paper is as follows. 
In Section 2, we overview the survey of ALCS and the data sets. 
%n RXCJ0600-2007.  
Section 3 outlines the methods of 
the source extraction, flux measurements, redshift estimates, magnification estimates, corrections for the flux measurements and completeness through Monte-Carlo simulations, and survey area estimates. 
We characterize our ALCS sources in Section~\ref{sec:prop}. 
In Section 5, we present the 1.2-mm number counts and the contribution to the CIB. 
The redshift evolutions of IR LFs and the cosmic SFRD are discussed in Section 6. 
A summary is presented in Section 7. 
Throughout this paper, we assume a flat universe with 
$\Omega_{\rm m} = 0.3$, 
$\Omega_\Lambda = 0.7$, 
%$\sigma_8 = 0.8$, 
and $H_0 = 70$ km s$^{-1}$ Mpc$^{-1}$. 
We adopt a search radius of $1\farcs0$ in cross-matching catalogs unless otherwise specified\footnote{
While \cite{gomez2021} suggest that $\leq0\farcs4$ is the robust counterpart search radius, we adopt the relatively large search radius as a default because the spatial offsets among the different wavelengths can be enhanced in highly magnified sources.   
}. 
We take the cosmic microwave background (CMB) effect in submm/mm observations into account \citep[e.g.,][]{dacunha2013, pallottini2015, zhang2016, lagache2018}, following the recipe presented by \cite{dacunha2013}.

%%%%%%%%%%%%%%%%%%%%%%%%%%%%%%%%%%%%%%%%%%%%%%%%%%%%%%%%%%%%%%%%%
\section{Observations \& Data Processing} 
\label{sec:data}

\subsection{Survey Design}
\label{sec:survey}

ALCS is a cycle-6 ALMA large program (Project ID: 2018.1.00035.L; PI: K. Kohno) 
to map high-magnification regions in 33 massive galaxy clusters at 1.2-mm in Band 6. 
The sample is selected from the best-studied clusters drawn from {\it HST} treasury programs, i.e. {\it Hubble} Frontier Fields (HFFs; \citealt{lotz2017}), the Cluster Lensing And Supernova Survey with {\it Hubble} (CLASH; \citealt{postman2012}), and the Reionization Lensing Cluster Survey (RELICS; \citealt{coe2019}).  
Observations were carried out between December 2018 and December 2019 in compact array configurations of C43-1 and C43-2, where 26 and 7 clusters were observed in Cycles 6 and 7, respectively. 
We adopt two frequency setups with the sky frequencies of 259.4 GHz and 263.2 GHz, accomplishing a 
15-GHz wide spectral scan in ranges of 250.0--257.5 GHz and 265.0--272.5 GHz to enlarge the survey volume for line-emitting galaxies. 
For the five HFF clusters visible with ALMA, 
ALCS only performed observations in one of the two frequency setups, 
because observations for the other frequency setup had been taken in previous ALMA HFF surveys (2013.1.00999.S \& 2015.1.01425.S; e.g., \citealt{gonzalez2017c}). 
The observations were carried out in mosaic mode, covering highly magnified regions spanning $\sim1$--9 arcmin$^{2}$ sky areas per cluster. 
The survey description is also presented in \cite{kohno2023}. 

\subsection{Data reduction, calibration, and imaging}
\label{sec:reduction}

The ALMA data were reduced and calibrated with the Common Astronomy Software Applications package versions 5.4.0 and 5.6.1 (CASA; \citealt{mcmullin2007}) for the data taken in Cycles 6 and 7, respectively, with the pipeline script in the standard manner. 
For the HFF data taken in the previous surveys, we used CASA versions 4.2.2--4.5.3 in order to use the pipeline scripts from the previous cycles. 
We created several data products (measurement sets, maps, and cubes) for every ALCS cluster. 
We imaged the calibrated visibilities with natural weighting, a pixel scale of $0\farcs16$, and a primary beam limit of 0.2 with the CASA task {\sc tclean}. 
For continuum maps, 
the {\sc tclean} routines were executed down to the 2$\sigma$ level with a maximum iteration number of 100,000 in the automask mode.\footnote{We determined sub-parameters of tclean in the automask mode based on recommendations from the ALMA automasking guide: https://casaguides.nrao.edu/index.php/Automasking\_Guide} 
For cubes, 
we adopted common spectral channel bins of 30 km~s$^{-1}$ and 60 km~s$^{-1}$ and created the cubes without the CLEAN iteration. We did not identify bright emission per channel in 31 cluster cubes, where we use the cubes in the following analysis. In the other two clusters of MACSJ0553 and MACS1931, we identified significant signals in each channel due to bright line emitters serendipitously detected in the cubes (Section \ref{sec:line}). We thus performed the CLEAN algorithm for these two cluster cubes. 
When we found systematic stripe patterns in the products by visual inspection, we applied additional flaggings\footnote{
Field ID = 18 in uid\_\_A002\_Xd98580\_X44f4.ms was flagged before imaging, because the field had been observed for a calibration purpose, but wrongly named as a target field. 
} and/or performed {\sc tclean} using manual mode masking.\footnote{
We invoked manual mode in {\sc tclean} for the cubes of MACSJ0553 and MACS1931 by setting rectangle masks ($\sim10''\times10''$) around the bright line emitters. 
} 
The natural-weighted maps achieved full-width-half-maximum (FWHM) sizes of the synthesized beam between  $0\farcs94\times0\farcs65$--$1\farcs57\times1\farcs01$, with $1\sigma$ sensitivities 
of 46.9--91.6 $\mu$Jy~beam$^{-1}$ for the continuum and 848.1--1706.4 $\mu$Jy~beam$^{-1}$ for the 60~km~s$^{-1}$ width channel cube. 
We summarize the data properties of the continuum map and the cube in Table \ref{tab:data_prop}. 
We also produced lower resolution maps and cubes by applying a $uv$-taper parameter ($2''\times2''$) to recover spatially extended, low-surface brightness emission associated with large, resolved galaxies or due to gravitational lensing effects. 
We refer to our ALMA maps (cubes) without and with the $uv$-taper as natural and tapered maps (cubes), respectively. 
The ALCS products, including the $uv$-tapered maps and cubes, are publicly available via the dedicated page\footnote{\url{http://www.ioa.s.u-tokyo.ac.jp/ALCS/}} and ALMA science data archive. 

%The ALCS products, including the $uv$-tapered maps and cubes, are publicly available.\footnote{
%\url{http://www.ioa.s.u-tokyo.ac.jp/ALCS/}
%}} 

\setlength{\tabcolsep}{10pt}
\begin{deluxetable*}{lcccccc} 
\tablecaption{ALCS Data Properties for 33 Lensing Clusters}
\tablehead{
\colhead{Cluster Name} & \colhead{Area} & \colhead{Beam}   & \colhead{$\sigma$ (cont)}  & \colhead{$\sigma$ (cube, tune1)} & \colhead{$\sigma$ (cube, tune2)} & \colhead{$N$}  \\
  & \colhead{arcmin$^{2}$}   & \colhead{arcsec}  &  \colhead{$\mu$Jy/beam}    &  \colhead{$\mu$Jy/beam}    & \colhead{$\mu$Jy/beam}&  \colhead{}   \\
\colhead{(1)} & \colhead{(2)} & \colhead{(3)} & \colhead{(4)} & \colhead{(5)} & \colhead{(6)} & \colhead{(7)}}
\startdata
\multicolumn{7}{c}{HFF} \\ \hline
A370    & 5.44 & 1.12 $\times$ 0.89 & 47 & 673 & 848 & 7 \\
A2744   & 5.13 & 0.94 $\times$ 0.65 & 51 & 839 & 643 & 11 \\
A1063   & 5.01 & 1.15 $\times$ 0.96 & 53 & 765 & 974 & 4 \\
M0416.1 & 5.24 & 1.48 $\times$ 0.85 & 55 & 879 & 1158 & 7 \\
M1149.5 & 5.52 & 1.24 $\times$ 1.1  & 64 & 778 & 975  & 4  \\ \hline
\multicolumn{7}{c}{CLASH} \\ \hline
A209 & 1.20 & 1.31 $\times$ 1.07 & 63 & 946 & 1193 & 1 \\
A383 & 1.39 & 1.13 $\times$ 0.87 & 61 & 989 & 1245 & 3 \\
M0329 & 3.03 & 1.57 $\times$ 1.01 & 71 & 1083 & 1354 & 1 \\
M0429 & 1.27 & 1.58 $\times$ 0.96 & 92 & 1397 & 1706 &  3 \\
M1115 & 1.67 & 1.16 $\times$ 1.06 & 63 & 925 & 1166 & 5 \\
M1206 & 2.87 & 1.13 $\times$ 0.98 & 53 & 789 & 1008 & 8  \\
M1311 & 1.38 & 1.10 $\times$ 0.98 & 62 & 906 & 1159 & 2 \\
M1423 & 1.88 & 1.34 $\times$ 1.07 & 65 & 951 & 1266 & 4 \\
M1931 & 2.64 & 1.34 $\times$ 1.03 & 56 & 817 & 1023 & 6 \\
M2129 & 2.54 & 1.23 $\times$ 0.95 & 47 & 773 & 940 & 3  \\
R2129  & 1.01 & 1.26 $\times$ 1.04 & 40 & 720 & 876 & 2 \\
R1347  & 3.57 & 1.15 $\times$ 1.01 & 53 & 795 & 1037 & 6 \\ \hline
\multicolumn{7}{c}{RELICS} \\ \hline
A3192 & 5.17 & 1.44 $\times$ 0.96 & 73 & 1125 & 1472 & 7 \\
A2163 & 2.14 & 1.23 $\times$ 1.07 & 50 & 751 & 953 & 1 \\
A2537 & 3.03 & 1.34 $\times$ 1.09 & 69 & 971 & 1262 & 5  \\
A295 & 4.63 & 0.90 $\times$ 0.81 & 74 & 1219 & 1481 & 2  \\
AC0102 & 6.02 & 1.17 $\times$ 0.91 & 72 & 1048 & 1336 & 14  \\
M0035 & 3.38 & 1.42 $\times$ 1.03 & 52 & 846 & 1016 & 4 \\
M0159 & 3.25 & 1.38 $\times$ 1.08 & 63 & 942 & 1192 & 4 \\
M0257 & 2.54 & 1.46 $\times$ 0.94 & 83 & 1238 & 1587 & 1 \\
M0417 & 6.54 & 1.47 $\times$ 0.95 & 84 & 1248 & 1566 & 8 \\
M0553 & 9.28 & 1.21 $\times$ 0.92 & 65 & 1057 & 1275 & 14 \\
P171  & 5.40 & 1.32 $\times$ 0.99 & 73 & 1231 & 1595 & 4 \\
RJ2211 & 6.83 & 1.33 $\times$ 1.07 & 78 & 1078 & 1389 & 4 \\
R0032 & 8.17 & 1.17 $\times$ 1.13 & 71 & 1030 & 1306 & 20 \\
R0600 & 7.91 & 1.22 $\times$ 0.95 & 57 & 822 & 1043 & 5 \\
R0949 & 3.62 & 1.22 $\times$ 1.18 & 62 & 903 & 1144 &  7 \\
SM0723 & 2.52 & 1.01 $\times$ 0.77 & 66 & 969 & 1242 & 4 \\ \hline
AVERAGE & 3.98 & 1.26 $\times$ 0.98 & 63 & 955 & 1195 & 5.5 \\
TOTAL & 131.22 & \nodata           & \nodata & \nodata & \nodata & 180 \\ \hline
\enddata
\tablecomments{
(1) Names of the 33 ALCS target clusters. We shorten the prefixes of Abell(S), MACS(J), PCLKG, RX(C)J, and SMACSJ to A, M, P, R, and SM, respectively. 
The numbers of declinations in the names are omitted. 
(2) Sky area observed in ALCS within the relative sensitivity above 30\% to the deepest part of the mosaic.  
(3) FWHM of the synthesized beam in the natural-weighted map.
(4) Data depth evaluated by the standard deviation of the pixel with the CASA task {\sc imstat} with the byweight algorithm. 
\tcb{(5) Data depth evaluated in the same manner as (4), but for the 60-km~s$^{-1}$ width cube obtained in the frequency setup with the central sky frequency at 259.4~GHz.  
(6) Same as (5), but for the cube with the central sky frequency at 263.2~GHz. 
(7) Number of continuum sources identified.} 
}
\label{tab:data_prop}
\end{deluxetable*}

\section{Data analysis}
\label{sec:analysis}

\subsection{Source Extraction}
\label{sec:source_ext}

%%%%%%%%%%%%%%%%%%%%%%%
\begin{figure}
\includegraphics[trim=0cm 0cm 0cm 0cm, clip, angle=0,width=0.5\textwidth]{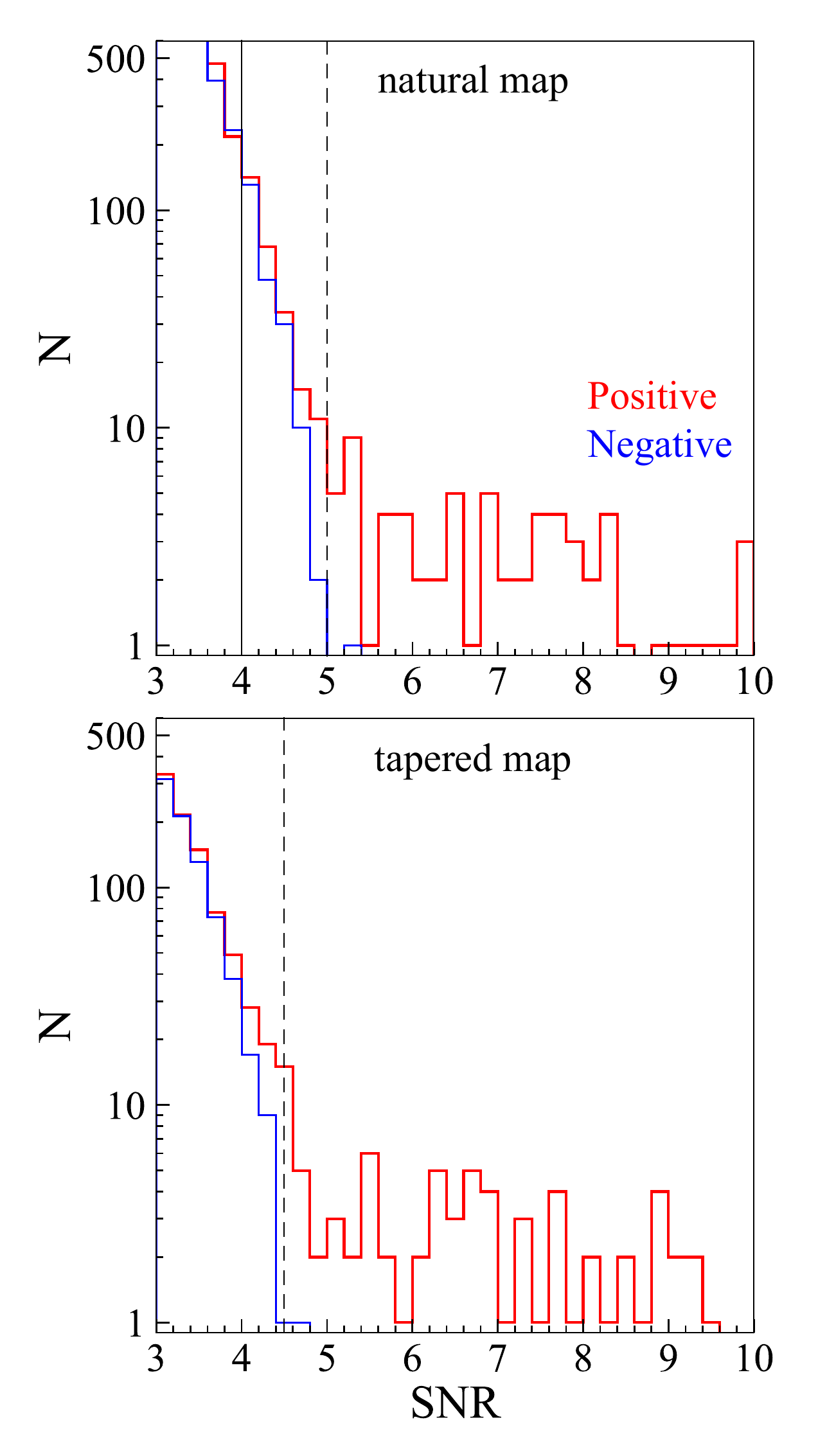}
 \caption{
Differential number of positive (red) and negative (blue) sources as a function of peak SNR, summed over the 33 ALCS fields.
Based on the excess of the positive to the negative sources, a total of 177$^{+12}_{-14}$ sources are expected to be real down to SNR = 4.0 in the natural map (black solid line).  The purity $p$ exceeds 0.99 at SNR = 5.0 and 4.5 in the natural and tapered maps, respectively (black dashed line). 
We identify 141 sources at $p>0.99$, referred to as the blind catalog. 
\label{fig:pn_hist}}
\end{figure}
%%%%%%%%%%%%%%%%%%%%%%% 

We conduct a blind two-dimensional source extraction with {\sc sextractor} version 2.5.0 \citep{bertin1996}. 
We use the natural and tapered maps before primary beam correction for the source extraction and apply the correction to the flux measurements in the following analysis. 
We extract sources with positive peak counts above 2.5$\sigma$ confidence, referred to as a peak catalog. 
Because the peak catalog should include a large number of spurious sources, after confirming that the noise remains Gaussian and centered at zero flux, 
we also conduct negative peak analysis \citep[e.g.,][]{hatsukade2013,ono2014,carniani2015,fujimoto2016} to evaluate the expected number of real sources. 
We create inverted maps by multiplying $-1$ to both the natural and tapered maps, run {\sc sextractor} again, and extract sources with negative peak counts above 2.5$\sigma$ level. 
In both original and inverted maps, we select sources that are identified in high-sensitivity regions, 
\tcb{where the relative sensitivity to the deepest part of the mosaic is greater than 30\%.} 

In Figure \ref{fig:pn_hist}, we show histograms of the positive and negative sources in our ALMA maps from 33 ALCS fields as a function of the signal-to-noise ratio (SNR) at the peak count.
Note that this differs from the histogram of all pixels in the map, as an island of the emission is counted as a single source. 
In these histograms, the excess of the positive to the negative sources indicates the expected number of the real sources, known as the purity \citep[e.g.,][]{gonzalez2020,gomez2021}, defined as
\begin{eqnarray}
\label{eq:purity}
p = \frac{N_{\rm pos}-N_{\rm neg}}{N_{\rm pos}}, 
\end{eqnarray}
where $N_{\rm pos}$ and $N_{\rm neg}$ represent the number of positive and negative sources at a given SNR, respectively.  
We find that $p$ shows $>$ 0.99 at SNR $\geq5.0$  and $\geq4.5$ in the natural (SNR$_{\rm nat}$) and tapered maps (SNR$_{\rm tap}$), respectively. 
The difference is explained by less areas relative to the beam size in the tapered maps, where the significant count caused by the noise fluctuation should be reduced. 
We also find that $p$ maintains an excess above zero down to SNR$_{\rm nat}$ = 4.0, 
indicating that a certain number of sources are likely real down to SNR$_{\rm nat}$ = 4.0.    
Based on these procedures, we obtain a source candidate catalog consisting of 399 objects  with SNR$_{\rm nat}$ $\geq$ 4.0 from the peak catalog. 
In this candidate catalog, the number of the real sources down to SNR$_{\rm nat}$ $= 4.0$ is expected to be 177$^{+12}_{-14}$ (where the uncertainty is calculated from Poisson statistics) by summing the excess of the positive to the negative sources 
down to the SNR limit.

\subsection{Catalog}
\label{sec:catalog}

We analyze the source candidate catalog consisting of the 399 objects to construct a reliable source catalog. Because the positive and negative source histograms of Figure \ref{fig:pn_hist} show a rapid decrease of purity below SNR$_{\rm nat}$ = 5.0 and SNR$_{\rm tap}$ = 4.5, 
we make two catalogs from blind and prior-based approaches. 
First, in the blind approach, we adopt thresholds of SNR$_{\rm nat}$ $\geq$ 5.0 or SNR$_{\rm tap}$ $\geq$ 4.5 in the source candidate catalog, yielding 141 ALMA sources that achieve a purity of $p>0.99$. 
We refer to these 141 sources as the blind catalog. 
Second, in the prior-based approach, we perform source extraction in {\it Spitzer}/IRAC channel 2 maps and retrieve real sources from the source candidates with SNR$_{\rm nat}$ down to 4.0 by cross-matching them with the IRAC sources. 
This is because, both in previous ALMA studies and in the ALCS, it is found that the vast majority ($\simeq93$\%) of real ALMA sources (e.g., from the blind sample with $p=1.0$) have IRAC ch2 counterparts.
We use secure IRAC sources with \tcb{SNR$_{\rm IRAC}$} $\geq5$, which corresponds to a typical detection limit of $\sim$23 mag among the IRAC maps taken in the ALCS clusters. 
The details of the IRAC data reduction and related source extraction are described in \cite{kokorev2022} and \cite{sun2022}, respectively. 
We associate 40 ALMA sources with \tcb{SNR$_{\rm nat}$} = 4--5 with IRAC counterparts.  
Among these, the probability of the chance projection is estimated to be 4.6\% (0.05\%) at the offset of $1\farcs0$ ($0\farcs1$) \citep{downes1986} based on the typical number density in the IRAC ch2 map of 0.015 arcsec$^{-2}$ \citep{sun2021}. 
More precisely, we calculate the probability for each 40 ALMA source according to its spatial offset from the associated IRAC source and obtain the expected number of chance projections to be $\sim$1 among the 40 ALMA sources. 
In fact, from visual inspection, we find that one source is likely caused by the chance projection, and we do not include the source in the following analysis. 
We refer to the remaining 39 sources as the prior sample.  
From both approaches and the visual inspection, a total of 180 ($=141+39$) ALMA sources are identified as reliable sources, which is in excellent agreement with the expected number of the real sources of 177$^{+12}_{-14}$ estimated from the excess of the positive to negative sources down to our SNR limit (Section \ref{sec:source_ext}). 
We list these 180 ALMA sources in Table \ref{tab:alcs_catalog}, referred to as ALCS continuum source catalog. 

\setlength{\tabcolsep}{3pt}
\startlongtable
\begin{deluxetable*}{lllccccccc}
\tablecaption{Example of ALCS Continuum Sources}
\tablehead{
\colhead{ID} & \colhead{R.A.} & \colhead{Dec}   & \colhead{SNR$_{\rm nat}$} (SNR$_{\rm tap}$) & \colhead{PB} & \colhead{$S_{\rm nat}$} & \colhead{$S_{\rm tap}$}  &  \colhead{$S_{\rm ap}$} & \colhead{$S_{\rm imf}$} & \colhead{flag} \\
  \colhead{}  & \colhead{(deg)}   & \colhead{(deg)}  &  \colhead{}    &  \colhead{}    & \colhead{(mJy/beam)}&  \colhead{(mJy/beam)} &   \colhead{(mJy)}    &  \colhead{(mJy)}     &  \colhead{}  \\
\colhead{(1)} & \multicolumn{2}{c}{(2)} & \colhead{(3)} & \colhead{(4)} & \colhead{(5)} & \colhead{(6)} & \colhead{(7)} & \colhead{(8)} & \colhead{(9)}}
\startdata
\multicolumn{10}{c}{Blind Catalog ($N=141$)} \\   \hline
AC0102-C11 & 15.7477863 &$ -49.2904001 $& 5.4 (3.1) & 0.96 & 0.39 $\pm$ 0.07 & 0.34 $\pm$ 0.11 & 0.28 $\pm$ 0.12 & 0.35 $\pm$ 0.1 &$ 0 $\\
AC0102-C22 & 15.7626821 &$ -49.2864566 $& 7.9 (7.4) & 0.79 & 0.71 $\pm$ 0.09 & 0.97 $\pm$ 0.13 & 1.23 $\pm$ 0.14 & 0.96 $\pm$ 0.19 &$ 2 $\\
AC0102-C50 & 15.7475095 &$ -49.2821233 $& 5.8 (5.6) & 0.99 & 0.41 $\pm$ 0.07 & 0.58 $\pm$ 0.1 & 0.66 $\pm$ 0.11 & 0.67 $\pm$ 0.16 &$ 2 $\\
AC0102-C52 & 15.7561196 &$ -49.2823856 $& 5.8 (3.2) & 0.99 & 0.41 $\pm$ 0.07 & 0.33 $\pm$ 0.11 & 0.04 $\pm$ 0.11 & 0.4 $\pm$ 0.12 &$ 0 $\\
AC0102-C118 & 15.7128798 &$ -49.2607919 $& 34.3 (26.5) & 0.92 & 2.63 $\pm$ 0.08 & 2.99 $\pm$ 0.11 & 3.29 $\pm$ 0.12 & 2.93 $\pm$ 0.14 &$ 2 $\\
AC0102-C160 & 15.7508527 &$ -49.2677044 $& 4.7 (4.9) & 0.98 & 0.34 $\pm$ 0.07 & 0.52 $\pm$ 0.11 & 0.72 $\pm$ 0.12 & 0.87 $\pm$ 0.23 &$ 2 $\\
AC0102-C215 & 15.7288441 &$ -49.2540865 $& 42.4 (32.7) & 0.99 & 3.0 $\pm$ 0.07 & 3.4 $\pm$ 0.1 & 3.83 $\pm$ 0.11 & 3.6 $\pm$ 0.18 &$ 2 $\\
AC0102-C223 & 15.7051647 &$ -49.2524655 $& 7.9 (6.7) & 0.96 & 0.58 $\pm$ 0.07 & 0.72 $\pm$ 0.11 & 1.01 $\pm$ 0.12 & 0.78 $\pm$ 0.26 &$ 2 $\\
AC0102-C224 & 15.7320137 &$ -49.2525191 $& 91.8 (78.4) & 0.98 & 6.57 $\pm$ 0.07 & 8.24 $\pm$ 0.11 & 9.5 $\pm$ 0.12 & 8.99 $\pm$ 0.26 &$ 2 $\\
AC0102-C241 & 15.742206 &$ -49.2489458 $& 5.4 (4.5) & 0.45 & 0.85 $\pm$ 0.16 & 1.05 $\pm$ 0.23 & 1.08 $\pm$ 0.25 & 0.53 $\pm$ 0.16 &$ 1 $\\
AC0102-C251 & 15.7294397 &$ -49.2386615 $& 5.1 (4.4) & 0.63 & 0.56 $\pm$ 0.11 & 0.72 $\pm$ 0.16 & 0.97 $\pm$ 0.18 & 0.75 $\pm$ 0.24 &$ 2 $\\
AC0102-C276 & 15.7053793 &$ -49.2439317 $& 9.9 (8.6) & 0.81 & 0.86 $\pm$ 0.09 & 1.09 $\pm$ 0.13 & 1.41 $\pm$ 0.14 & 1.16 $\pm$ 0.18 &$ 2 $\\
AC0102-C294 & 15.7056994 &$ -49.2514717 $& 14.6 (19.1) & 0.97 & 1.06 $\pm$ 0.07 & 2.03 $\pm$ 0.11 & 3.1 $\pm$ 0.12 & 3.16 $\pm$ 0.27 &$ 2 $\\
M0417-C46 & 64.3888354 &$ -11.9175841 $& 48.0 (36.7) & 0.97 & 4.04 $\pm$ 0.08 & 4.26 $\pm$ 0.12 & 4.29 $\pm$ 0.12 & 4.3 $\pm$ 0.13 &$ 1 $\\
M0417-C49 & 64.4172807 &$ -11.9168995 $& 36.1 (28.6) & 0.71 & 4.18 $\pm$ 0.12 & 4.57 $\pm$ 0.16 & 4.61 $\pm$ 0.16 & 3.32 $\pm$ 0.17 &$ 2 $\\
M0417-C58 & 64.3985559 &$ -11.9147976 $& 30.4 (24.3) & 0.95 & 2.63 $\pm$ 0.09 & 2.9 $\pm$ 0.12 & 3.18 $\pm$ 0.12 & 2.89 $\pm$ 0.14 &$ 2 $\\
M0417-C121 & 64.4042424 &$ -11.9055111 $& 20.7 (16.3) & 0.89 & 1.91 $\pm$ 0.09 & 2.07 $\pm$ 0.13 & 2.14 $\pm$ 0.13 & 1.93 $\pm$ 0.14 &$ 2 $\\
\enddata
\tablecomments{
The full list is presented in Table \ref{tab:alcs_catalog_full}. 
(1) Source ID, including the prefix of the cluster name + ``-C''. 
(2) Source coordinate of the continuum peak in the natural-weighted map.
(3) Signal-to-noise ratio of the peak pixel in the natural-weighted map. 
(4) Primary beam sensitivity in the mosaic map. 
(5) Source flux density at 1.2 mm measured by the peak pixel count in the natural-weighted map.  
(6) Source flux density at 1.2 mm measured by the peak pixel count in the $uv$-tapered ($2''\times2''$) map. 
(7) Source flux density at 1.2 mm measured by a $2.0''$-radius aperture in the natural-weighted map.  
(8) Source flux density at 1.2 mm measured by 2D elliptical Gaussian fitting with CASA {\sc imfit} with the natural maps.
(9) Flag of the {\sc imfit} output (0: point source, 1: marginally resolved \tcb{(i.e., resolved in only in one direction)}, 2: fully resolved, $-$1: fitting error, $-2$: near the edge of the map). 
}
\label{tab:alcs_catalog}
\end{deluxetable*}

\subsection{Flux Measurement}
\label{sec:flux}

%%%%%%%%%%%%%%%%%%%%%%%
\begin{figure*}[t!]
\includegraphics[trim=0cm 0cm 0cm 0cm, clip, angle=0,width=1.0\textwidth]{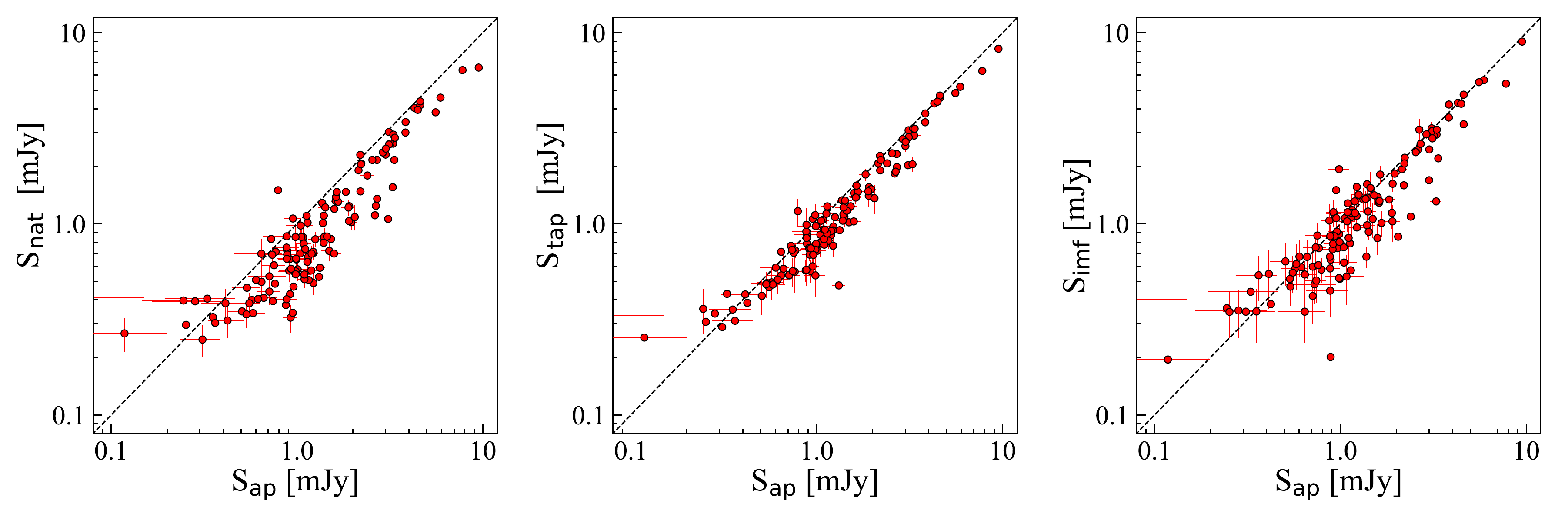}
 \caption{
Comparison between our various photometric methods described in Section \ref{sec:flux} for SNR $\geq5$ sources. 
The red dots are our measurements and the dashed line is the one-to-one relation. 
%\textbf{Left}, middle and left panels}: 
From left to right, we show the comparison between the $2\farcs0$-radius aperture flux ($S_{\rm ap}$) and the peak flux in the natural-weighted map ($S_{\rm nat}$), the peak flux in the $uv$-tapered map ($S_{\rm tap}$), and the 2D-fit flux with {\sc imfit} ($S_{\rm imf}$).  
%The mean flux ratio and its standard deviation are shown in the panels.  
\label{fig:flux_comp}}
\end{figure*}
%%%%%%%%%%%%%%%%%%%%%%% 

We estimate the flux density for our 180 ALCS sources in several different ways. 
The peak count in the natural maps ($S_{\rm nat}$) is obtained from the source extraction procedure (Section \ref{sec:source_ext}). 
When the source is unresolved, the peak count (Jy/beam) should equal the total flux density. 
However, as ALCS sources may be spatially resolved, we further measure the flux density in the following three methods.  
In the first method, we measure the peak count in the tapered map ($S_{\rm tap}$). 
Here we regard a peak above 3$\sigma$ level in the $uv$-tapered map identified within a radius of $1\farcs0$ from the peak in the natural map as the corresponding peak count of the source in the tapered map. If we do not identify any peaks above 3$\sigma$ level within the $1\farcs0$ radius, we use the count in the tapered map at the peak pixel position in the natural map. 
In the second method, we measure an enclosed flux density with an aperture radius of $2\farcs0$ with the natural map ($S_{\rm ap}$). 
We analyze the enclosed flux density growth curve as a function of aperture radius for our ALCS sources to confirm that the $2\farcs0$ radius sufficiently exceeds the peak of the growth curve for the majority of our ALCS sources. 
In the third method, we measure a spatially integrated flux density with the 2D elliptical Gaussian fitting with the CASA task {\sc imfit} ($S_{\rm imf}$). In the {\sc imfit} fitting, we use the peak count, position, and beam size in the natural map for the initial values, but all parameters are ultimately fitted as free parameters. We limit the fitting area to $5''\times5''$. 
Table \ref{tab:alcs_catalog} summarizes all flux density measurements described above for our ALCS sources.

In Figure \ref{fig:flux_comp}, we compare the flux density measurements.  
To understand secure trends, we only show the ALCS sources with SNR$_{\rm nat}$ $\geq$ 5.  
We find that the peak count measurement in the natural map is generally lower than the $2\farcs0$-aperture measurement. 
This trend implies that most of the ALCS sources are spatially resolved.
Given that the typical beam size of $\sim1\farcs0$ in our natural maps is sufficiently large for the emission from high-redshift galaxies \tcb{(FWHM$\gtrsim0\farcs2$--$0\farcs3$; see \citealt[e.g.,][]{ikarashi2015, simpson2015a, hodge2016, fujimoto2017, fujimoto2018, gonzalez2017, gomez2021})}, 
this trend may at least be partially related to elongation from gravitational lensing. 
We also find no significant difference among the latter three measurements. 
To obtain secure results, we adopt $S_{\rm imf}$ for the source with SNR$_{\rm nat}\geq10$, but with no fitting errors in {\sc imfit} (e.g., fitting not converged, significant residual). 
We use $S_{\rm ap}$ for the remaining sources, while we adopt $S_{\rm tap}$ or $S_{\rm nat}$ in several cases if the aperture encloses the emission from nearby sources or $S_{\rm ap}$ shows negative due to the increased noise fluctuation in the aperture measurement. We choose $S_{\rm tap}$ or $S_{\rm nat}$, which provides a higher SNR than the other.

\subsection{Serendipitous Line Detection}
\label{sec:line}

The 15-GHz frequency coverage of the ALCS data sets sometimes detects emission lines from the sources serendipitously, which could affect the continuum flux measurements. 
To investigate this possibility, we extract the 15-GHz spectra with $1\farcs0$-diamter aperture for our ALCS sources in the two ALCS cubes with velocity channel widths of 30 km~s$^{-1}$ and 60 km~s$^{-1}$, respectively. 
We systematically calculate integrated fluxes with integration velocity ranges of 120--1200 km~s$^{-1}$ for all channels, where the contribution of the continuum is subtracted by determining it with the median of the spectrum.  
We identify 13 line emitters whose integrated fluxes exceed the $5\sigma$ level in both velocity-channel spectra. 
Among these is the strong \cii\ line emitter at $z=6.0719$ reported in recent ALCS studies \citep{fujimoto2021,laporte2021}. 
We evaluate the equivalent width (EW) of these emission lines and find they contribute to the 15-GHz width continuum flux measurement at the $\sim$1--12\% level. 
We summarize the line properties and their contributions to the continuum flux density in \tcb{Appendix Table \ref{tab:line_prop}}. 
We correct the continuum flux measurements for the 15 ALCS sources with detected line emission by subtracting these line contributions in the following analyses.

\begin{figure}%[h]
\includegraphics[trim=0cm 0cm 0cm 0cm, clip, angle=0,width=0.5\textwidth]{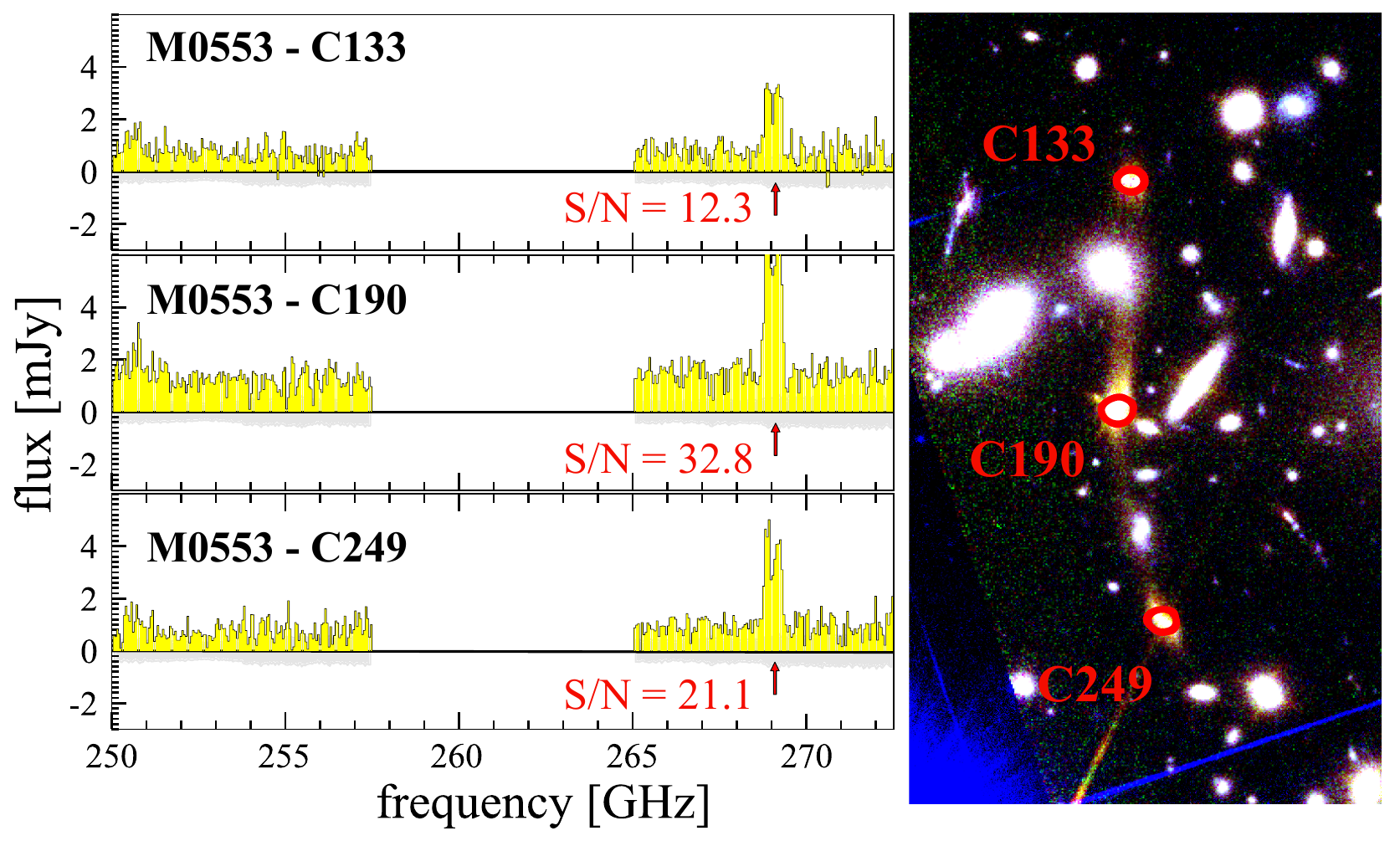}
 \caption{
Example of the line emitters serendipitously detected among the ALCS continuum sources. 
We evaluate the EW and subtract the line contributions to the continuum flux measurement in our analysis (Section \ref{sec:line}).  
{\it Left}: 
ALCS 15-GHz wide spectrum in ALMA Band~6. 
Lines are clearly detected at $\sim$269~GHz among three line emitters identified in the M0553 cluster. 
{\it Right}: \hst\ color image for M0553-C133/190/249, using F814W (blue), F125W (green), and F160W (red) filters.  
Red contours show the line intensity drawn at 6$\sigma$, 8$\sigma$, and 10$\sigma$. 
The counterparts of the line emitters are a triplet multiple-imaged system spectroscopically confirmed at $z=1.14$ \citep{ebeling2017}, concluding that the mm-line is CO(5-4). 
\label{fig:line_eg}}
\end{figure}

In Figure \ref{fig:line_eg}, we show three of these line emitters detected in the M0553 field as an example. 
The left panel presents the 15-GHz spectra, where the three lines are consistently detected at $\sim269$ GHz. 
The right panel shows an {\it HST} color map overlaid with the line intensity in red contours. 
The counterparts of these three-line emitters are triplet multiple images spectroscopically confirmed at $z=1.14$ \citep{ebeling2017}, such that the mm line corresponds to CO(5-4). 
The line properties and spectra of the remaining line candidates are summarized in Appendix \ref{sec:appendix_line}. 
In addition to lines associated with ALCS continuum source positions, we also identify dozens of "blind" line candidates using a blind line search algorithm \citep[e.g.,][]{fujimoto2021}, which will be presented in a separate paper.

%%%%%%%%%%%%%%%%%%%%%%%
%%%%%%%%%%%%%%%%%%%%%%% 

\subsection{Source Redshift}
\label{sec:redshift}

We estimate the source redshift for our ALCS sources. 
\tcb{We adopt five categories for our redshift estimates and describe the details of those categories in the following subsections.}

\subsubsection{\tcb{Spectroscopic redshift}}

\tcb{The first category 1) is spectroscopic redshift ($z_{\rm spec}$). 
We cross-match the ALCS sources with spectroscopic catalogs in the literature \citep{biviano2013, caputi2021, caminha2019, cerny2018, mirka2017,ebeling2009, ebeling2017, fujimoto2021, gomez2012, kokorev2022, richard2021, munoz-arancibia2022, sand2004, schmidt2014, soucail1999, treu2015, wang2015, wold2012},  
identifying 51 ALCS sources in this category.
} 
We also newly determine $z_{\rm spec}$ for 8 sources from the serendipitous line detections in 13 ALCS sources (Section \ref{sec:line}). 
In addition, single or multiple lines have been successfully detected in line scan follow-up observations, and we further obtain $z_{\rm spec}$ for another 8 sources. 
For all sources with only single-line detections, we adopt the most plausible line associated with their $z_{\rm phot}$ distribution function. 
\tcb{In total, 67 (= 51+8+8) sources have been regarded as $z_{\rm spec}$ and categorised as this category.}
In Appendix \ref{sec:appendix_line}, we individually describe the process of the redshift determination for these line-detected sources. 

\subsubsection{\tcb{Photometric redshift}}

\tcb{Because of the difficulty of the heterogeneous availability of optical--NIR data among our ALCS sources (e.g., sources fall outside \hst\ footprints), we divide the photometric redshift ($z_{\rm phot}$) further into four categories (2, 3, 4, 5). As a summary, 
\begin{itemize}
\item 2) 7 sources with robust multiple images that refine $z_{\rm photo}$ from their spatial configurations with the lens model. 
\item 3) 80 sources with $z_{\rm photo}$ from \cite{kokorev2022}, constrained by the optical--NIR spectral energy distribution (SED) analysis for the sources detected in both \hst\ and IRAC. 
\item 4) 18 and 4 sources that are not detected in \hst\ and fall outside the \hst\ footprints, respectively, and whose $z_{\rm phot}$ are constrained by a SED template fitting. 
\item 5) 4 sources have no \hst\ nor IRAC, so $z_{\rm photo}=3.0\pm2.0$ is assumed.
\end{itemize}
}

For 2), we cross-match the ALCS sources with catalogs of the multiple images used in constructing the lens models in the literature. 
We identify \tcb{7} ALCS sources in this category. 
Here we adopt the same redshift uncertainty of $\Delta z=0.5$ used to constrain the lens models. 

For 3), 
We cross-match them with the {\it HST}+IRAC catalog of the ALCS 33 fields \citep{kokorev2022}. 
We identify 80 ALCS sources cross-matched with the {\it HST}+IRAC catalog
and adopt the $z_{\rm phot}$ estimate obtained from the SED fitting using the {\sc eazy} code \citep{brammer2008}.
This catalog uses a template set consisting of 12 templates derived from the Flexible Stellar Population Synthesis (FSPS) library \citep{conroy2009, conroy2010}, which enables to reproduce a much larger library spanning a range of dust attenuation, ages, mass-to-light ratios, and realistic star-formation histories (e.g., bursty, slowly rising, slowly falling).
The fitting details are presented in \cite{kokorev2022}. 
Because the mosaics for all available {\it HST} filters are combined and used for the detection image in this {\it HST}+IRAC catalog, the resulting cross-matches indicate that these 80 ALCS sources have counterparts in one or more {\it HST} band(s). 

For 4), 
Of the remaining sources, 18 lack counterparts in {\it HST}/WFC3 bands to faint limits, 
and are considered so-called NIR-dark (or OIR-dark) objects \citep[e.g.,][]{fujimoto2016, franco2018, twang2019, yamaguchi2019, williams2019, casey2019, romano2020, fudamoto2021, gomez2021, talia2021, fujimoto2022, xiao2022, manning2022,barrufet2023a,guilietti2023}, 
while the sky positions for the remaining 8 fall outside the {\it HST}/WFC3 footprints. 
For these 26 sources, 
we extract IRAC photometry using a $3\farcs0$-diameter aperture forced at the ALMA continuum positions. To mitigate contamination from nearby sources, we use the residual IRAC maps, whereby sources detected in the HST maps are modeled with the IRAC PSF and subtracted \citep{kokorev2022}. 
Given the limited optical--NIR information, we also include {\it Herschel} photometry obtained in \cite{sun2022} and evaluate $P(z)$ for these sources by calculating $\chi^{2}$ with a composite SED model obtained from 707 high-redshift dusty star-forming galaxies \citep{dudzeviciute2020}. 
Based on these procedures, the redshift estimates fall within a range of 0.45--4.25 for the remaining ALCS sources lacking {\it HST} counterparts/data. 
In Appendix \ref{sec:app_znir}, we show an example of $P(z)$ estimate with the composite SED model for these unique sources, based on upper limits along with our ALMA detection.

For 5), among the 26 sources in the category of 4), four blind-sample sources remain undetected ($<$ 2$\sigma$--3$\sigma$) in both {\it Herschel} photometry and the forced-aperture photometry in the residual IRAC maps, despite their secure ALMA detections. 
These sources are likely either associated  with a very faint dusty system at $z\sim1$--2 \citep[e.g.,][]{aravena2020}, or 
a significantly dust-attenuated system at $z\sim3$--5 \citep[e.g.,][]{umehata2020, smail2021}, or very high-redshift dusty system at $z>7$ \citep{fudamoto2021, fujimoto2022}. 
Based on these possibilities, we assume source redshifts of $z=3.0\pm2.0$ for these four sources in the following analyses. 
We list our redshift estimate of our ALCS sources in Table \ref{tab:alcs_catalog2}.

\setlength{\tabcolsep}{2pt}
\startlongtable
\begin{deluxetable*}{lcccccccc}
\tablecaption{Example of ALCS Continuum Source Properties}
\tablehead{
\colhead{ID}  & \colhead{$z_{\rm spec}$} & \colhead{$z_{\rm phot}$} & \colhead{$S_{\rm obs}$} & \colhead{$L_{\rm IR}$} & \colhead{$\mu$}  & \colhead{$z$ type} & \colhead{Ref.} & \colhead{Note} \\
\colhead{}    &   \colhead{}         &  \colhead{}               &  \colhead{(mJy)}           & \colhead{($L_{\odot}$)}        & \colhead{}    & \colhead{}       & \colhead{}     & \colhead{}     \\
\colhead{(1)}  & \colhead{(2)}  & \colhead{(3)} & \colhead{(4)} & \colhead{(5)} & \colhead{(6)}      & \colhead{(7)}  & \colhead{} & \colhead{} }
\startdata
\multicolumn{9}{c}{Primary Catalog ($N=125^{\dagger}$)} \\   \hline
AC0102-C11 &    \nodata         & $ (3.0\pm2.0) $ & $ 0.30\pm 0.10 $ & $ 12.00\pm0.19 $ & 1.6$_{-0.6}^{+0.7}$ & 5 &  \nodata     & No counterparts in {\it HST}, IRAC, {\it Herschel}\\
AC0102-C22 &    \nodata         & $ 3.45^{+0.17}_{-0.17}$ & $ 1.11\pm0.13 $ & $ 12.53\pm0.19 $ & 2.1$_{-0.8}^{+0.8}$ & 4 & App.\ref{sec:app_znir}  & No counterparts in {\it HST} \\
AC0102-C50 &    \nodata          & $ 2.38^{+0.10}_{-0.11}$ & $ 0.60\pm0.10 $ & $ 12.85\pm0.07 $ & 2.4$_{-1.0}^{+1.0}$ & 3 & K22 &  \\
AC0102-C52 &    \nodata          & $ 2.02^{+0.35}_{-0.16}$ & $ 0.30\pm0.09 $ & $ 12.36\pm0.11 $ & 2.3$_{-0.9}^{+1.0}$ & 3 & K22 &  \\
AC0102-C160 &    \nodata          & $ 1.99^{+0.18}_{-0.14}$ & $ 0.65\pm0.10 $ & $ 12.61\pm0.13 $ & 2.3$_{-0.9}^{+0.9}$ & 3 & K22 &  \\
AC0102-C224 & 4.320 &  \nodata & $ 3.19\pm0.15 $ & $13.07\pm0.14$ & 5.1$_{-3.6}^{+3.6}$ & 1  & C21 & multiple images (AC0102-C118/215/224)  \\
AC0102-C241 &    \nodata          & $ 2.82^{+0.05}_{-0.67}$ & $ 0.97\pm0.23 $ & $ 12.70\pm0.14 $ & 2.0$_{-0.8}^{+0.8}$ & 3 & K22 &  \\
AC0102-C251 &    \nodata         & $ (3.0\pm2.0) $ & $ 0.90\pm 0.16 $ & $ 11.84\pm0.51 $ & 1.0$_{-0.0}^{+0.4}$ & 5 &  \nodata     & No counterparts in {\it HST}, IRAC, {\it Herschel}\\
AC0102-C276 &    \nodata          & $ 3.46^{+0.30}_{-0.43}$ & $ 1.41\pm0.14 $ & $ 12.95\pm0.10 $ & 5.2$_{-2.7}^{+2.6}$ & 3 & K22 &  \\
AC0102-C294 &    \nodata         & $ 4.0^{+0.5}_{-0.5}$ & $ 2.12\pm0.28 $ & $ 12.54\pm0.13 $ & 3.7$_{-2.1}^{+2.1}$ & 2 & Ce18 & multiple images  (AC0102-C223/294)  \\
M0417-C49 &    \nodata          & $ 3.16^{+0.10}_{-0.10}$ & $ 4.69\pm0.24 $ & $ 13.19\pm0.25 $ & 2.0$_{-0.8}^{+0.8}$ & 3 & K22 &  \\
M0417-C121 & 3.652 &  \nodata & $ 3.05\pm0.15 $ & $ 13.09\pm0.14 $ & 4.0$_{-2.3}^{+2.3}$ & 1  & App.\ref{sec:appendix_line} & multiple images (M0417-C46/58/121)  \\
M0417-C204 &    \nodata         & $ 1.80^{+0.22}_{-0.22}$ & $ 0.94\pm0.13 $ & $ 12.57\pm0.16 $ & 2.8$_{-1.2}^{+1.1}$ & 4 & App.\ref{sec:app_znir} & Outside of {\it HST}/WFC3 \\
M0417-C218 &    \nodata         & $ 2.00^{+0.07}_{-0.07}$ & $ 1.42\pm0.18 $ & $ 12.76\pm0.29 $ & 4.8$_{-1.9}^{+1.9}$ & 4 & App.\ref{sec:app_znir}  & No counterparts in {\it HST} \\
\hline
\enddata
\tablecomments{
The full list is presented in Table \ref{tab:alcs_catalog2_full}. 
(1) ALCS continuum source ID. 
(2) Spectroscopic redshift (Section \ref{sec:redshift}). 
(3) Photometric redshift (Section \ref{sec:redshift}) 
(4) Observed flux density, including the corrections for the primary beam, 
Eddington bias (Section \ref{sec:simulation}), and the line contamination (Section \ref{sec:line}), 
\tcb{but without the lens correction.}
The errors in the flux measurement and corrections are propagated.  
For the multiple-image system, we show an observed flux density calculated by multiplying the average values of the intrinsic flux densities and magnifications. 
(5) Magnification factor based on our fidcual lens model, where the error is evaluated from the propagation of the redshift uncertainty and the systematic uncertainty from different lens models (Section \ref{sec:lens}).  
(6) Redshift category.    
1: $z_{\rm spec}$, 
2: $z_{\rm phot}$ + lens model constraint from the multiple image positions in the literature, 
3: $z_{\rm phot}$ with {\sc eazy}, 
4: $z_{\rm phot}$ based on the composite SED from 707 dusty galaxies in AS2UDS for the sources without {\it HST} counterparts/data (Appendix \ref{sec:app_znir}), 
5: $z=3.0\pm2.0$ assumed for four sources with no counterparts in all bands other than ALMA).  
(7): Reference: 
C21 \citep{caputi2021},
Ce18 \citep{cerny2018}, 
K22 \citep{kokorev2022}, 
and App.~\ref{sec:appendix_line} (Appendix~\ref{sec:appendix_line}). \\
\tcb{$\dagger$ This is after correcting the lensing effect for the multiply imaged sources, resulting in a slightly lower number than that in Table~\ref{tab:alcs_catalog}. }
}
\label{tab:alcs_catalog2}
\end{deluxetable*}

\subsection{Lens Model and Magnification Correction}
\label{sec:lens}

We construct lens models for the ALCS 33 lensing clusters. 
We select cluster member galaxies as well as multiple images behind clusters based on the photometric redshift, colors, and morphology of galaxies in the {\it HST} images and apply them to independent algorithms including {\sc glafic} \citep{oguri2010,kawamata2016, kawamata2018, okabe2020}, {\sc Lenstool} \citep{jullo2007, caminha2016, caminha2017, caminha2017b, caminha2019, richard2014, niemiec2020}, Light-Traces-Mass (Zitrin-LTM; \citealt{zitrin2013,zitrin2015}), and Pseudo-Isothermal Elliptical Mass Distribution plus elliptical Navarro–Frenk–White (Zitrin-dPIEeNFW; \citealt{zitrin2013,zitrin2015}). 
In Appendix \ref{sec:app_model_all}, we summarize the number of multiple images, the accuracy of the model predictions of the multiple images, and lens models available in each of the ALCS 33 lensing clusters. 
In this paper, we adopt the lens model of {\sc glafic} as a fiducial model for our analyses, although we also use other available models to evaluate uncertainties in magnification factors. 
The lens model of {\sc glafic} is constructed in the same manner as \cite{kawamata2016}, wherein interested readers can find more specific lens modeling procedures using {\sc glafic}.

We apply lensing corrections to all ALCS sources located behind their respective clusters, defined as sources with photometric (spectroscopic) redshift estimates exceeding the cluster redshift by more than 0.2 (0.1). 
Among the ALCS sample, six groups of the sources have been spectroscopically confirmed as multiple images. 
Based on source positions, lens model predictions, and similar SED properties, we also classify ACT0102-C223/294 and R0032-C127/131/198 as multiple images at $z=4.0\pm0.5$ and $z_{\rm spec}=2.391$, respectively. 
After removing the cluster member galaxies and correcting the counts for the multiple images, we identify 146 out of 180 ALCS sources in the blind+prior sample as the unique sources behind the clusters. 
We then evaluate the magnification factor based on the peak pixel position of the ALMA continuum emission and the redshift estimate. 
We then calculate the intrinsic flux density by dividing the observed flux density by this magnification factor.  
For the multiply imaged systems, we adopt the average value of the intrinsic flux densities among the multiple images.
Table \ref{tab:alcs_catalog2} includes these 146 ALCS sources and their intrinsic flux density after the lens correction. 
The median magnification is 2.70 for the unique ALCS sources behind the clusters. 

To evaluate the systematic uncertainty of the lens model, we calculate $\Delta\mu/\mu$, where $\mu$ is the magnification factor from the fiducial model and $\Delta\mu = |\mu_{\rm other} - \mu|$ is the difference between magnification factors of the fiducial and other available lens models ($\mu_{\rm other}$) at random positions in the cluster field. 
Because the accuracy of the lens model generally depends on the richness of the multi-wavelength data in the cluster field for identifying the multiple images and the cluster member galaxies, we separately evaluate the $\Delta\mu/\mu$ values for HFF, CLASH, and RELICS clusters. 
In Appendix \ref{sec:app_models}, we compare the magnification factors between our fiducial and other lens models at random pixel positions of the maps. 
We find a trend of increasing $\Delta\mu/\mu$ towards high $\mu$ from $\sim$20--40\% at $\mu=3$ to $\sim$50--60\% at $\mu=10$, where the trends in CLASH and RELICS fields are almost comparable. Among the CLASH and RELICS clusters, the total number of multiple images is generally the same, while the fraction of spectroscopic redshifts among the multiple images in CLASH is higher(see Table~\ref{tab:model_all}). 
\tcb{The small differences between the CLASH and RELICS trends are consistent with the fact that the magnification uncertainty is more driven by the total number of multiple images rather than the spectroscopic redshift fraction \citep{johnson2016}.} 
We also confirm that the increasing trend of the magnification uncertainty is generally consistent with the previous HFF lens model comparison results \citep{mineghetti2017}, where $\Delta\mu/\mu$ is estimated to be $\sim10\%$ at $\mu=3$ and $\sim30\%$ at $\mu=10$. 
Based on our comparison results, we adopt 20\% (40\%), 50\%, 60\%, and 80\% for 
the systematic uncertainty in $\mu$ among the different lens models 
at $\mu\leq5$, $5<\mu\leq10$, $10<\mu\leq30$, and $\mu >30$ in HFF (CLASH \& RELICS).

We caution that choosing the median magnification from multiple lens models dismisses highly magnified sources. This is because even slight differences in the critical curve positions among different lens models will smear out the highly magnified regions when the median is sampled. 
In Appendix \ref{sec:app_models}, we also show the survey area by producing the median magnification map in AS1063 and confirm that the effective survey area is underestimated at $\mu\gtrsim10$ than the areas of each model individually. 
To retain the full range of the lensing magnification, we thus use the fiducial lens model and include the systematic uncertainties from the different models in the following analysis. 
We further discuss potential systematics in our results in Section \ref{sec:caveats}.

\subsection{Simulations for flux measurements and completeness}
\label{sec:simulation}

We perform Monte-Carlo (MC) simulations to investigate potential systematics in our flux density measurements due to Eddington bias or other unknown factors.  
First, we produce a pure noise mosaic map to inject artificial sources. 
In ALMA data cubes, multiplying every other channel by a factor of $-1$ will make the continuum emission disappear in the collapsed map.
We thus collapse the 60~km~s$^{-1}$ width cube after multiplying by ($-1$)$^{i}$ for every $i$-th channel and regard this as the noise map. 
Second, we create 1,000 artificial sources with uniform distributions in the total flux density and the source size for each ALCS cluster field. 
For the total flux density, we adopt a range of 2.5--80 times larger than $\sigma_{\rm cont}$, where $\sigma_{\rm cont}$ is the continuum data depth in each field. For the source size, potential correlations between the IR-emitting region size and the IR luminosity have been reported. However, it is still under whether the correlation is positive, negative, or null \citep[e.g.,][]{gonzalez2017, tadaki2018, fujimoto2017, fujimoto2018, smail2021}. 
Given the lack of a definitive conclusion, we fix the intrinsic source size at a typical value in the literature of $0\farcs15$ with a circularly symmetric Gaussian morphology. 
Third, we inject the artificial sources at random positions into the noise map and calculate the expected source distortion in the observed frame based on our fiducial lens model (Sec. \ref{sec:lens}), assuming a source redshift of $z=2$.  
Fourth, we perform mock observations for the artificial sources in the observed frame with CASA {\sc simobserve}. 
To obtain the same beam profile according to each ALCS cluster, we set the same configuration and sky coordinate as the observations for each ALCS cluster in {\sc simobserve} and produce the $uv$-visibility of the artificial source. 
We then create CLEANed maps of the artificial sources with the same CLEAN parameters as our ALCS maps and inject them at random positions in the noise map. 
We run these procedures for both the natural and tapered maps. 

%%%%%%%%%%%%%%%%%%%%%%%
\begin{figure}
\begin{center}
\includegraphics[trim=0cm 0cm 0cm 0cm, clip, angle=0,width=0.47\textwidth]{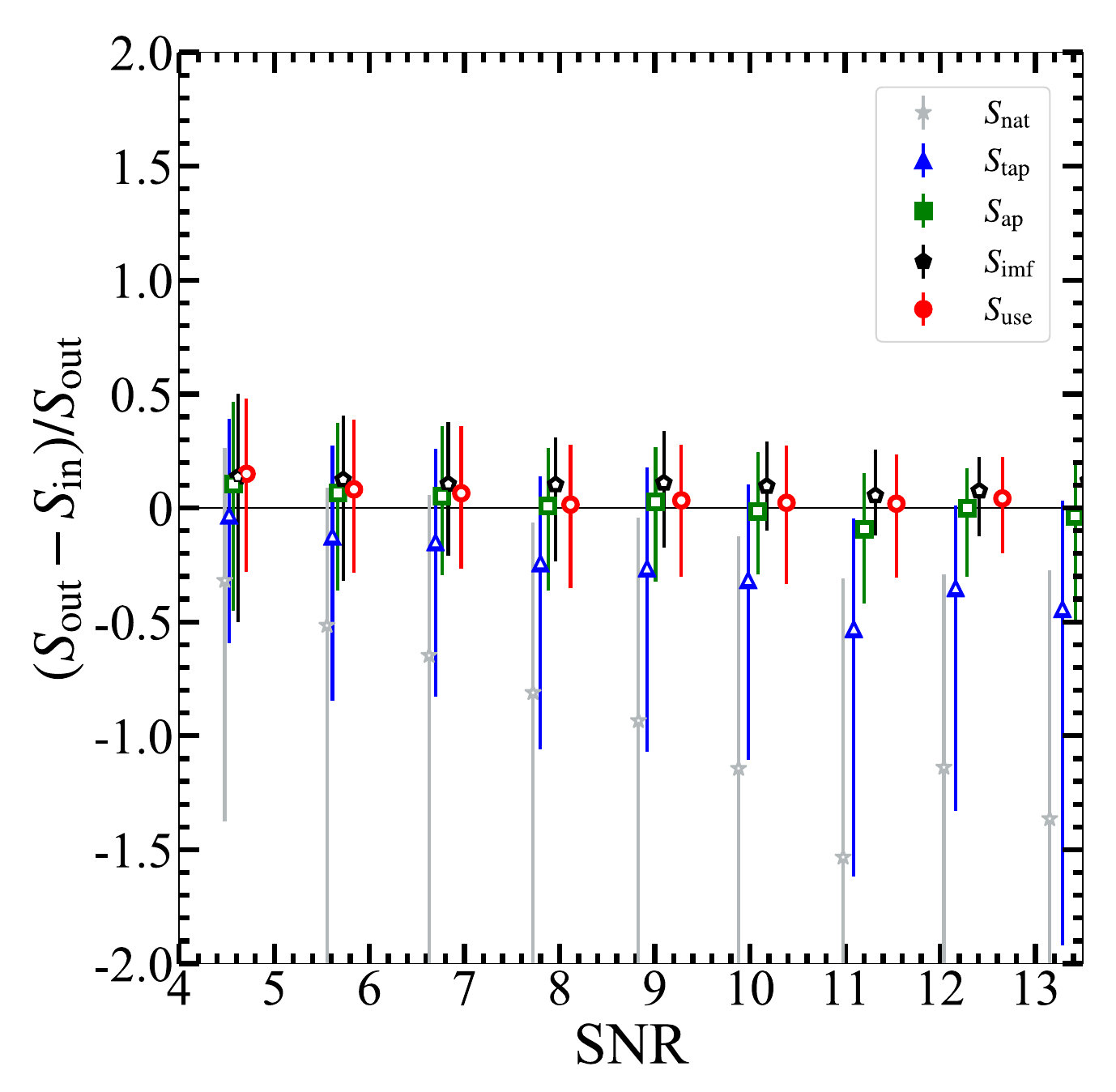}
\caption{
MC simulation results for input and output flux densities as a function of output SNR$_{\rm nat}$ to correct the effects of the flux boosting and Eddington bias. 
We compare the four types of flux measurements ($S_{\rm nat}$, $S_{\rm tap}$, $S_{\rm ap}$, and $S_{\rm imf}$) as well as the photometry finally used in our analysis ($S_{\rm use}$) (Section \ref{sec:flux}). 
We find that $S_{\rm use}$ deviates from the input value by $\sim10$\% below SNR$_{\rm nat}=8$, where we apply the correction of 10\% for our flux density measurement.  
\label{fig:Sout-in}}
\end{center}
\end{figure}
%%%%%%%%%%%%%%%%%%%%%%% 

In Figure \ref{fig:Sout-in}, we show the comparison between output and input flux densities as a function of SNR$_{\rm nat}$ from the MC simulations. For the output, we show the four different photometries of $S_{\rm nat}$, $S_{\rm tap}$, $S_{\rm ap}$, and $S_{\rm imf}$, in addition to the final photometry employed in those combinations, $S_{\rm use}$ (Section \ref{sec:flux}). The error bars represent the 16-84th percentile in the MC simulations. 
In general, $S_{\rm nat}$ and $S_{\rm tap}$ underestimate the input value, while $S_{\rm ap}$ and $S_{\rm imf}$ are consistent with the input value within the 1$\sigma$ range. 
We find that $S_{\rm use}$ is in excellent agreement with the input value down to SNR $=8$. Although $S_{\rm use}$ still agrees with the input value within the 1$\sigma$ range below SNR = 8, $S_{\rm use}$ overestimates the input value by $\sim10$\% probably due to the Eddington bias.   
We thus apply a correction to our flux density measurement for the sources with SNR $<$ 8 by 10\%. 

%%%%%%%%%%%%%%%%%%%%%%%
\begin{figure}
\begin{center}
\includegraphics[trim=0cm 0cm 0cm 0cm, clip, angle=0,width=0.47\textwidth]{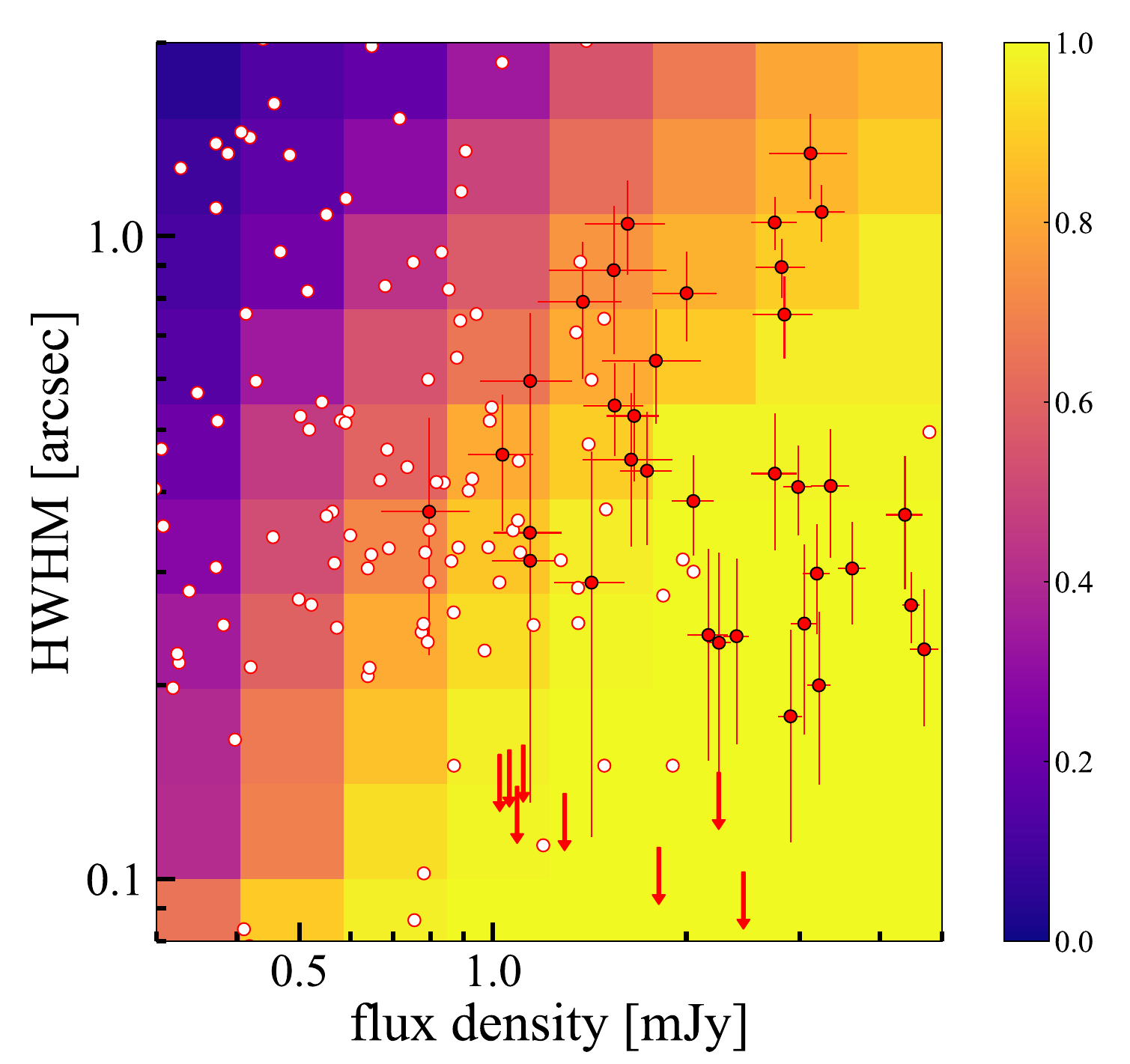}
\caption{
MC simulation results for completeness. 
The red circles and arrows present the spatial size in half width at half maximum (HWHM) and upper limit evaluated from CASA {\sc imfit}, respectively, for the ALCS sources \tcb{whose structure is well resolved with SNR$_{\rm nat}\geq10$}. 
The white circles show the remaining ALCS sources whose spatial sizes are estimated from the lensing distortion based on the source redshifts and our fiducial lens models by assuming the intrinsic effective radius of $0\farcs15$.
As an example, the background color scale denotes the completeness of our sample selection for the blind sample (SNR$_{\rm nat}\geq5.0$ or SNR$_{\rm tap}\geq4.5$; Section \ref{sec:source_ext}) based on the ALCS map taken in M2129 whose data depth and synthesized beam size are the typical value among our 33 ALCS lensing clusters (Tabe \ref{tab:data_prop}).
\label{fig:completeness}}
\end{center}
\end{figure}
%%%%%%%%%%%%%%%%%%%%%%% 

We also evaluate the completeness of our source detection, which is affected by the flux density and the spatial size of the sources in the observed frame. 
The background color in Figure \ref{fig:completeness} shows the completeness estimated with the output from the above MC simulations.  
We regard the sources recovered in the MC simulations if the artificial sources show a positive peak count within $1\farcs0$ ($\approx$beam size) from the injected position with SNR$_{\rm nat} \geq 5.0$ or SNR$_{\rm tap}\geq$ 4.5 for the blind sample and SNR$_{\rm nat} \geq 4.0$ for the blind+prior sample. 
For comparison, Figure \ref{fig:completeness} also shows our ALCS sources in the blind sample. 
\tcb{For the sources whose structure is well resolved with SNR$_{\rm nat}$ $\geq 10$ (red filled circles)}, the {\sc imfit} results are used for the source size estimate, but for the other sources (white circles), we use the expected source sizes via the lensing distortion according to the source position, our fiducial lens model, the source redshift, and the intrinsic source size assumption of $0\farcs15$. 
We find that the majority of our ALCS sources fall in the parameter space with the completeness of $> 50\%$, 
but that a few sources show very low completeness of $< 10\%$. 
Given the potential uncertainty in the completeness estimate, 
and in order not to make our results significantly affected by the few sources with very low completeness, we perform the following analyses by using 10\% as the bottom value for the completeness. 
\tcb{Note that we confirm the predicted sizes in the above method generally being consistent with the observed sizes by {\sc imfit} within $\sim40\%$ by comparing these estimates among the spatially-resolved high SNR sources shown in red-filled circles.}

\subsection{Survey Area}
\label{sec:survey_area}

%%%%%%%%%%%%%%%%%%%%%%%
\begin{figure}
\begin{center}
\includegraphics[trim=0.2cm 0cm 0cm 0cm, clip, angle=0,width=0.47\textwidth]{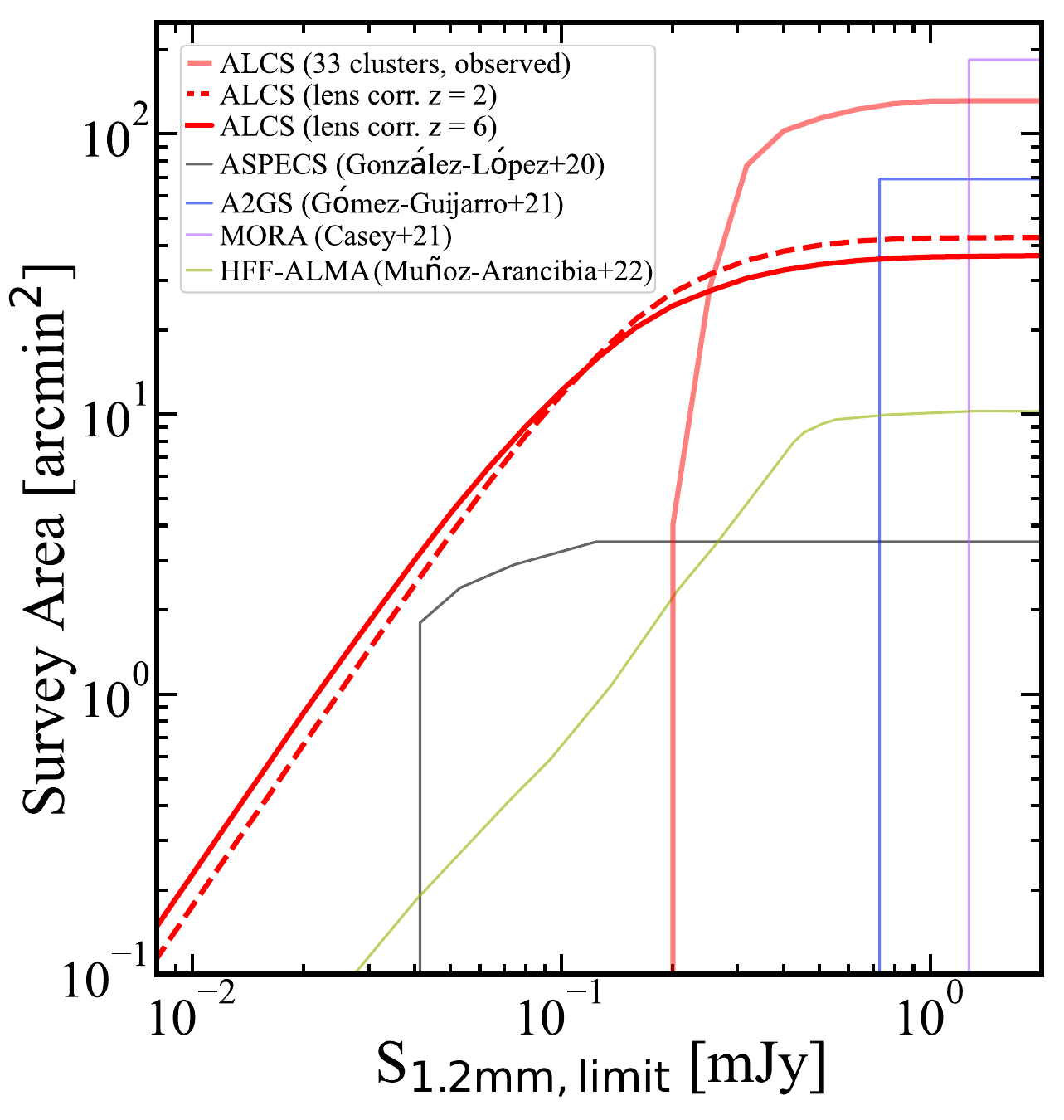}
 \caption{
Survey areas of ALCS (SNR$_{\rm nat}\geq4.0$, PB $\geq30$\%) and other large ALMA surveys in the literature \citep{gonzalez2020, gomez2021, casey2021, munoz-arancibia2022}. 
The light red line denotes the survey area before the lensing correction. 
The red dashed and solid lines represent the effect survey area after the lensing correction, assuming the redshift at $z=2.0$ and $z=6.0$, respectively, with our fiducial lens models. 
ALCS explores the unique parameter space towards faint and wide regimes, compared to previous ALMA surveys, demonstrating the power of the lensing support. 
In addition to the little difference in the results between $z=2.0$ and $z=6.0$, 
we confirm that the difference in the survey areas with different lens models is negligible in Appendix \ref{sec:app_models}.  
\tcb{We scale the flux densities using the methods described in Section~\ref{sec:nc_comp_pre}.}
\label{fig:area}}
\end{center}
\end{figure}
%%%%%%%%%%%%%%%%%%%%%%% 

We estimate the survey areas of each ALCS field by counting the areas with the PB sensitivity in the mosaic maps of $\geq$30\%, where we perform the source extraction (Section \ref{sec:source_ext}).  
Figure \ref{fig:area} presents the total survey areas of the 33 ALCS lensing cluster fields before and after the lens correction based on our fiducial lens models for $z=2$ and $z=6$. 
Since the sensitivity and the magnification are not spatially uniform across the ALMA maps, the survey area varies according to the intrinsic flux density. 
Although the magnification at a given sky position should change as a function of redshift, 
the total survey areas show a negligible change between redshifts $z=2$ and $z=6$.  
In Appendix \ref{sec:app_models}, we also evaluate the systematic uncertainty of the survey area from the choice of the lens model, which is also confirmed to be negligible. 
With our fiducial model, we create the magnification maps for each cluster from $z=0$ to $z=6$, with a step of 0.5, and calculated the effective survey area at each redshift. In the following analysis, we use the closest redshift calculation result in each ALCS source, while we adopt the $z=0$ result for the sources in the foreground of the clusters. 

For comparison, we also present survey areas of other ALMA blind surveys in previous studies \citep{gonzalez2020, gomez2021, casey2021, munoz-arancibia2022}. 
Critically, we find that the ALCS survey area after the lens correction (red line) explores the widest and deepest parameter spaces among the ALMA blind surveys so far performed (color lines). 
For example, the ALCS survey area decreases at $z=6$ down to $\simeq$ 20~arcmin$^{2}$ and $\simeq$ 4~arcmin$^{2}$ at detection limits of 0.1~mJy and 0.04~mJy, which are still larger than those in ASPECS (grey line) at the same detection limits by factors of $\simeq$ 4.8 and 2.0, respectively. 
This demonstrates the power of gravitational lensing, which allows us to identify unique objects that have been missed in previous ALMA surveys \citep[e.g.,][]{caputi2021,fujimoto2021,laporte2021}.

\section{Properties of ALCS sources}
\label{sec:prop}

%%%%%%%%%%%%%%%%%%%%%%%
\begin{figure*}
\begin{center}
\includegraphics[trim=0cm 0cm 0cm 0cm, clip, angle=0,width=0.9\textwidth]{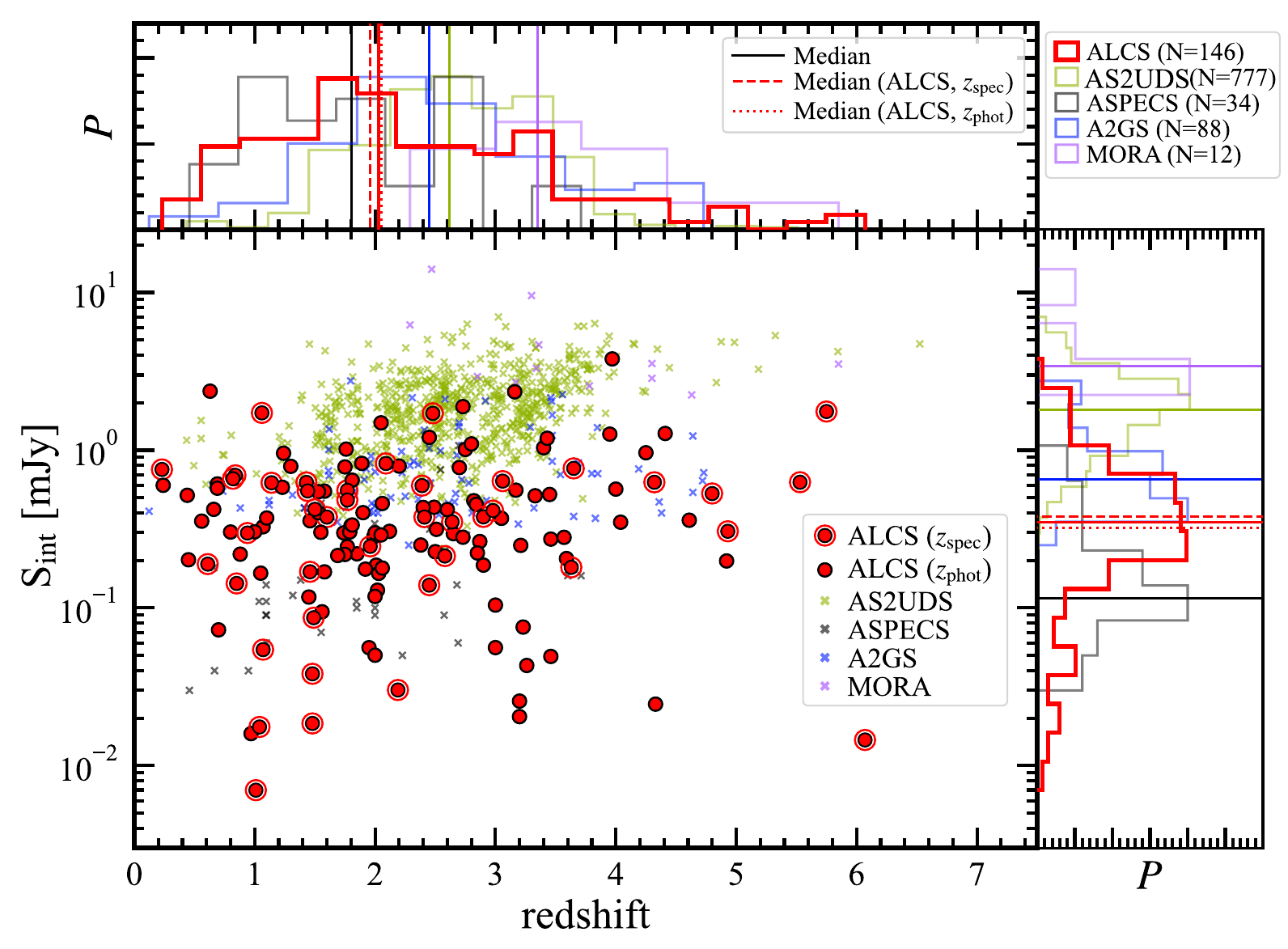}
 \caption{
Intrinsic ALMA flux density ($S_{\rm int}$) and redshift distributions of the ALCS sources after the lens correction.  
The red-filled circles indicate the 146 ALCS sources after removing the cluster member galaxies and adopting the average intrinsic flux density for multiply imaged systems.
The spectroscopically confirmed sources are marked with red open circles. 
The color crosses denote ALMA sources identified in other recent ALMA surveys --  
707 sources from Band~7 (870~$\mu$m) snapshot follow-up observations for SCUBA2 sources in UDS (AS2UDS; \citealt{dudzeviciute2020}), 34 sources from Band~6 deep blind observations in HUDF (ASPECS; \citealt{gonzalez2020}), 88 sources from Band~6 wide blind observations in GOODS-S (A2GS; \citealt{gomez2021}), and 12 sources from Band~4 (2~mm) wide blind observations in COSMOS (MORA; \citealt{casey2021}). For the AS2UDS and MORA sources, we scale their flux density by assuming a modified black body based on their best-fit $T_{\rm d}$ and $\beta_{d}$ values presented in the literature. 
The top and right panels present the probability distribution for the redshift and intrinsic flux density, 
where the peaks are normalized.  
The solid lines are the median value for each sample. 
The dashed and dotted lines show the median values for the spectroscopic and photometric samples in the ALCS sources, respectively, which show good agreement with each other and indicate that their potential difference is negligible owing to the lensing support.  
\label{fig:prop}}
\end{center}
\end{figure*}
%%%%%%%%%%%%%%%%%%%%%%% 

Based on the analyses presented in Section~\ref{sec:analysis}, we characterize our ALCS sources by comparing them with other recent ALMA sources identified in similarly large surveys (total observation times $\gtrsim 40$~hrs).
Figure \ref{fig:prop} displays the intrinsic flux density ($S_{\rm int}$) and redshift distributions of our ALCS sources (red circles).
To obtain unbiased results, we only show the 146 ALCS sources after removing the cluster member galaxies and correcting for duplicated counts among the multiply imaged systems (Section~\ref{sec:lens}). 
We also present the ALMA sources identified in other recent ALMA surveys \citep{dudzeviciute2020, gonzalez2020, gomez2021, casey2021}.
The normalized probability distributions of $z$ and $S_{\rm int}$ for each sample are drawn in the top and right panels, with the median value indicated by the solid line.

Compared to the other ALMA samples \tcb{highlighted here}, we find that the ALCS sources are the most widely distributed in both parameter spaces, spanning $S_{\rm int}\simeq0.007$--3.8~mJy and $z\simeq0$--6. 
The median $S_{\rm int}$ and $z$ values are estimated to be $S_{\rm int} =$0.35~mJy and $z=2.03$, respectively. These median $S_{\rm int}$ and $z$ values of our ALCS sources fall between those of the ASPECS and A2GS samples, but our ALCS sources explore fainter $S_{\rm int}$ and higher $z$ parameter spaces than those two samples.   
The ALCS survey provides the largest sample number ($N=146$) 
\tcb{among the aforementioned} blind surveys ($N$=12--88). 
Our results demonstrate the capability of the lensing boost to increase sensitivity even at high redshifts, as well as the efficiency of the wide lensing survey scheme (Section~\ref{sec:intro}) 
The median values of $S_{\rm int}$ and $z$ are $S_{\rm int}=$ 0.32~mJy and 0.37~mJy and $z=2.05$ and 1.96, for the photometric and spectroscopic ALCS samples, respectively. The small differences between these two samples suggest a general agreement of their physical properties, strengthening our interpretations for the photometric sample in the statistical sense. 
Overall, our ALCS sources represent an optimal statistical sample to study the faint mm population in wide flux and redshift ranges. 

Comparing the median values among the different samples, we find a general positive correlation between the source flux and redshift. A similar positive correlation is reported in the bright SMG population with the submm flux density \citep[e.g.,][]{stach2018, simpson2020}, indicating that this correlation also exists among fainter mm sources than SMGs down to $S_{\rm 1.2mm}\sim0.1$--0.01~mJy. These correlations are likely in line with the increasing luminosity evolution of the IR LFs towards high redshifts reported in previous studies \citep[e.g.,][]{gruppioni2013, koprowski2017}. 
We further discuss the redshift evolution of IR LFs in Section~\ref{sec:irlf}.

%%%%%%%%%%%%%%%%%%%%%%%
\begin{figure*}[t!]
\begin{center}
\includegraphics[trim=0cm 0cm 0cm 0cm, clip, angle=0,width=0.98\textwidth]{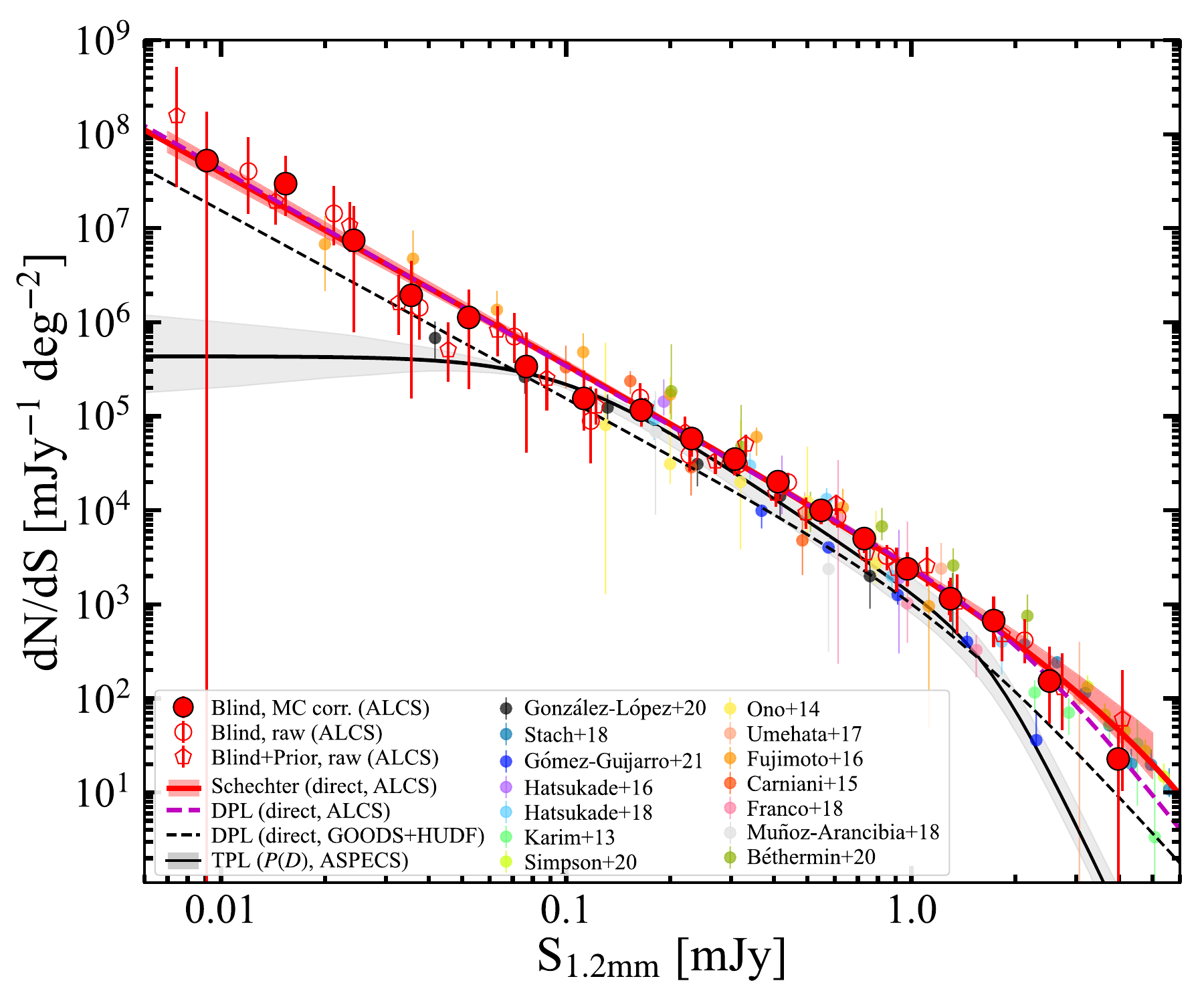}
 \caption{
Differential number counts at 1.2 mm.
The red open circles and pentagons are the number counts derived from our ALCS blind and blind+prior samples without the MC simulations, respectively, where the error bars only take Poisson uncertainties into account.  
The filled red circles represent the number counts corrected by the MC simulations that implement all relevant uncertainties, such as redshift, magnification, and flux density measurements, where the error bars indicate the 16--84th percentile in the 1,000 MC realizations. 
The solid red line and the shaded region denote our fiducial Schechter function estimate and the associated $1\sigma$ error, while the dashed magenta line presents our fiducial DPL function estimate. 
In the fitting, we use our number count estimates alone to avoid potential systematic uncertainty in different measurements from different surveys.   
For comparison, we also show previous ALMA submm/mm number counts in the literature \citep{gonzalez2020, stach2018, gomez2021, hatsukade2014, hatsukade2018, karim2013, ono2014, umehata2017, fujimoto2016, carniani2015, franco2018, arancibia2018, munoz-arancibia2022, bethermin2020}. 
The black dashed line is the best-fit Schechter function using the results obtained in GOODS-S \citep{gomez2021} and HUDF (ASPECS; \citealt{gonzalez2020}). 
The black solid line is taken from \cite{gonzalez2020} obtained from the $P(D)$ analysis that indicates the presence of the flattening below $\sim0.1$~mJy, the black shaded region denotes the associated 1$\sigma$ error. 
When comparing previous measurements of the number counts at different wavelengths to our 1.2~mm band, 
we scale the flux densities using the methods described in Section~\ref{sec:nc_comp_pre}.
\label{fig:nc_dif}}
\end{center}
\end{figure*}
%%%%%%%%%%%%%%%%%%%%%%% 

\section{Number counts \& CIB}
\label{sec:result1}

\subsection{Number counts at 1.2 mm}
\label{sec:number_counts}

We derive number counts at 1.2 mm based on the most common method: directly counting the sources, correcting their purity and completeness, and obtaining the number density per survey area \citep[e.g.,][]{hatsukade2013,ono2014,carniani2015,fujimoto2016,aravena2016,dunlop2017,umehata2017,hatsukade2018,arancibia2018,franco2018,gonzalez2020,gomez2021}. 
Note that the lensing effect is corrected in the measurements of the flux density and the survey area. 
A total of 104 and 146 sources are accounted for in our number counts analysis based on the blind and blind+prior samples, after removing the sources that are regarded as the cluster member galaxies and correcting the counts from the multiple images (Section \ref{sec:lens}). 

A contribution to the number counts from an identified source, $\xi$, is given by 
\begin{equation}
\label{eq:ef_num}
\xi(S) = \frac{p(S)}{C(S, D)A_{\rm eff}(S)},
\end{equation}
where $S$, $C$, \tcb{$D$}, $A_{\rm eff}$, and $p(S)$ are the intrinsic flux density, completeness, \tcb{intrinsic source size}, survey area, and purity (equation~\ref{eq:purity}), respectively. 
Then, a sum of the contributions for each flux bin is computed by,
\begin{eqnarray}
n(S) = \frac{\Sigma\xi(S)}{\Delta S},
\end{eqnarray}
where $\Delta S$ is the bin width with the unit of mJy to normalize the difference due to the bin size.  
To evaluate the uncertainty of $n(S)$, we include Poisson statistical errors of the source counts per bin and the uncertainty of the intrinsic flux density estimate. 
For the Poisson error, we use the values presented in \cite{gehrels1986}
that are applicable to a small number of statistics. 
For the intrinsic flux density uncertainty, 
we take the following contributions into account: the random noise, the error of the absolute accuracy of the ALMA Band~6 flux calibration, and errors of the lens correction that are contributed by uncertainties from the lens model and the source redshift. 
\tcb{Given the negligible impact of the Poisson statistical errors on the purity estimate at our high purity cut (Section~\ref{sec:source_ext}), the purity uncertainty is not included in our procedure.}

We perform MC simulations to include all the uncertainties described above and derive realistic number counts.
We make a mock catalog of the ALCS sources whose flux densities follow Gaussian probability distributions. 
The standard deviations of the Gaussians are given by the combination of the random noise and the uncertainties of the absolute flux accuracy and lens correction. 
We adopt the measurement uncertainty for the random noise and 10\% for the absolute flux accuracy.\footnote{
\url{https://almascience.eso.org/documents-and-tools/cycle6/alma-proposers-guide}
} 
We evaluate the 1$\sigma$ uncertainty of the lens correction by propagating the magnification error from the source redshift uncertainty (Section \ref{sec:redshift}) and \tcb{the systematic lens model uncertainty among different models} (Section \ref{sec:lens}). 
We produce 1,000 mock catalogs and derive the number counts for each catalog in the same manner. 
We then evaluate the average and the 16--84th percentile of the number counts per bin, where the uncertainty of the intrinsic flux density estimate is fully taken into account. 
Propagating the Poisson error per bin based on the median in the MC iterations, we finally obtain the differential number counts and the associated 1$\sigma$ uncertainties. 

%%%%%%%%%%%%%%%%%%%%%%%
\begin{figure*}[t!]
\begin{center}
\includegraphics[trim=0cm 0cm 0cm 0cm, clip, angle=0,width=1\textwidth]{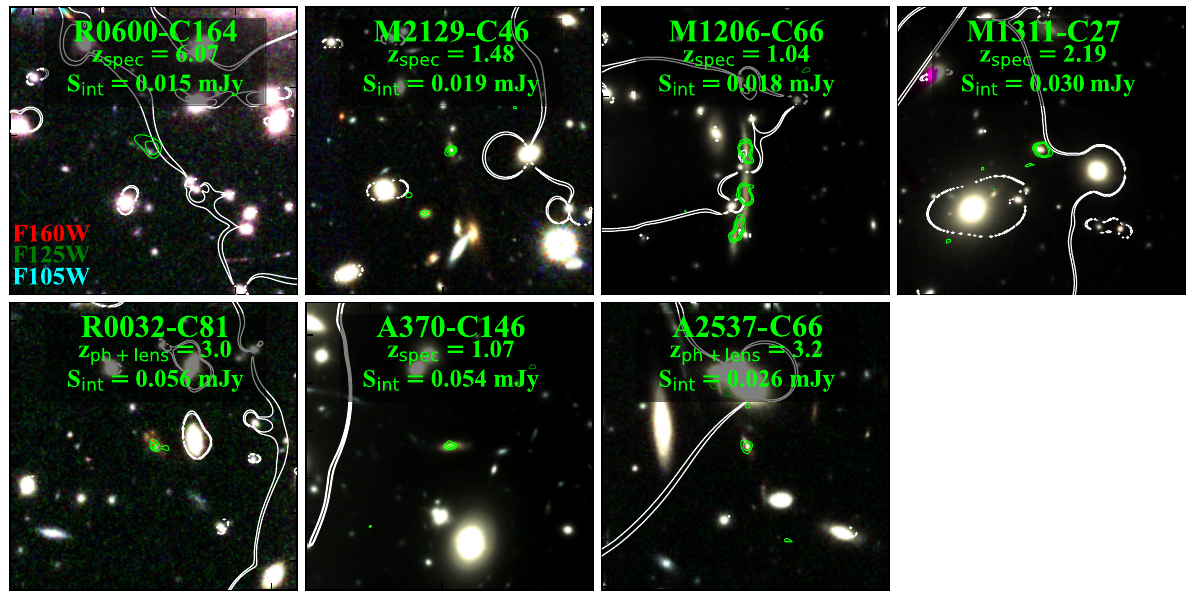}
 \caption{
HST/F160W $30''\times30''$ cutout images for seven ALCS sources falling in the intrinsically faintest mm regime of 0.01--0.06~mJy in the blind sample (SNR$_{\rm nat}>5.0$ or SNR$_{\rm tap}>4.5$), whose redshifts are spectroscopically confirmed or securely constrained by the lens models from its multiple image positions and their photometric redshifts. 
These secure redshift estimates mitigate the magnification uncertainty and do not support any strong flattening at $\sim$0.1~mJy (see Section~\ref{sec:nc_comp_pre}).  
The green contour indicates the ALMA Band~6 continuum at 2$\sigma$, 3$\sigma$, 4$\sigma$, and 5$\sigma$. 
The white curves represent the contours for the lensing magnification of $\mu=200$.
\label{fig:highmu}}
\end{center}
\end{figure*}
%%%%%%%%%%%%%%%%%%%%%%% 

In Figure \ref{fig:nc_dif}, 
we show our differential number count estimates based on the blind sample and the blind+prior sample with the red open circle and pentagon, respectively. 
The estimate after applying the MC simulations based on the blind sample is also shown in the red-filled circle. 
We list all the estimates in Table~\ref{tab:nc_value}. 
Figure \ref{fig:nc_dif} shows that our study successfully covers 2.5-dex in flux density and explores the 1.2-mm number counts down to $\sim$7 $\mu$Jy, owing to the wide-area mapping of ALCS towards 33 massive lensing clusters.  
We find that our number-count estimates based on the blind and blind+prior samples are consistent within the 1$\sigma$ uncertainties in the wide flux range.

To characterize the shape of our number counts, 
we fit Schechter \citep{schechter1976} and double power law (DPL) functions to our differential and cumulative number counts. 
The Schechter function form is given by 
\begin{equation}
\label{eq:schchter}
\phi(S_{\nu})= \phi_{0}\left(\frac{S_{\nu}}{S_{0}}\right)^{\alpha} {\rm exp}\left(-\frac{S_{\nu}}{S_{0}} \right), 
\end{equation}
where $\phi_{0}$, $S_{0}$, and $\alpha$ are the normalization, characteristic flux density, and faint-end slope power-law index, respectively, 
and the DPL function form is given by 
\begin{equation}
\phi(S_{\nu})= \phi_{0}\left[\left(\frac{S_{\nu}}{S_{0}}\right)^{\alpha} + \left(\frac{S_{\nu}}{S_{0}}\right)^{\beta}  \right]^{-1}, 
\end{equation}
where the definition of $\phi_{0}$, $S_{0}$, and $\alpha$ are the same as those of Equation \ref{eq:schchter}, and $\beta$ is the bright-end slope. 
We search the best-fit model with the MCMC method using {\sc emcee} \citep{foreman-mackey2013}. 
Given the statistically sufficient number in our ALCS sample, we only use our ALCS sample for the fitting to remove systematic uncertainties due to different measurement approaches among the studies. 

We show the best-fit Schechter and DPL functions with the solid red and dashed magenta curves in Figure \ref{fig:nc_dif}, and summarize the best-fit parameters with the $\chi^{2}$/dof values in Table \ref{tab:nc_param}. 
We also perform the Schechter and DPL function fitting in the 1,000 realizations of the MC simulations, which yield the red and magenta shaded regions representing the 16-84th percentile of the 1,000 best-fit Schechter and DPL functions.  
Figure \ref{fig:nc_dif} and the $\chi^{2}$/dof values indicate that the number counts are well represented by Schechter and DPL. 
In the following comparison and analyses, we adopt the best-fit Schechter function for the entire 33 ALCS clusters without any weights among the clusters as our fiducial measurement. We further discuss the impact of the faint-end slope estimate with different assumptions, sub-samples, and weights according to the different qualities of the lens model in Section \ref{sec:impact_comp} and Section \ref{sec:caveats}.

%%%%%%%%%%%%%%%%%%
\setlength{\tabcolsep}{16pt}
\begin{table}
\begin{center}
\caption{Differential Number Counts at 1.2~mm}
\label{tab:nc_value}
\vspace{-0.2cm}
\begin{tabular}{ccc} \hline \hline
$S$ &  $\log$(dN/dS) &  $<N>$ \\
 mJy & deg$^{-2}$~mJy$^{-1}$ &  \\
    (1)  &  (2)  & (3) \\ \hline 
0.007--0.012 & 7.76$_{-99.99}^{+0.24}$ & 0.8  \\
0.012--0.020 & 7.51$_{-0.25}^{+0.15}$ & 2.7  \\
0.020--0.029 & 6.89$_{-1.04}^{+0.25}$ & 1.8  \\
0.029--0.043 & 6.29$_{-1.03}^{+0.29}$ & 1.7  \\
0.043--0.063 & 6.00$_{-0.26}^{+0.19}$ & 2.8  \\
0.063--0.093 & 5.47$_{-0.95}^{+0.24}$ & 1.9  \\
0.093--0.136 & 5.16$_{-0.47}^{+0.20}$ & 3.3  \\
0.14--0.20 & 5.06$_{-0.19}^{+0.13}$ & 9.7 \\
0.20--0.27 & 4.76$_{-0.20}^{+0.15}$ & 11.5 \\
0.27--0.36 & 4.54$_{-0.19}^{+0.13}$ & 15.0 \\
0.36--0.47 & 4.27$_{-0.18}^{+0.12}$ & 14.6 \\
0.47--0.63 & 3.99$_{-0.16}^{+0.11}$ & 13.6 \\
0.63--0.84 & 3.73$_{-0.12}^{+0.09}$ & 12.0 \\
0.84--1.12 & 3.38$_{-0.12}^{+0.11}$ & 7.7 \\
1.12--1.50 & 3.03$_{-0.20}^{+0.16}$ & 4.8 \\
1.50--2.00 & 2.79$_{-0.26}^{+0.13}$ & 3.7 \\
2.00--3.16 & 2.16$_{-0.30}^{+0.17}$ & 2.0 \\
3.16--5.00 & 1.36$_{-1.36}^{+0.30}$ & 0.5 \\
\hline \hline
\end{tabular}
\end{center}
\vspace{-0.4cm}
\tablecomments{
(1) Flux range used in the differential number counts. 
(2) Differential number counts ($dN/dS$) in the logarithm scale. 
The 1$\sigma$ uncertainties are estimated from the combination of the number count Poisson statistical errors and the uncertainty of the intrinsic flux density estimate associated with the measurement, redshift, and magnification errors (see the text).
(3) Average source number in each flux bin among the 1,000 MC realizations. 
}
\end{table}
%%%%%%%%%%%%%%%%%%

%%%%%%%%%%%%%%%%%%%%%%%
\begin{figure}[t!]
\begin{center}
\includegraphics[trim=0cm 0cm 0cm 0cm, clip, angle=0,width=0.5\textwidth]{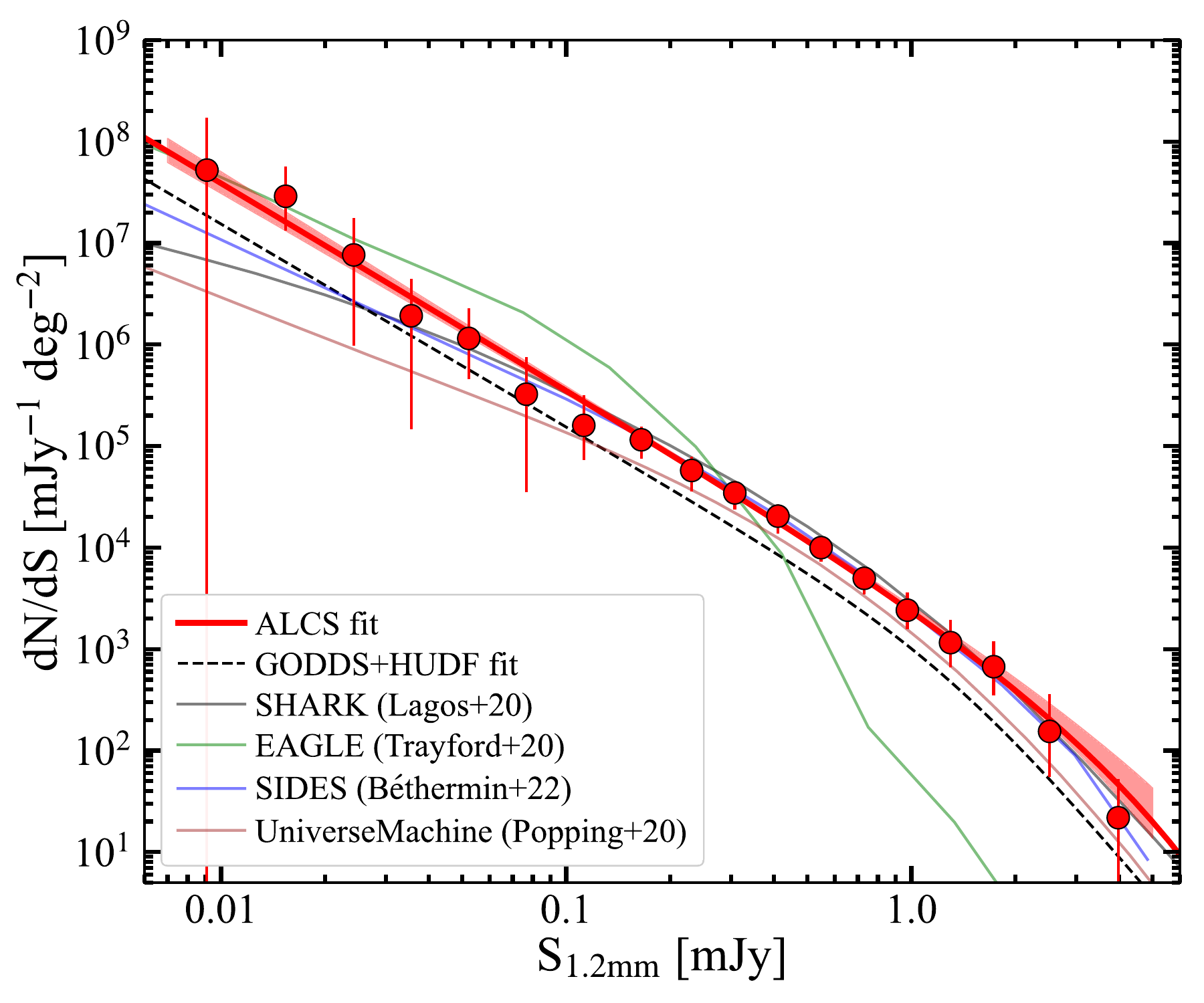}
 \caption{
Comparison of differential number counts at 1.2~mm with the simulation predictions. 
The red-filled circles, red line, red shaded region, and black dashed line are assigned as in Figure \ref{fig:nc_dif}. 
The other colored lines show several model predictions for number counts \citep{lagos2020, trayford2020, popping2020, bethermin2022}. 
These model predictions are generally consistent with the individual data points of ALCS within 1--2$\sigma$.
\label{fig:nc_comp}}
\end{center}
\end{figure}
%%%%%%%%%%%%%%%%%%%%%%% 

\subsection{Comparison with previous ALMA observations}
\label{sec:nc_comp_pre}

In Figure \ref{fig:nc_dif}, we also show previous ALMA measurements of submm--mm number counts in the literature to compare with our estimate \citep{hatsukade2013,ono2014,carniani2015,fujimoto2016,oteo2016,aravena2016,dunlop2017,umehata2017,hatsukade2018,zavala2018,arancibia2018,franco2018,gonzalez2020,gomez2021}. 
For measurements observed at different wavelengths from 1.2 mm, we scale the flux density by assuming a typical FIR SED shape based on a single modified black body with dust temperature $T_{\rm dust}$=35\,K, spectral index $\beta_{\rm d}=1.8$, and $z=2$. 
For several measurements at 870~$\mu$m in Band~7 \citep{karim2013,stach2018,simpson2020,bethermin2020}, we use the predicted 870 and 1150~$\mu$m fluxes from the {\sc magphys} modeling of the 707 870-$\mu$m selected SMGs in the AS2UDS survey from \cite{dudzeviciute2020} to derive a relation between $S_{870\mu \rm m}/S_{1150\mu \rm m}$ and $S_{870\mu \rm m}$ of the form:  $S_{870\mu \rm m}/S_{1150\mu \rm m} = 2.3 - 0.04 \times S_{870\mu \rm m}$. This flux ratio variation is primarily driven by the evolution in the median redshift of the SMG population with submm flux density \citep[e.g.,][]{stach2018,simpson2020}. 

\tcb{Although some uncertainty may remain in the flux conversion to 1.15~mm,} we find that our measurements are consistent with previous studies within the errors in general.
At $S_{\rm 1.2mm}> 1.0$ mJy, our measurements are in excellent agreement with the previous results from the follow-up ALMA observations for bright SMGs at 870$\mu$m \citep{karim2013, stach2018, simpson2020}. 
For intermediate flux densities of $S_{\rm 1.2mm}\simeq$ 0.1--1.0 mJy, the ALCS number counts fall between the scatter of previous studies and are mostly consistent with the previous measurements within the 1$\sigma$ uncertainties. 
At $S_{\rm 1.2mm}<$ 0.1 mJy, our measurements remain consistent with the results obtained in the HUDF (ASPECS; \citealt{gonzalez2020}) down to $\simeq$ 0.04~mJy and with \cite{fujimoto2016} to $\simeq$ 0.02~mJy, within their 1$\sigma$ uncertainties. 

To compare the shape of the number counts, we also fit the DPL function in the same manner as results observed in the latest ALMA blind surveys towards GOODS-S (A2GS; \citealt{gomez2021}) and HUDF (ASPECS; e.g., \citealt{gonzalez2020}) whose area is a part of GOODS-S.
For a fair comparison, we fit the results that are also obtained by the direct counts, instead of the $P(D)$ analysis presented in ASPECS.  
The black dashed line shows the best-fit DPL function obtained from the previous GOODS+HUDF results. 
Over the entire flux range probed, the best-fit shape of our number counts is higher than that in the GOODS+HUDF. 
Around the characteristic luminosity of $\sim1$~mJy, the number counts in \tcb{A2GS} are a factor of $\approx2$ lower than our estimate. 
\cite{popping2020} evaluate the effects of cosmic variance on the number count measurements for the same wavelength and area as the ASPECS 1.2-mm image \citep{gonzalez2020}to have a 2$\sigma$ scatter of 1.5. However, the calculation in \cite{popping2020} uses the survey area of 4.2 arcmin$^{2}$, where the PB sensitivity decreases down to 0.1. 
When the same calculation is performed based on the survey area of 1.8 (2.4) arcmin$^{2}$ with a PB sensitivity of $\geq0.9$ ($\geq0.7$), 
the 2$\sigma$ scatter increases to factors of $\sim$5 ($\sim$2). 
Notably, previous UVLF and AGN studies also suggest that the galaxies are underdense in HUDF \citep[e.g.,][]{cowie2002, moretti2003, bauer2004, oesch2007}. 
Thus we attribute the differences to cosmic variance. 

Over the entire flux regime, in general, 
the larger survey area and the 33 independent lines of sight in our ALCS survey help to mitigate the cosmic variance compared to the previous surveys, likely making our number counts fall between the scatter among previous studies. 
Moreover, our conservative purity cut of $0.99$, compared to typical purity cuts of $\sim$0.3--0.5 in previous studies \citep[e.g.,][]{hatsukade2013, ono2014, carniani2015, fujimoto2016, oteo2016, gonzalez2020}, makes our results relatively immune to purity uncertainties. Additionally, our intrinsic source size assumption, instead of a point source, allows us to perform realistic completeness corrections. 
These aspects also likely contribute to converging within the scatter of the previous studies.
We further discuss the impact of the different source size assumptions on the completeness correction in Section \ref{sec:impact_comp}.

\cite{gonzalez2020} report that the faint-end slope has a flattening shape below $S_{\rm 1.2mm}\simeq 0.1$~mJy from the triple power law (TPL) function fit based on the $P(D)$ analysis, which we plot in Fig.\ref{fig:nc_dif}. 
This TPL from the $P(D)$ analysis predicts fewer number counts than the DPL from the direct counts analysis by almost 2-dex, even with the same data, suggesting that it remains challenging to conclude the existence of any flattening at the faint-end of the 1.2-mm number counts and that some caution should be exercised in the choice of the methodology. 
Given the average sensitivity of our ALCS maps ($\sim60$~$\mu$Jy) and the secure detection cut of SNR$_{\rm nat}=5.0$, we require magnifications in the range of $\mu\sim3$--30 to detect sources as faint as 0.01--0.1 mJy in our survey. This requirement still falls within the regime of lower systematic uncertainty in lens corrections, as reported in previous UVLF studies \citep[e.g.,][]{bouwens2017b}. 
Although the redshift uncertainty in photometric sources enhances the magnification uncertainty, 
we account for both uncertainties through MC simulations in our estimates.

We note that there are seven sources with secure redshift estimates ($z_{\rm spec}$ or $z_{\rm phot}$ + constraints from lens models due to their multiple images) below 0.06~mJy that mitigate the magnification uncertainty. 
In Figure~\ref{fig:highmu}, we summarize these seven sources labeled with their redshift and intrinsic source flux. 
All seven sources show distorted morphology in the HST maps whose shear orientation agrees with the predicted gravitational lensing effects, supporting the high magnification estimates. 
In the prior sample, we also identify two sources (A383-C50 and A2537-C24) whose redshifts are similarly and securely estimated, falling in the same faintest regime with high magnification estimates ($\mu>10$) supported by strongly distorted morphology in the \hst\ maps (see Figure~\ref{fig:app_postage2}). 
For example, R0600-C164 is spectroscopically confirmed at $z=6.072$, and its local and global scale magnification factors of 163$^{+27}_{-13}$ and 29$^{+4}_{-7}$ are securely ensured by flux ratios and positions of its multiple images with three independent models \citep{laporte2021, fujimoto2021}. 
The intrinsic flux density is estimated to be $\sim$0.01~mJy, such that 
even if we only count R0600-C164 in the flux density bin at 0.01~mJy, the number count estimate and its 2$\sigma$ lower limit from the single-sided Poisson uncertainty \cite{gehrels1986} 
are estimated to be $\sim$ $3\times10^{7}$~mJy$^{-1}$~deg$^{-2}$ and $1.5\times10^{6}$~mJy$^{-1}$~deg$^{-2}$, respectively; this source by itself 
rules out the predicted flattening shape at $S_{\rm 1.2mm}\sim$ 0.01~mJy by $>$ 0.5~dex.
\tcb{Naturally, when considering all seven robustly highly magnified sources summarized in Figure~\ref{fig:highmu}, the gap from flattening shape becomes even more significant.}

Overall, even accounting for possible lensing magnification uncertainties, our results disfavor the scenario whereby the 1.2-mm number counts start flattening at $S_{\rm 1.2mm}\sim$ 0.1~mJy. 
In \cite{gonzalez2020}, the $P(D)$ analysis is performed with the dirty image. Given the sparse density of dusty galaxies and the superb resolution of ALMA, the pixel counts below the direct detection limits might be more affected by the side lobes remained in the dirty map and the noise fluctuation rather than the weak signals from faint sources, which could be one of the causes in the different faint-end slope derived from the direct counts and the $P(D)$ analysis. 
The point source assumption for injected sources in the $P(D)$ analysis may also be another possible reason, making the completeness overestimated and thus favoring a shallow faint-end slope. 

%%%%%%%%%%%%%%%%%%
\setlength{\tabcolsep}{6pt}
\begin{table}
\begin{center}
\caption{Best-fit parameters for our 1.2-mm number counts}
\label{tab:nc_param}
\vspace{-0.4cm}
\begin{tabular}{ccccc}
\hline \hline
\multicolumn{5}{c}{Schechter} \\ \hline 
$\alpha$ &     & $\log(S_{\star})$ &  $\log(\phi_{\star}$) & $\chi^{2}$/dof  \\
                 &    &      (mJy)       &      (deg$^{-2}$)          &                     \\ 
(1)           &    &        (2)          &    (3)                            &     (4)       \\ \hline
$-2.05_{-0.10}^{+0.12}$ &  & $0.60_{-0.33}^{+0.38}$ & $2.85_{-0.52}^{+0.50}$ & 4.9/14 \\ \hline
\multicolumn{5}{c}{DPL} \\ \hline 
$\alpha$ & $\beta$  & $\log(S_{\star})$  & $\log(\phi_{\star}$)  & $\chi^{2}$/dof  \\  
                   &                   &      (mJy)          &      (deg$^{-2}$)          &                     \\ 
(5)      &   (6)   &    (7)          & (8)             &                 \\ \hline
$2.12_{-0.13}^{+0.15}$ & $3.81_{-1.65}^{+2.60}$ & $2.97_{-0.59}^{+0.89}$ & $0.44_{-0.59}^{+0.48}$ & 4.2/13 \\ \hline
\end{tabular}
\end{center}
\vspace{-0.4cm}
\tablecomments{
(1)--(3): Best-fit parameter set for the Schechter function. 
(4): $\chi^{2}$ over the degree of freedom. 
(5)-- (8): Best-fit parameter set for the DPL function.  
}
\end{table}
%%%%%%%%%%%%%%%%%%

\subsection{Comparison with simulations}
\label{sec:nc_comp_sim}

In Figure \ref{fig:nc_comp}, we also display theoretical predictions from cosmological simulations that do not include the tweak in the IMF. 
We show semi-analytical simulations from SHARK \citep{lagos2018, lagos2020} and SIDES \citep{bethermin2017, bethermin2022}, and a hydrodynamical simulation from EAGLE \citep{schaye2015}. 
We also show a data-driven model from UniverseMachine \citep{behroozi2019}, using empirical scaling relations to connect the SFR and stellar mass of galaxies to their dust continuum emission \citep{popping2020}. 
We note that SHARK and EAGLE are galaxy formation physics simulations. Hence, the IR number counts and luminosity functions are predictions of these models. In contrast, SIDES and UniverseMachine, being empirical models, are tuned to fit some of the observations in the literature.
\citet{cowley2019} reported that EAGLE simulations underestimate the abundance of bright dusty galaxies, which we confirm at $S_{\rm 1.2mm}\gtrsim0.2$ mJy. In Appendix \ref{sec:app_eagle}, we describe how we calculate the 1.2-mm flux density from the outputs of the EAGLE simulation.

Apart from the underestimate of the EAGLE simulation, 
these simulations show overall good agreement within 1--2$\sigma$ errors with the individual data points of our 1.2-mm number counts. 
On the other hand, below $\sim0.06$~mJy we find that our best-fit Schechter function (red line) exceeds all these simulations, except for EAGLE (green line). 
These offsets might indicate missing physical mechanisms in the current simulations to reproduce a steeper faint-end slope in the 1.2-mm number counts, such as a different dependence between the dust mass and metallicity at the low-mass regime. 
Another possibility is that the slope at the faintest end does not have the flattening at $\sim$0.1~mJy (Section~\ref{sec:nc_comp_pre}), but might be slightly shallower, and these offsets might be dismissed, as the uncertainties in the redshift estimates and lens models are reduced in future observations. We further discuss potential additional uncertainties in the faint-end slope measurement in Section~\ref{sec:impact_comp} and Section~\ref{sec:caveats}. 

Note that these simulations also exceed the best-fit DPL obtained with the data points in GOODS+HUDF (solid black line) down to $\sim$0.2~mJy, supporting the argument that the GOODS+HUDF region is underdense. 
We also note that the flattening below $\sim0.1$~mJy is not reproduced by any of these current simulations.

\subsection{Resolving CIB}
\label{sec:cib}

%%%%%%%%%%%%%%%%%%%%%%%
\begin{figure}
\begin{center}
\includegraphics[trim=0cm 0cm 0cm 0cm, clip, angle=0,width=0.5\textwidth]{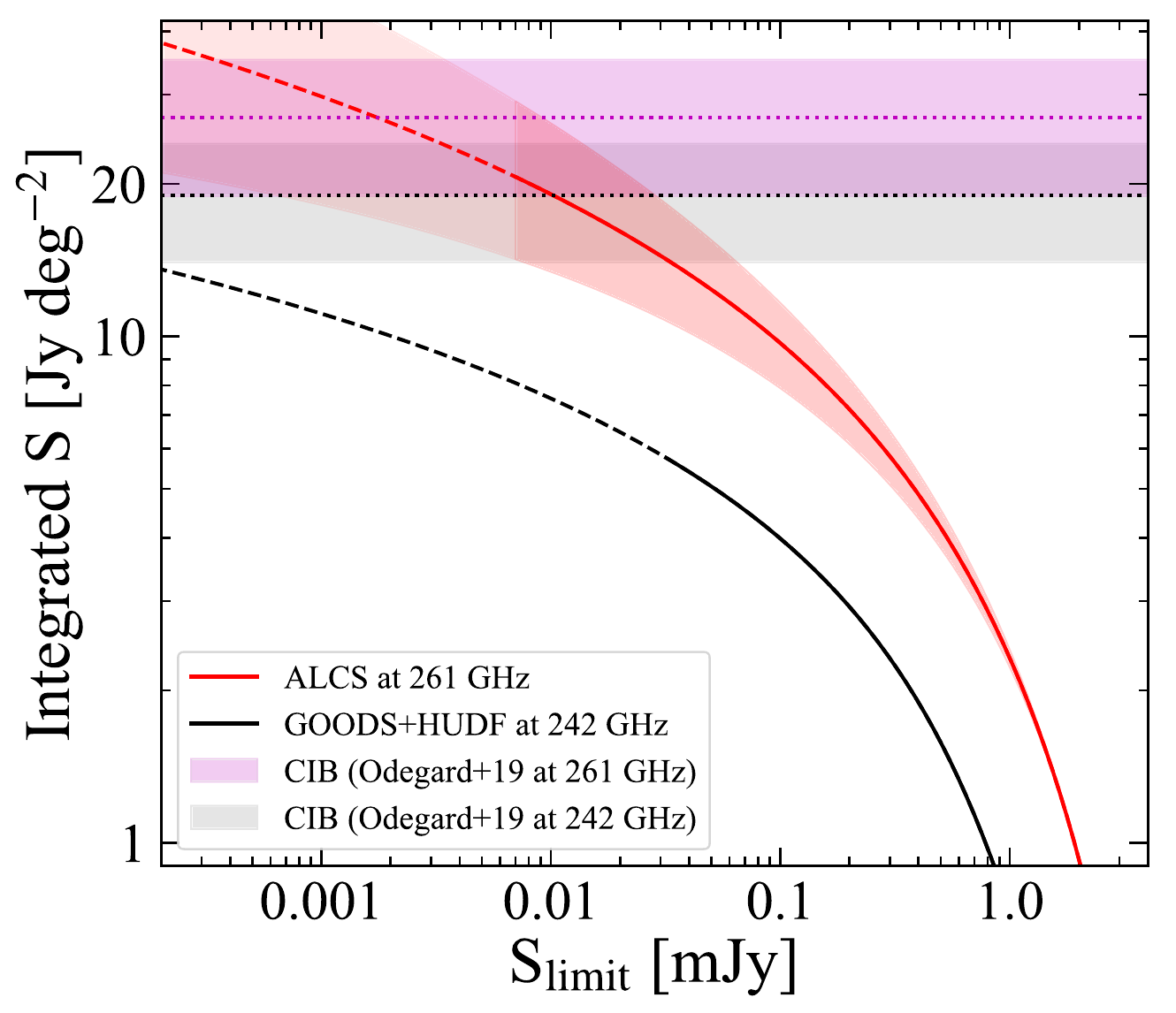}
 \caption{
Integration of the 1.2-mm number counts as a function of integration limit $S_{\rm ALMA, limit}$. 
The red (black) solid and dashed lines indicate the integration of our fiducial Schechter function estimate with the ALCS (GOODS+HUDF) number counts down to the nominal detection limit of $S_{\rm limit}=7~\mu$Jy (40~$\mu$Jy) and its extrapolation beyond the detection limit, respectively. 
The magenta (black) horizontal dotted line denotes the CIB estimate in \cite{odegard2019} by interpolating {\it COBE}/FIRAS measurements at 217~GHz and 353~GHz according to the central frequency of the ALCS (GOODS+HUDF) survey after subtracting the Galactic foreground from a linear combination of Galactic H{\sc i} and H$\alpha$. 
The shaded region shows the possible range of the interpolation of the CIB by taking 
the estimates from the lower and upper limits of the observed frequency range in the ALMA surveys. 
\label{fig:cib}}
\end{center}
\end{figure}
%%%%%%%%%%%%%%%%%%%%%%% 

%%%%%%%%%%%%%%%%%%%%%%%
\begin{figure*}
\begin{center}
\includegraphics[trim=0cm 0cm 0cm 0cm, clip, angle=0,width=1.0\textwidth]{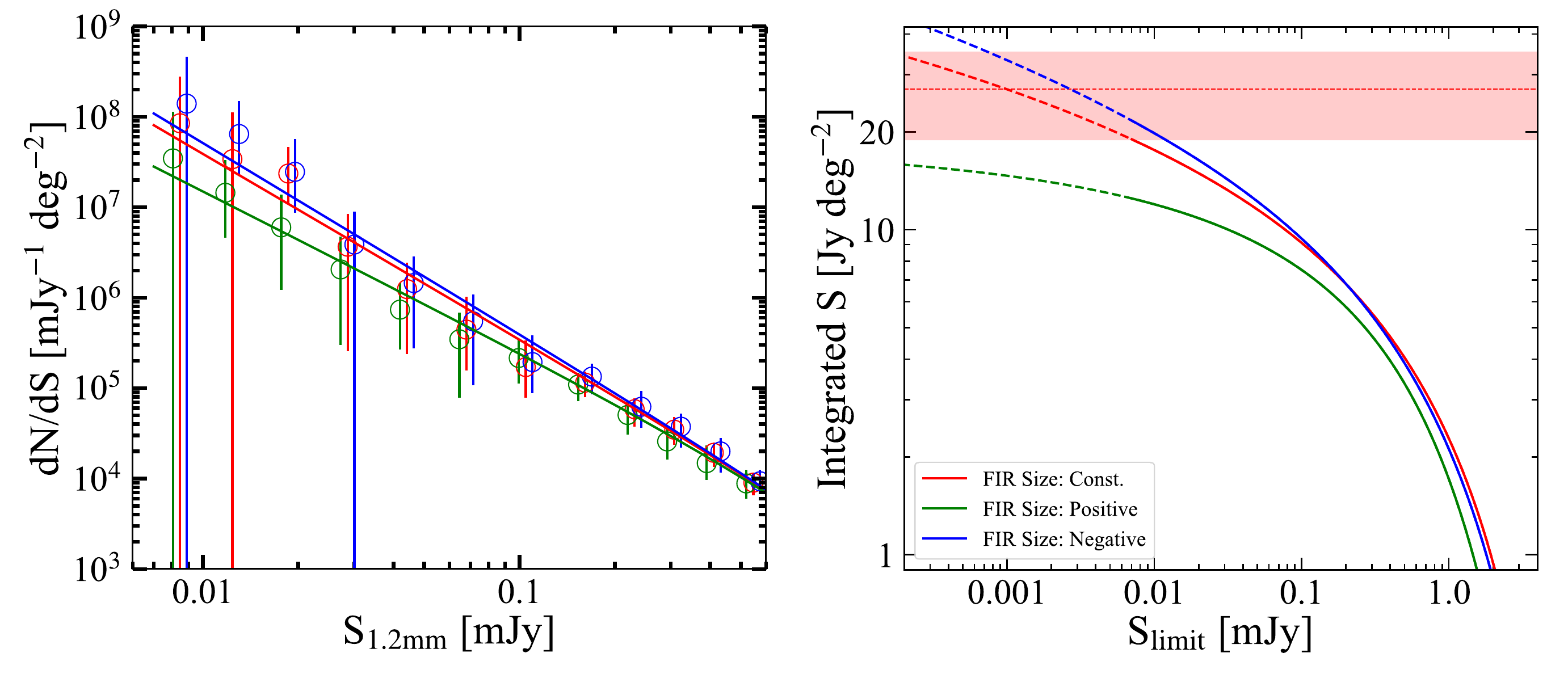}
 \caption{
Impact of the completeness correction from different intrinsic size distributions.
{\it Left}: Faint-end of the number counts. The red open circles and solid line are the same assignments as Figure \ref{fig:nc_dif} based on the constant size assumption for the dust continuum emission.
The blue and green circles and solid lines show the number counts and the best-fit Schechter functions obtained by re-running the same MC simulations for the corrections of the completeness and flux measurements with the assumptions of the negative \citep[e.g.,][]{gonzalez2017,smail2021} and positive \citep{fujimoto2017, fujimoto2018} correlations between $L_{\rm IR}$ and the dust continuum size, respectively. 
{\it Right}: Integration of the 1.2-mm number counts down to our detection limit according to the best-fit Schechter functions with the different size assumptions for the dust continuum emission, suggesting that the resolved fraction of the CIB at 1.2~mm could change by factors of $\sim$0.6--1.1 due to the different size assumptions. 
\label{fig:nc_three}}
\end{center}
\end{figure*}
%%%%%%%%%%%%%%%%%%%%%%% 

With our fiducial Schechter parameter set (Section \ref{sec:number_counts}), we calculate the integrated flux densities, $\int^{\infty}_{S_{\rm limit}}\it \phi(S)dS$, down to the flux limit of $S_{\rm limit}$. We adopt $S_{\rm limit}=7$~$\mu$Jy from the faintest bin of our number count data (Figure \ref{fig:nc_dif}). 
We calculate the integrated flux density to be 20.7$^{+8.5}_{-6.5}$~Jy~deg$^{-2}$ at 1.2 mm. 
Figure \ref{fig:cib} shows the integrated flux density as a function of $S_{\rm limit}$. With the integrated value, we also evaluate the fraction contributed to the CIB measurement at 1.2~mm by interpolating the {\it COBE}/FIRAS measurements at 217~GHz and 353~GHz where the Galactic foreground emission is subtracted with a linear combination of Galactic H{\sc i} and H$\alpha$ \citep{odegard2019}. 
We find that the individual sources, down to $S_{\rm limit}$ of the ALCS sample, contribute to \tcb{76.5$^{+31.4}_{-24.0}$\%} of the CIB at 1.2~mm.  
In Section \ref{sec:impact_comp}, we discuss the impact of the different source size assumptions on this resolved fraction of the CIB. 

As discussed in \cite{gonzalez2020}, we caution that the uncertainty of the CIB itself is large due to the model dependence of the CMB and Galactic dust emission, which are subtracted from the observations to estimate the CIB. 
Still, the same interpolation of the {\it COBE}/FIRAS measurements and the same $S_{\rm limit}$ suggest that $\sim40$\% of the CIB is resolved with the best-fit function for the GOODS+HUDF data points.  
This relatively small resolved fraction supports the interpretation that the GOODS+HUDF regions are underdense. 

The uncertainty of the CIB measurement notwithstanding, our best-fit shape of the number counts ($\alpha\sim-2.0$) indicates that integration exceeds the upper limit of the CIB from $0.2\mu$Jy, instead of reaching down to 0~Jy. This implies that the faint-end slope of the 1.2-mm number counts may flatten or turnover within $\sim$2~dex of our detection limit \citep[e.g.,][]{fujimoto2016, gonzalez2020}. 
\tcb{
If any, the flattening or turnover flux density ($S_{\rm turn}$) is expected to be below $\sim$20~$\mu$Jy (and above 0.2~$\mu$Jy), and we explore the physical origins of the potential flattening/turnover as follows. 
Based on the typical modified blackbody assumption at the median redshift of our ALCS sources ($z\sim2$, Section~\ref{sec:prop}), the 0.2--20$\mu$Jy flux corresponds to a dust mass $M_{\rm dust}$ of $\simeq10^{6.5-9.5},M_{\odot}$.
Assuming a typical dust-to-stellar mass ratio of 0.001 \citep[e.g.,][]{santini2014} and the $M_{\rm star}$--metallicity relation \citep[e.g.,][]{mannucci2010, sanders2016, iyer2018}, this $M_{\rm dust}$ range corresponds to 12+$\log$(O/H) $\sim$ 7.5--8.5. Interestingly, the gas-to-dust mass ratio (GDR) is known to increase at low metallicities of 12+$\log$(O/H) $\lesssim8.0$ \citep[e.g.,][]{Asano2013, remy-ruyer2014}, suggesting that the observed $S_{\rm turn}$ of 0.2--20~$\mu$Jy may be related to the break in the GDR--metallicity relation.}
On the other hand, an integration with a shallower faint-end slope of $\alpha \gtrsim -1.8$ does not diverge and thus does not require flattening nor turnover. Our fiducial Schechter estimate of $\alpha=-2.05^{+0.12}_{-0.10}$ is still consistent with $\alpha=-1.8$ within $\sim2\sigma$. 
If we adopt a different size assumption for the dust continuum in the completeness correction (Section~\ref{sec:impact_comp}) and add more weight to the cluster fields with higher quality lens models in the number-count calculations (Section~\ref{sec:caveats}), the $\alpha$ measurement becomes close to $\alpha\sim-1.8$.  
Therefore, our results remain consistent with the possibility of $\alpha \sim -1.8$ and no turnover/flattening.

\subsection{Impact from the dust continuum size distribution}
\label{sec:impact_comp}

In Section \ref{sec:number_counts}, we derive the number counts based on the completeness estimate with the assumption that the intrinsic source size is constant (Section \ref{sec:simulation}). 
However, recent ALMA studies suggest that there exists a negative \citep[e.g.,][]{gonzalez2017,tadaki2020, smail2021} or positive \citep{fujimoto2017, fujimoto2018} correlation between the dust continuum size and IR luminosity.  
Interestingly, recent deep ALMA follow-up observations for a strongly ($\mu\sim9$) lensed galaxy at $z=7.13$ also report hints of a very extended dust continuum structure beyond the stellar continuum observed in the rest-frame UV \citep{akins2022}. 
The different assumptions of the intrinsic source size distribution affect the completeness correction, especially for the strongly lensed galaxies \citep[e.g.,][]{kawamata2018}, which may have an impact on the shape of the 1.2-mm number counts and the contribution to the CIB. 
Therefore, we also derive the 1.2-mm number counts with the following two different assumptions for the intrinsic source size distribution based on the literature: 1) negative correlation of $r_{\rm e, dust}\propto L_{\rm IR}^{-0.53}$, and 2) positive correlation of $r_{\rm e, dust}\propto L_{\rm IR}^{0.28}$, where $L_{\rm IR}$ is the IR luminosity and $r_{\rm e, dust}$ is the effective radius of the dust continuum emission. We carry out the same MC simulations for correcting the flux measurements, completeness (Section \ref{sec:simulation}), and associated uncertainties in the derivation of the number counts (Section \ref{sec:number_counts}) with these different size assumptions. To convert the 1.2-mm flux density to $L_{\rm IR}$, we assume a single modified black body with dust temperature $T_{\rm dust}=35$~K \citep[e.g.,][]{coppin2008} and $\beta_{\rm d}=1.8$ \citep[e.g.,][]{planck2011}.

In Figure \ref{fig:nc_three}, we show the 1.2-mm number counts with the different size assumptions. 
The red, green, and blue circles with the error bars and curves represent the average and 16-84th percentile from the MC simulation and the best-fit Schechter functions for the results that are all obtained in the same manner as Section \ref{sec:number_counts}, but with the size assumptions of constant, negative, and positive correlations, respectively. 
We find that the different size assumptions can change the faint-end slope. 
This is because, in the positive (negative) correlation, the fainter source is more compact (extended), which requires the completeness correction less (more) than that in the constant size assumption case. 
We also find that the different size assumption provides negligible effects on the other parameters of the DPL function. 

With the best-fit Schechter parameters, we also integrate the 1.2-mm number counts and evaluate the contributions to the CIB with the different size assumptions. 
We find that the resolved fraction reaches $\sim$90\% and $\sim$50\% with the negative and positive correlation cases, respectively. Compared to our fiducial estimate with the constant size assumption, we find that the resolved fraction of the CIB can be changed by factors of $\sim$0.6--1.1 by the different size assumption. 
To evaluate the faint-end of the 1.2-mm number counts with less than the above precision, our results suggest the importance of constraining the source size distribution for the dusty galaxies, where little has been still known at $L_{\rm IR}\lesssim10^{12}L_{\odot}$ \citep[e.g.,][]{gonzalez2017, fujimoto2017, fujimoto2018, tadaki2020, smail2021, gomez2021}.  

\subsection{Potential Caveats}
\label{sec:caveats}

%%%%%%%%%%%%%%%%%%%%%%%
\begin{figure}[t!]
\begin{center}
\includegraphics[trim=0cm 0cm 0cm 0cm, clip, angle=0,width=0.5\textwidth]{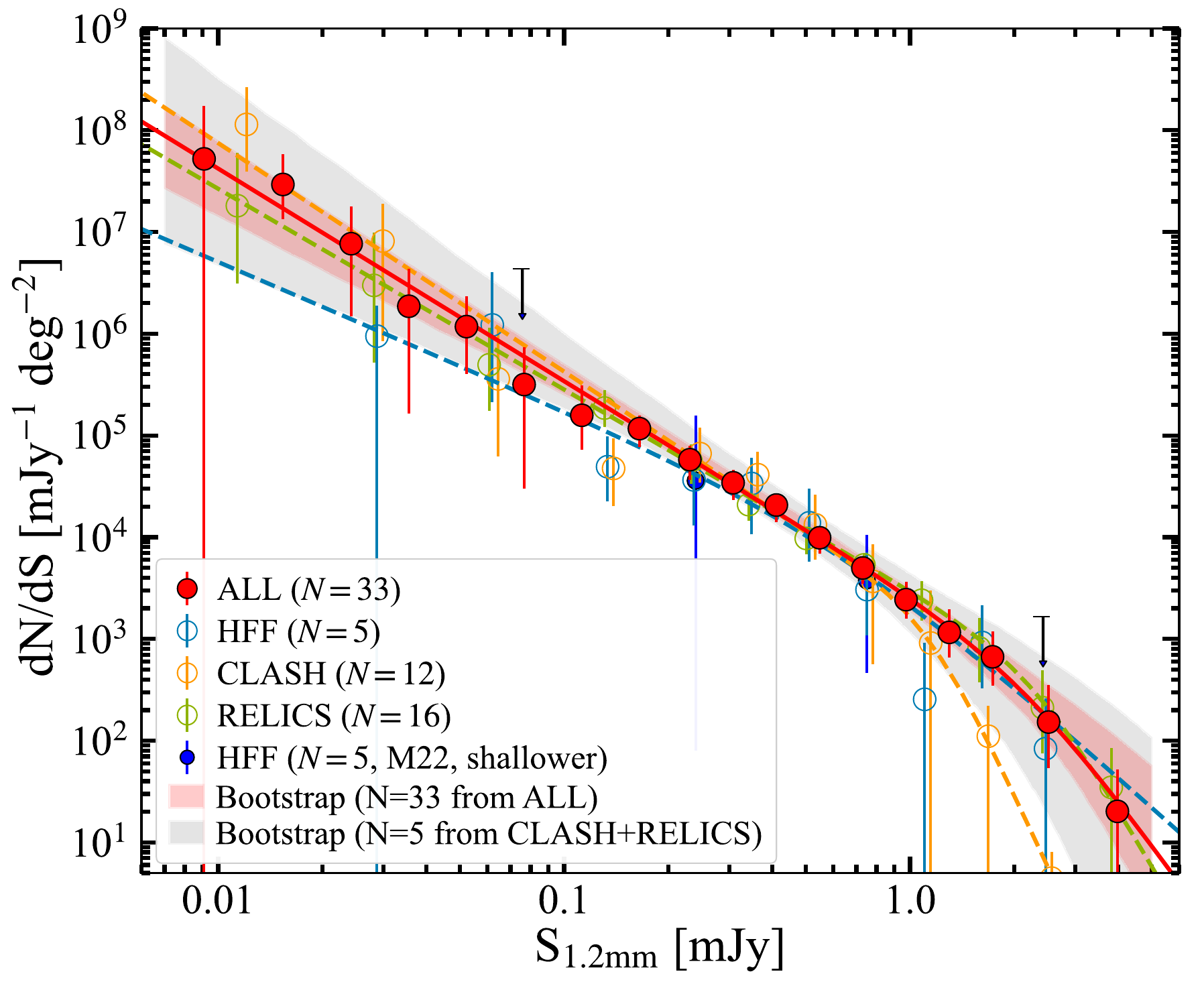}
 \caption{
Comparison of differential number counts at 1.2~mm among different subsamples of ALCS clusters.  
The red circles and line are the same assignments with Figure~\ref{fig:nc_dif}. 
The open blue, orange, and green circles and lines indicate the number counts and their best-fit Schechter functions derived only with the HFF, CLASH, and RELICS clusters, respectively. 
The red and grey shaded regions represent the $1\sigma$ areas evaluated with the Bootstrap tests by sampling 33 clusters from all ALCS clusters (including the duplication) and only 5 clusters from the CLASH and RELICS clusters, respectively. 
The blue circles and upper limits denote the previous ALMA Band~6 survey results in HFF \citep{munoz-arancibia2022}, where the data is shallower than ALCS. 
\label{fig:different_clusters}}
\end{center}
\end{figure}
%%%%%%%%%%%%%%%%%%%%%%% 

Although we evaluate the magnification uncertainty separately in the HFF, CLASH, RELICS clusters (Section~\ref{sec:lens}) and include these magnification uncertainties via the MC method in the number counts analysis (Section~\ref{sec:number_counts}), we caution that there may remain further systematic uncertainties related to the different qualities of the lens models among the clusters. 
In general, as more multiple images are identified in deeper data, the lens models have more granularity and small-scale structure, which in turn increases the estimated source plane area at high magnification becomes larger, compared to a less detailed model based on shallower data and fewer multiple images of the same cluster. In fact, \cite{jauzac2015} compare the surface areas in the source plane above a given threshold magnification in one of the HFF clusters of A2744 and find that the lens models constructed before the HFF underestimate the surface area compared to the lens models constructed with the HFF data by factors of $\sim$1.4--1.8 at $\mu\gtrsim3$ (see Figure~4 in \citealt{jauzac2015}). 
This indicates that the survey areas at high magnification in the CLASH and RELICS clusters ($\approx$ shallower \hst\ data than HFF) are likely underestimated. 
On the other hand, the increased area at high magnifications may lead to high magnification estimates in some galaxies currently with moderate magnification estimates. 
Therefore, the number density of intrinsically faint sources might not be changed much, and the impact on the measurement of the faint-end slope is uncertain.

To examine the potential impact from the different qualities of the lens model among the clusters, Figure~\ref{fig:different_clusters} shows the 1.2~mm number counts derived in the same manner as Section~\ref{sec:number_counts}, but separately with the HFF, CLASH, and RELICS clusters. 
The dashed line indicates the best-fit Schechter function for each subsample, 
and its faint-end slope is summarized in Table~\ref{tab:alpha_test}. 
The faint-end slopes are steeper in CLASH and RELICS ($\alpha_{\rm CLASH}=-2.24_{-0.20}^{+0.24}$, $\alpha_{\rm RELICS}=-1.91_{-0.14}^{+0.17}$) than that in HFF ($\alpha_{\rm HFF}=-1.71_{-0.29}^{+0.34}$), while the large uncertainty in the HFF results still makes all these results consistent within 1$\sigma$ due to its small statistics from the small survey area. 
This indicates that the underestimate of the high magnification area in less detailed lens models might be related to the steeper faint-end slope in CLASH and RELICS than in HFF, while the effect from the cosmic variance might still be a more dominant factor. In fact, the faintest data point in HFF deviates from the best-fit Schechter function with all 33 ALCS clusters (red line) by a factor of $\sim6$.  
The sum of the survey areas of the CLASH and RELICS clusters contributes to $\simeq$ 80\% of the total 33 clusters (Appendix~\ref{sec:app_models}), and the practical factor of the underestimate is $\sim$1.1--1.5. 
This is much smaller than the above factor to fill the gap,  
suggesting that the steeper faint-end slope results in CLASH, RELICS, and the entire 33 clusters than in HFF are unlikely solely due to the underestimate of the high magnification area.

%%%%%%%%%%%%%%%%%%
\setlength{\tabcolsep}{10pt}
\begin{table}
\begin{center}
\caption{
Faint-end slope $\alpha$ in various data and assumption
}
\vspace{-0.4cm}
\label{tab:alpha_test}
\begin{tabular}{ccc}
\hline \hline
Data set             & FIR Size     & $\alpha$  \\
(1)                  &  (2)     &   (3)     \\ \hline
ALL ($N=33$)         & Const.   &  $-2.05_{-0.10}^{+0.12}$        \\
ALL ($N=33$, w/ weight) & Const. & $-1.96_{-0.13}^{+0.14}$         \\
HFF only   ($N=5$)   & Const.   & $-1.71_{-0.29}^{+0.34}$         \\
CLASH only ($N=12$)  & Const.   & $-2.24_{-0.20}^{+0.24}$          \\
RELICS only ($N=16$) & Const.   & $-1.91_{-0.14}^{+0.17}$          \\
Bootstrap ($N=5$)    & Const.   & $-2.02_{-0.15}^{+0.16}$        \\
Bootstrap ($N=33$)   & Const.   & $-2.05_{-0.13}^{+0.13}$          \\
$\mu\lesssim10$ only & Const.   & $-1.94_{-0.15}^{+0.18}$          \\ 
$\mu\lesssim3$ only  & Const.    & $-2.00_{-0.23}^{+0.34}$         \\ 
ALL ($N=33$) & $\propto L_{\rm IR}^{a\dagger}$ & $-1.86_{-0.14}^{+0.16}$   \\ 
ALL ($N=33$) & $\propto L_{\rm IR}^{-b\dagger}$ & $-2.12_{-0.11}^{+0.13}$  \\ 
\hline \hline
\end{tabular}
\end{center}
\vspace{-0.4cm}
\tablecomments{
(1) Various data sets for the $\alpha$ measurement.  
(2) FIR size assumption in the completeness correction estimate. 
(3) Derived faint-end slope of $\alpha$ in the Schechter function. \\
$\dagger$ Assuming positive and negative correlations with $L_{\rm IR}$ (see Sec.~\ref{sec:impact_comp}). 
}
\end{table}
%%%%%%%%%%%%%%%%%%

To evaluate the cosmic variance effect in the 5 HFF clusters, we also perform the Bootstrap test by randomly sampling 5 clusters from CLASH and RELICS. 
In Figure~\ref{fig:different_clusters}, the grey shaded region shows the 1$\sigma$ region from the 1,000 realizations of the Bootstrap test. For comparison, we also show the red shaded region, corresponding to the 1$\sigma$ region obtained from another Bootstrap test by sampling 33 clusters among all ALCS clusters, including the duplication. 
We find that the best-fit Schechter function in HFF is included in the 1$\sigma$ region from the 5 random CLASH+RELICS clusters. 
We thus conclude that we cannot rule out the possibility that the different faint-end slope results among different clusters are still dominated by the cosmic variance. 

As an alternative approach to take the different qualities of the lens models among the clusters, we also add weights to the number counts according to the quality of the lens model. 
Based on the number of the multiple images ($N_{\rm img}$) identified in each cluster field (Table~\ref{tab:model_all}), we classify the five HFF clusters and the R1347 and M1206 as a group of the good model with $N_{\rm img}>100$. Following the results of the relative magnification uncertainty among the lens models as a function of the total number of the multiple presented in \citep{johnson2016}, we add four times more weights on the number counts of the good model clusters than the others. This yields $-1.96_{-0.13}^{+0.14}$, which is consistent with other results within 1$\sigma$. 

To reduce any systematic uncertainty from lensing effects, we also estimate the faint-end slope only with the data points at $>0.03$~mJy and $>0.1$~mJy from all 33 clusters, which are equal to using only the sources with $\mu\lesssim10$ and $\mu\lesssim3$, given the typical 5$\sigma$ detection limit of 0.3~mJy ($=5\times0.06$~mJy; see Table~\ref{tab:data_prop}). 
We obtain the best-fit $\alpha$ of $-1.94_{-0.15}^{+0.18}$ and $-2.00_{-0.23}^{+0.34}$ for the sources only with $\mu\lesssim10$ and $\mu\lesssim3$, respectively. 
Both $\alpha$ measurements are consistent with our fiducial estimate ($-2.05_{-0.10}^{+0.12}$) down to $\sim0.01$~mJy within 1$\sigma$.

Table~\ref{tab:alpha_test} summarizes the $\alpha$ measurement for each subsample above. We also list the $\alpha$ measurements by adopting the different FIR size assumptions with all 33 ALCS clusters (Section~\ref{sec:impact_comp}). 
We find that all these $\alpha$ measurements are consistent within the $\sim1$--2$\sigma$ errors, regardless of the subsample and the FIR size assumptions. 
These various test results suggest the $\alpha$ measurement with a conservative uncertainty would be $\simeq -2.0^{+0.2}_{-0.3}$.

\section{Luminosity Function \& Cosmic SFR Density}
\label{sec:discuss}

\subsection{IR Luminosity Function at $z\simeq$ 1--8}
\label{sec:irlf}

%%%%%%%%%%%%%%%%%%
\setlength{\tabcolsep}{8pt}
\begin{table*}
\caption{ALCS IR LFs}
\vspace{-0.8cm}
\label{tab:irlf}
\begin{center}
\begin{tabular}{lcccccc}
\hline \hline
$\log(L_{\rm IR}) [L_{\odot}]$ &  \multicolumn{6}{c}{$\log(\Phi)\,[\rm Mpc^{-1}dex^{-1}]$} \\
                            &  $0.6 \leq z < 1.0$ & $1.0 \leq z < 2.0$ & $2.0 \leq z < 3.0$ & $3.0 \leq z < 4.0$ & $4.0 \leq z < 6.0$ & $6.0 \leq z < 7.5$  \\ \hline 
9.9--10.4 & $\cdots$ & $-0.85^{+0.52}_{-0.76}$ & $\cdots$ & $\cdots$ & $\cdots$ & $-2.49^{+0.52}_{-0.76}$ \\
10.4--10.9 & $-3.72^{+2.35}_{-99.9}$ & $-1.93^{+0.3}_{-0.34}$ & $\cdots$ & $-1.94^{+0.3}_{-0.34}$ & $-3.24^{+0.52}_{-0.76}$ & $\cdots$ \\
10.9--11.4 & $-2.13^{+0.4}_{-1.24}$ & $-2.54^{+0.2}_{-0.22}$ & $-2.86^{+0.3}_{-0.34}$ & $-2.96^{+0.25}_{-0.28}$ & $\cdots$ & $\cdots$ \\
11.4--11.9 & $-3.38^{+0.54}_{-0.25}$ & $-2.95^{+0.13}_{-0.15}$ & $-3.05^{+0.1}_{-0.12}$ & $-3.98^{+0.26}_{-0.28}$ & $-3.78^{+0.22}_{-0.25}$ & $\cdots$ \\
11.9--12.2 & $-3.59^{+0.57}_{-0.28}$ & $-3.16^{+0.1}_{-0.17}$ & $-3.59^{+0.12}_{-0.17}$ & $-3.68^{+0.2}_{-0.22}$ & $-4.3^{+0.3}_{-0.34}$ & $\cdots$ \\
12.2--12.6 & $-4.28^{+0.52}_{-99.9}$ & $-3.62^{+0.2}_{-0.22}$ & $-3.91^{+0.22}_{-0.25}$ & $-3.93^{+0.25}_{-0.28}$ & $-4.51^{+0.37}_{-0.45}$ & $\cdots$ \\
12.6--12.9 & $\cdots$ & $-4.59^{+0.52}_{-99.9}$ & $-4.59^{+0.52}_{-99.9}$ & $-4.27^{+0.37}_{-0.45}$ & $\cdots$ & $\cdots$ \\
12.9--13.2 & $\cdots$ & $-4.5^{+0.52}_{-0.76}$ & $-4.52^{+0.52}_{-99.9}$ & $\cdots$ & $\cdots$ & $\cdots$ \\
\hline
\end{tabular}
\end{center}
\end{table*}
%%%%%%%%%%%%%%%%%%

While the IR LFs have been extensively studied with  \textit{Herschel} \citep[e.g.,][]{gruppioni2013, magnelli2013, lwang2019}, 
the results are inevitably affected by the large beam size, which causes the source blending and the confusion limit in the source detection. 
The IR LFs have also been studied with recent ALMA observations \citep[e.g.,][]{koprowski2017, hatsukade2018, gruppioni2020, zavala2021}, 
while only limited luminosity and redshift range have been constrained by the small statistics due to the small FoV of ALMA. 

Taking advantage of the large blind sample of our ALCS sources (Section \ref{sec:prop}),   
we also analyze the IR LFs based on their secure redshift estimates that benefit from the homogeneous {\it HST} and IRAC data sets. 
Because we confirm the consistency between the blind and blind+prior samples in the number counts (Section \ref{sec:number_counts}), we use the blind+prior sample to obtain the statistically reliable results in the following analyses. 

\tcb{To calculate $L_{\rm IR}$ for our ALCS sources, we use a panchromatic SED fitting tool {\sc stardust} \citep{kokorev2021} based on the IRAC \citep{kokorev2022}, {\it Herschel} \citep{sun2022}, and ALMA photometry (Section~\ref{sec:flux}), including the upper limits ($2\sigma$). 
The {\sc stardust} code utilizes \cite{draine2007} (hereafter DL07) templates for the IR--mm SED with the dust emission model, including the additional updates from \cite{draine2014} (see also \citealt{aniano2020}). 
These models describe the contribution from warm dust and polycyclic aromatic hydrocarbon (PAH) features in the photodissociation regions, together with cold dust in the diffuse part of the ISM. 
To ensure our fits are physically meaningful, we restrict our distribution of the fitting parameters as follows. The minimum radiation field intensity ($U_{\rm min}$), acts as a proxy for $T_{\rm dust}$ in the DL07 set. 
We restrict $U_{\rm min}$ to lie in the $3.0 < U_{\rm min} < 50.0$ range, which corresponds to a range of 20~K~$\lesssim T_{\rm dust}\,\lesssim$~50~K. 
We additionally restrict the fraction of the total dust mass locked in PAHs ($q_{\rm PAH}$) between 0 \% and 10\%. We have fixed $U_{\rm max} = 10^6$ and $\alpha= 2$, following both \cite{magdis2012} and \cite{kokorev2021}. 
Among our 180 ALCS sources, 125 sources are detected at least with 1 band in {\it Herschel}/SPIRE or PACS \citep{sun2021}, providing moderately secure FIR SED models based on the ALMA 1.2-mm detection and the 2$\sigma$ upper limits in the non-detection bands. Although caution may arise, we also infer $L_{\rm IR}$ for the remaining sources without any {\it Herschel} detection based on the $2\sigma$ upper limits and the secure ALMA 1.2-mm detection. 
}

Based on the associated X-ray emission, three ALCS sources are reported to be lensed AGNs (A370-C110, M0416-C117, and M0329-C11; \citealt{uematsu2023}), and we use the $L_{\rm IR}$ estimates after subtracting the AGN component evaluated in \cite{uematsu2023} for these three ALCS sources. 
For the other ALCS sources, we do not take the AGN component into account in the IR--mm SED fitting with {\sc stardust}, while, after the fitting, we systematically subtract the typical fraction of 20\%, which is estimated from deep stacking and dedicated SED analysis for faint mm sources \citep{dunlop2017}. 
Our $L_{\rm IR}$ measurements are summarized in Table~\ref{tab:alcs_catalog_full}. We confirm that these measurements \tcb{show no systematic offsets from the estimates presented in \cite{sun2022} that use another fitting code of {\sc magphys} \citep{dacunha2015} within $1\sigma$ uncertainties. The consistency in the FIR SED fitting outputs between {\sc magphys} and {\sc cigale} \citep{boquien2019} has also been confirmed in our ALCS sources \citep{uematsu2024}.}

\begin{figure*}
\begin{center}
\includegraphics[trim=0cm 0cm 0cm 0cm, clip, angle=0,width=1.0\textwidth]{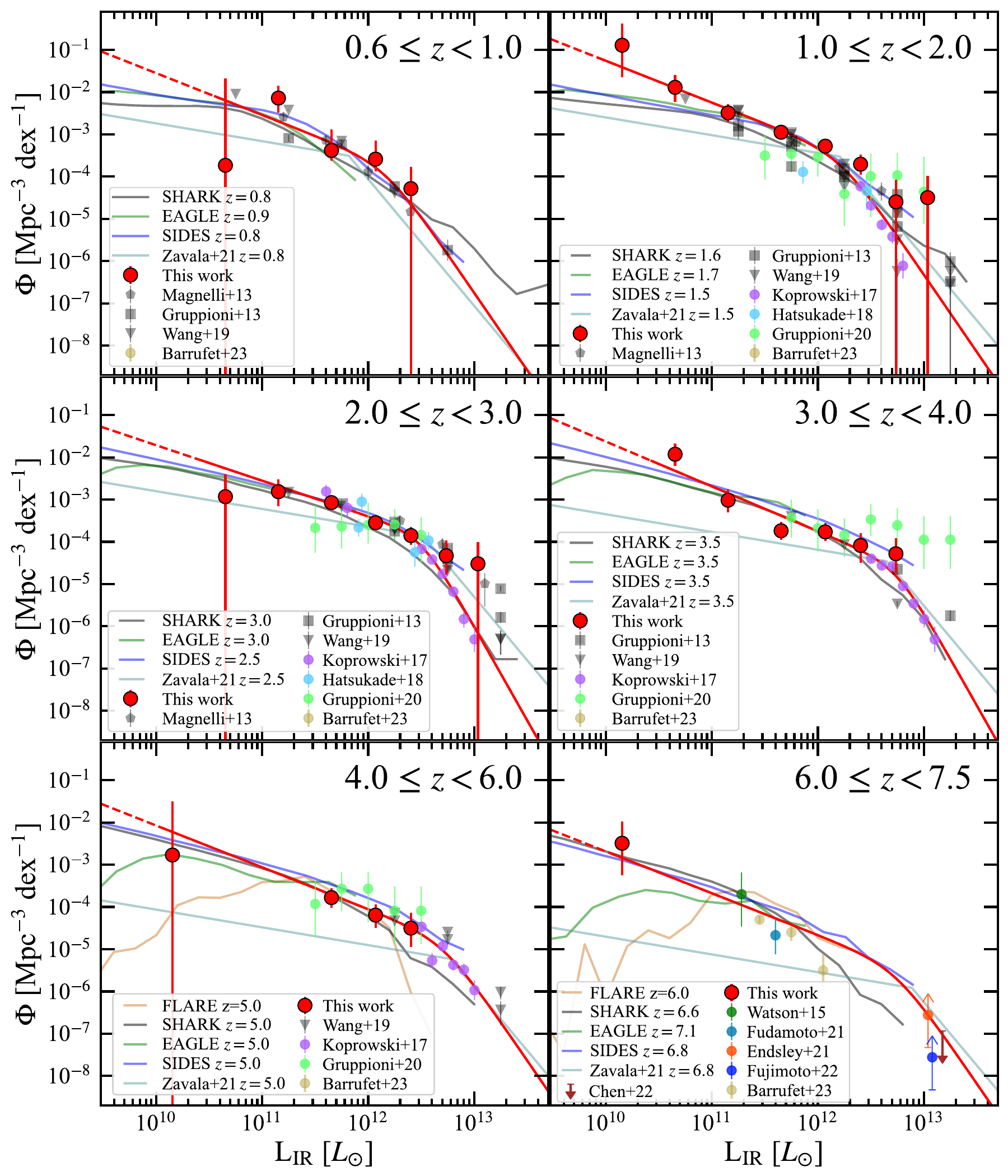}
 \caption{
ALCS IR LF measurements at $z=0.6$--7.5. 
The red-filled circles represent our ALCS sources corrected from the MC simulations that implement all relevant uncertainties, such as redshift, magnification, and flux density measurements, 
where the error bars indicate the 16-84th percentile in the 1,000 MC realizations \tcb{(see Section~\ref{sec:number_counts})}. 
Previous IR LF measurements with {\it Herschel} \citep{magnelli2013, gruppioni2013, lwang2019} and ALMA(+SCUBA2) are presented in grey and colored symbols, respectively. 
The red line presents the best-fit DPL functions, where we include our measurements and other ALMA blind survey results \citep{koprowski2017, hatsukade2018} in the fitting. 
The other color lines show the predictions from simulations \citep{lagos2020, trayford2020, vijayan2022, bethermin2022} and an empirically-calibrated model \citep{zavala2021}. 
\label{fig:irlf}}
\end{center}
\end{figure*}

We adopt the $V_{\rm max}$ method \citep{schmidt1968} to evaluate the IR LFs. 
This method uses the maximum observable volume of each source. 
The LF gives the number of the ALMA sources in a comoving volume per logarithm of luminosity, given by 
\begin{eqnarray}
\Phi(L, z) = \frac{1}{\Delta L} \Sigma \frac{1}{C V_{\rm max, i}}, 
\end{eqnarray}
where $V_{\rm max,i}$ is the maximum observable volume of the $i$th source, $C$ is the completeness, and $\Delta L$ is the width of the luminosity bin in the logarithm scale. We calculate $V_{\rm max}$ as the integration of co-moving volume spherical shells, given by 

\begin{eqnarray}
V_{\rm max,i} = \int^{z_{\rm max}}_{z_{\rm min}}\frac{\Omega(S_{i(z)})}{4 \pi}\frac{dV}{dz}dz,
\end{eqnarray}
where $S_{i}(z)$ is the flux density of $i$th source observed at $z$, $\Omega$ is the effective survey area that can detect the flux density of the $i$th source, and 
$z_{\rm min}$ and $z_{\rm max}$ are maximum and minimum redshifts of a redshift bin.
To study the faint ALMA sources that are gravitationally lensed behind the massive lensing clusters ($z\lesssim0.5$), we focus on the redshift range beyond $z=0.6$ and derive the IR LFs with the redshift bins of 
$0.6\leq z < 1.0$ ($N=13$), 
$1.0\leq z < 2.0$ ($N=46$), 
$2.0\leq z < 3.0$ ($N=39$), 
$3.0\leq z < 4.0$ ($N=30$), 
$4.0\leq z < 6.0$ ($N=12$), and 
$6.0\leq z < 7.5$ ($N=1$). 
To include all relevant uncertainties, such as the source redshift and the lensing magnification, we similarly perform the MC simulations as Section \ref{sec:number_counts}. We make a mock catalog of the ALCS sources whose flux densities and redshifts are randomly redistributed based on the Gaussian probability distribution and re-calculate the magnification according to the redshift. 
We divide each luminosity and redshift bin and repeat this procedure 1,000 times. 
In Figure \ref{fig:irlf}, we show the our IRLFs results at $z=0.6$--7.5. 
The red circles and error bars denote the average and 16-84 percentile from the MC realization.

For comparison, Figure \ref{fig:irlf} also presents previous IR LF measurements in the literature from {\it Herschel} \citep{magnelli2013, gruppioni2013, lwang2019} and ALMA observations \citep{koprowski2017, hatsukade2018, gruppioni2020, fudamoto2021, barrufet2023b}. 
We adopt the central values of the redshift ranges in the previous measurements to assign them to our redshift bins. 
When there are multiple measurements in a redshift bin from one literature, 
we show the previous measurement whose central value of the redshift range is the closest to our redshift bin. 

To obtain statistically stable results in the highest redshift bin of $z=6.0$--7.5, we also calculate the volume density of several IR-detected sources serendipitously identified in blind surveys in the literature. 
\cite{watson2015} report the faint dust emission from a Lyman-break galaxy at $z=7.13$ \citep{wong2022} in an ALMA blind continuum survey in Abell~1689. \cite{fujimoto2016} estimate the effective survey area of this ALMA survey in Abell~1689 in the same manner as Section \ref{sec:survey_area}, in which we use the survey volume calculation. 
\cite{endsley2022a} identify 41 luminous LBGs at $z\simeq$6.6--6.9 by combining the narrow and broad band filters in 1.5~deg$^{2}$ COSMOS field, where one of the LBGs turns out to be obscured AGN population and with $L_{\rm IR}$ of $9.0\times10^{12}\,L_{\odot}$ after subtracting the AGN contribution and spectroscopically confirmed at $z=6.853$ \citep{endsley2022b}. 
We compute the survey volume with the area of 1.5~deg$^{2}$ and the redshift range of $z=$ 6.6--6.9.   
\cite{fujimoto2022} report the discovery of a red quasar at $z=7.19$, GNz7q, enshrouded by a dusty starburst host in the systematic analysis of all publically available {\it HST} archive ($\sim$3 deg$^{2}$). After subtracting the AGN contribution, the host galaxy is estimated to have $L_{\rm IR}$ of 1.1$\times10^{13}$. 
Given that GNz7q is identified from the unambiguous Lyman-break feature in the \hst\ bands in the first place, we adopt the redshift range of $\Delta z=1.0$, which is a typical range of the dropout technique to identify $z\sim7$ galaxies, and survey area of 3~deg$^{2}$ to compute its survey volume. 
The errors of these estimates are obtained from the Poisson confidence limit of 84.13\% in \cite{gehrels1986}. 
Because the sources in \cite{endsley2022a} and \cite{fujimoto2022} are identified initially in the optical--NIR surveys, we place the lower limits for these estimates. 
\cite{chen2022} perform the ALMA line scan observations for the brightest submm sources at 850 $\mu$m ($S_{850\mu \rm m} > 5$~mJy) and identify no sources at $z>6$, despite the redshift range covered in the line scan includes $z=6.0$--7.2. We thus place the 84.13\% upper limit from Poisson statistic \citep{gehrels1986}, where $L_{\rm IR}$ is calculated by assuming the single modified black body at $z=6.5$ with $S_{\rm 850\mu m}=5$~mJy with the typical $T_{\rm dust}$ of 35~K \citep[e.g.,][]{coppin2008} and $\beta_{\rm d}=1.8$ \citep[e.g.,]{planck2011}. 

In Figure \ref{fig:irlf}, we find that our IR LF estimates explore the faint-end by $\sim$1--2 dex in most redshift bins, compared to the previous studies. We also find that our IR LF estimates are generally consistent with the previous studies within the 1$\sigma$ uncertainties. 

To evaluate the shape of IR LFs, we perform the DPL fit to our IR LF estimates. 
For the LF form, we define the DPL by 
\begin{eqnarray}
\label{eq:dpl}
\Phi(L) = \Phi_{\star} \left[\Bigl(\frac{L}{L_{\star}}\Bigr)^{\alpha}+\Bigl(\frac{L}{L_{\star}}\Bigr)^{\beta}\right]^{-1}, 
\end{eqnarray}
where $L_{\star}$ is the characteristic IR luminosity, 
and $\Phi_{\star}$ is the volume density at $L_{\star}$. 
In the fitting, 
we make the faint- and bright-end slopes $\alpha$ and $\beta$ as free parameters in the redshift bins at $z=0.6$--4.0.
On the other hand, we fix both of them at $z\geq4$ by using the best-fit results at $z=1.0$--2.0, where the IR LF shape is the most securely determined with the largest sample size.  
We include the data points in the fitting that are obtained from other ALMA blind surveys in the literature \citep{koprowski2017, hatsukade2018}.  
In the highest redshift bin at $z=6.0$--7.5, we also use measurements from \cite{fudamoto2021} and our calculation from \cite{watson2015} for the fitting, \tcb{given their serendipitous discovery nature, while we do not include the measurements from \cite{barrufet2023b} due to its targeted observation aspect.} 
In addition, because the bright end $z=6.0$--7.5 is also constrained both from the lower and upper limits \citep{endsley2022a, fujimoto2022, chen2022}, we also calculate the central value from these upper and lower limits and use it for the fitting, where we regard the difference between the average and the upper/lower limits as the errors. 

%%%%%%%%%%%%%%%%%%
\setlength{\tabcolsep}{2pt}
\begin{table}
\begin{center}
\caption{Best-fit DPL parameters of ALCS IR LFs}
\label{tab:irlf_dpl_param}
\vspace{-0.4cm}
\begin{tabular}{ccccc}
\hline \hline
$z$ range & $\alpha$ & $\beta$ & $\log(\Phi_{\star})$ & $\log(L_{\star})$ \\ \hline
(1) & (2) & (3) & (4) & (5) \\ \hline
0.6--1.0 &$ 0.94 \pm \tcb{0.39} $ & (3.72) &$ -3.65 \pm 0.12 $ &$ 12.15 \pm 0.06 $ \\
1.0--2.0 &$ 0.94 \pm 0.43 $ &$ 3.72 \pm 1.10 $ &$ -3.38 \pm 0.62 $ &$ 12.21 \pm 0.27 $ \\
2.0--3.0 &$ 0.93 \pm 0.26 $ & (3.72) &$ -3.86 \pm 0.3 $ &$ 12.52 \pm 0.12 $ \\
3.0--4.0 &$ 1.04 \pm 0.20 $ & (3.72) &$ -4.49 \pm 0.36 $ &$ 12.73 \pm 0.15 $ \\
4.0--6.0 & (0.94) & (3.72) &$ -4.77 \pm 0.19 $ &$ 12.72 \pm 0.09 $ \\
6.0--7.5 & (0.94) & (3.72) &$ -5.26 \pm 0.43 $ &$ 12.71 \pm 0.26 $ \\
\hline
\end{tabular}
\end{center}
\vspace{-0.4cm}
\tablecomments{
(1): Redshift bin range of ALCS IR LFs.  
(2): Faint-end slope in DPL function. 
It is fixed in the fitting in the redshift bins of $z=4.0$--6.0 and $z=6.0$--7.5.  
(3): Bright-end slope in DPL function. 
It is fixed in the fitting in the redshift bins, except for $z=1.0$--2.0.  
(4): Characteristic volume density in DPL function.
(5): Characteristic IR luminosity in DPL function. 
}
\end{table}
%%%%%%%%%%%%%%%%%%
In Figure \ref{fig:irlf}, we present the best-fit DPL functions with the red solid curves that are extrapolated with the dashed curves below the faintest luminosity bin. The best-fit parameters in each redshift bin are summarized in Table \ref{tab:irlf}. 
We find that our and previous IR LF measurements are well represented by the DPL functions. 
However, we find that our best-fit values of the faint-end slope, $\alpha\simeq$ 0.9--1.0, are higher than the previous reports of $\alpha\simeq$ 0.4--0.6 \citep{magnelli2013,gruppioni2013, koprowski2017, casey2018b, zavala2021}, 
while our results are close to the reports in the UVLF studies of $\simeq1.5$--2.0 at similar redshifts \citep[e.g.,][]{bouwens2022, finkelstein2022a}. 
A wide dynamic range is essential to determine the IR LF shapes securely. 
Compared to our measurements, the previous IR LF measurements generally cover only the part of the IR luminosity range close to $L_{\star}$, which could be the reason for the difference. 
We note for the previous {\it Herschel} studies that the faint-end slope is generally fixed with a canonical value estimated at their lowest redshift bin at $z\lesssim$ 0.3 \citep{magnelli2013, gruppioni2013}. 
The UV LF studies suggest that the faint-end slope becomes steeper towards high redshifts \citep[e.g.,][]{bouwens2022, finkelstein2022a}. Besides, the physical process of triggering the dusty star-forming activity may be much different between local and high-redshift universe due to the rapid decrease of the gas fraction towards lower redshifts \citep[e.g.,][]{casey2014, tacconi2020}. 
The superb sensitivity and spatial resolution of ALMA would also contribute to overcoming the confusion limit and reduce the impact of the incompleteness of identifying the faint sources. 
We thus think that the difference in the faint-end slope estimate would be explained by the combination of these reasons described above.

%%%%%%%%%%%%%%%%%%%%%%%
%%%%%%%%%%%%%%%%%%%%%%% 

For comparison, Figure \ref{fig:irlf} also shows predictions from recent simulations \citep{lagos2020, trayford2020, bethermin2022,vijayan2022} and an empirically-calibrated model \citep{zavala2021} in colored lines. 
Note that the turnover at faint luminosities in EAGLE (green) and FLARES (brown) simulations is caused by the incompleteness due to the mass resolution limit at $\sim10^{9}M_{\odot}$ in their calculations. 
We find that our measurements and the best-fit DPL functions are generally consistent with these recent simulations within the 1--2$\sigma$ errors, \tcb{except for the model of \tcb{\cite{zavala2021}}.
Since the \cite{zavala2021} model is calibrated to the faint-end 1.2-mm number counts only from the ASPECS results in HUDF \citep{gonzalez2020}, its deviation from our and other measurements and simulation predictions is likely explained by the underdense of HUDF (Section~\ref{sec:nc_comp_pre}). 
}
We also find that our best-fit DPL functions start to deviate from the predictions of the simulations at the faint end of $L_{\rm IR} < 10^{11}L_{\odot}$ in the redshift bins of $z=0.6$--4.0.
These trends are consistent with the results in the 1.2-mm number counts (Figure~\ref{fig:nc_comp}). 
These results might indicate that our current understanding of the dependence between dust mass and metallicity among low-mass galaxies might be different. Still, since the different size distribution may raise the same impact on the completeness correction in the faint-end regime as the 1.2-mm number counts (Section \ref{sec:impact_comp}), the positive correlation between the IR luminosity and size may mitigate the deviation by a factor of $\sim0.6$.

\subsection{Redshift Evolution of IR LF}
\label{sec:z_irlf}

%%%%%%%%%%%%%%%%%%%%%%%
\begin{figure}
\begin{center}
\includegraphics[trim=0.1cm 0cm 0cm 0cm, clip, angle=0,width=0.5\textwidth]{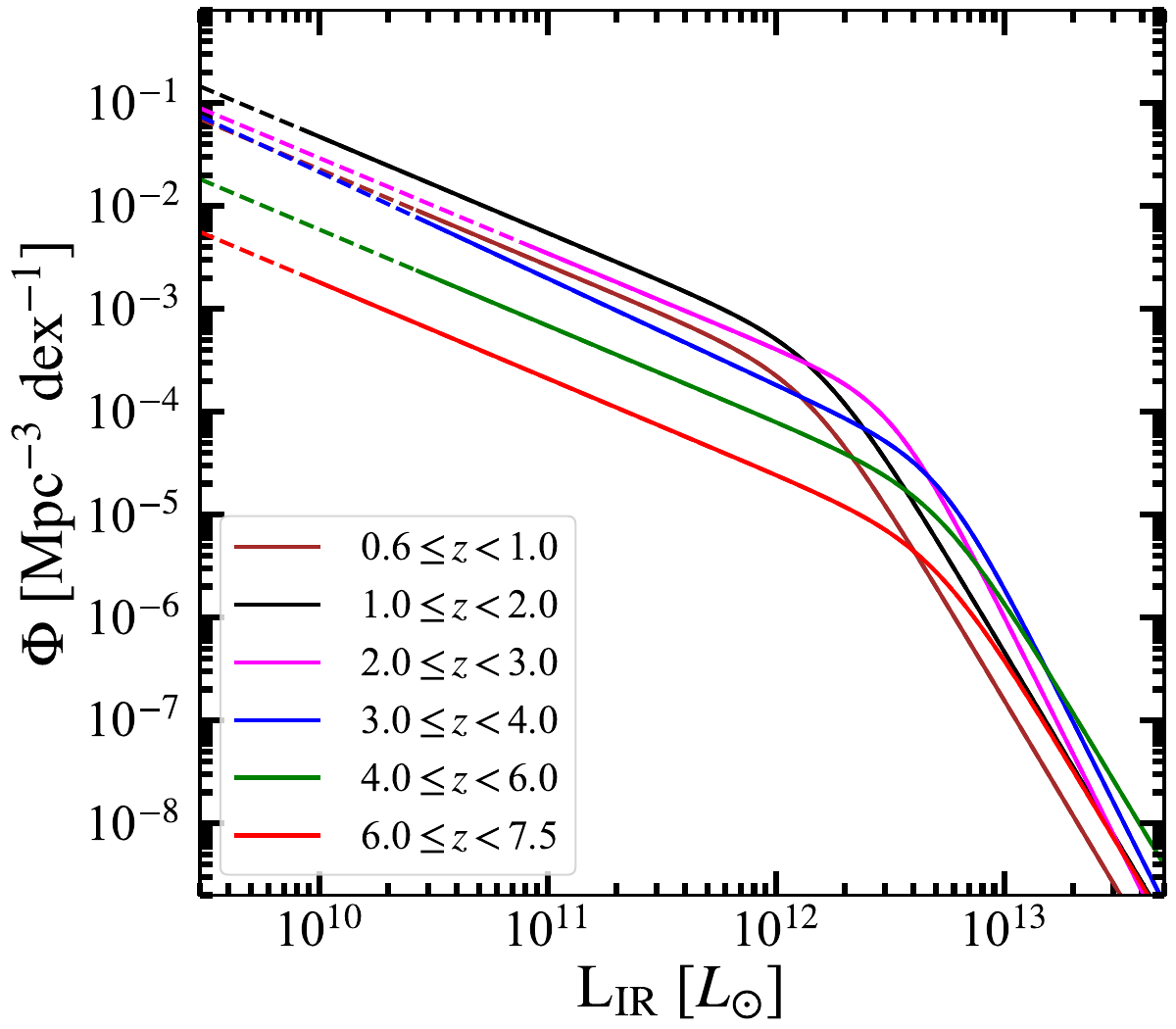}
 \caption{
Evolution of IR LFs based on our best-fit DPL functions presented in Figure \ref{fig:irlf}. 
We find the overall trends characterized by the positive luminosity evolution coupled with the negative density evolution. 
\label{fig:z_IRLF}}
\end{center}
\end{figure}
%%%%%%%%%%%%%%%%%%%%%%% 

%%%%%%%%%%%%%%%%%%%%%%%
\begin{figure*}
\begin{center}
\includegraphics[trim=0cm 0cm 0cm 0cm, clip, angle=0,width=1.0\textwidth]{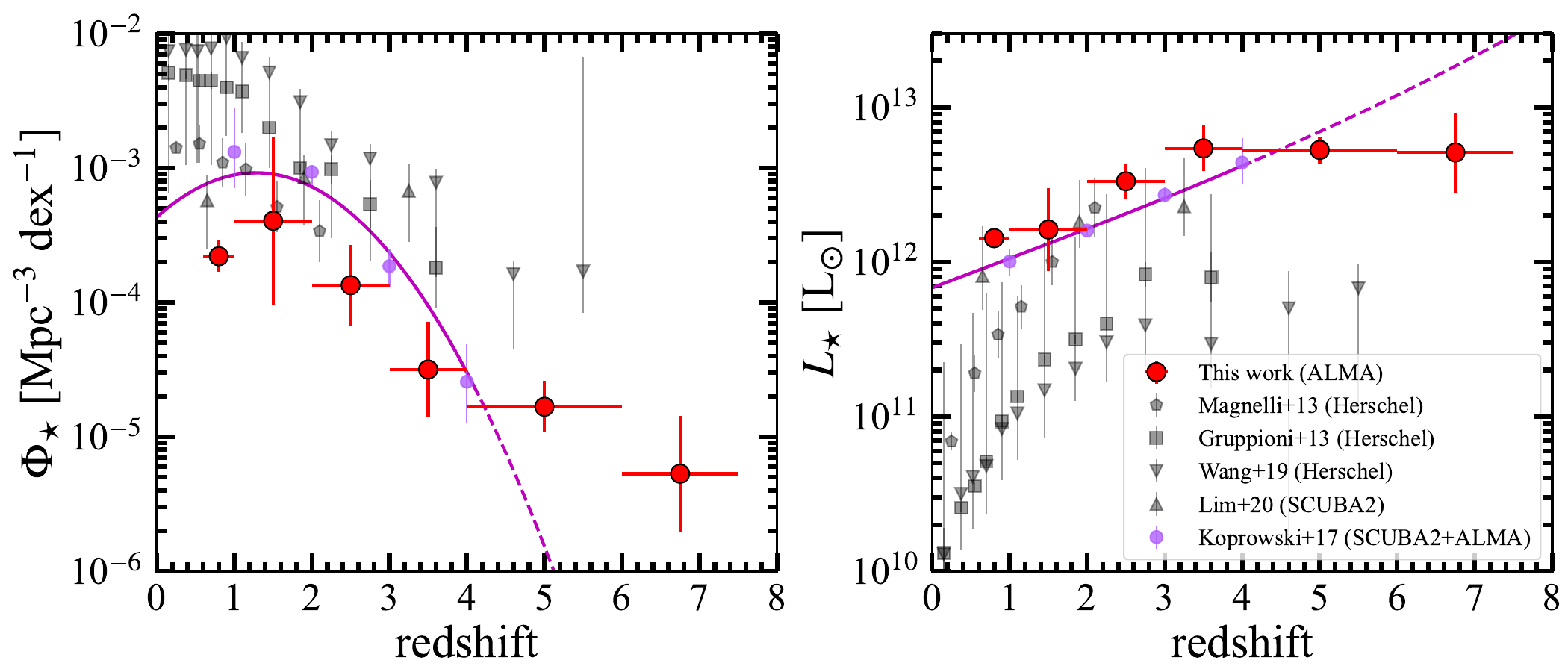}
 \caption{
Evolution of $\Phi_{\star}$ (left) and $L_{\star}$ (right). 
The red circles are our IR LF measurements with the best-fit DPL functions (Table \ref{tab:irlf_dpl_param}). We note in our measurements that all parameters in the DPL function are free in the fitting for the $z=$1.0--2.0 bin, while the bright-end (and faint-end) slope is fixed for the other bins.
For comparison, we also show the previous mesurments with {\it Herschel} \citep{magnelli2013, gruppioni2013, lwang2019}, SCUBA2 \citep{lim2020}, and SCUAB2+ALMA \citep{koprowski2017}. 
The magenta solid and dashed lines are the best-fit measurements presented in \cite{koprowski2017} and their extrapolations, respectively. 
\label{fig:z_Ls-phi}}
\end{center}
\end{figure*}
%%%%%%%%%%%%%%%%%%%%%%% 

In Figure \ref{fig:z_IRLF}, we compare the best-fit DPL functions from $z=0.6$ to 7.5. We find that $\Phi_{\star}$ and $L_{\star}$ decrease and increase towards high redshifts, respectively. 
These trends are thought to represent the apparent ``down-sizing'' of the mm source population \citep[e.g.,][]{koprowski2017}, which is consistent with the evolution in median redshift with the 1-mm flux density (Section~\ref{sec:prop}).  
In Figure \ref{fig:z_Ls-phi}, we also summarize the redshift evolution of $\Phi_{\star}$ and $L_{\star}$ to quantitatively analyze these trends. 
For comparison, we also show the previous measurements with {\it Herschel} and SCUBA2 in the grey symbols \citep{magnelli2013, gruppioni2013, lwang2019, lim2020} and with ALMA+SCUBA2 in the magenta symbol \citep{koprowski2017}. 
The solid magenta lines denote the best-fit empirical functions in \cite{koprowski2017}, where the dashed line presents their extrapolations. 
The blue lines also show the best-fit empirical function obtained in the latest UV LF study \citep{bouwens2022}. 

We confirm that our measurements are consistent with the general trends of the evolution of IR LFs characterized by positive $L_{\star}$ and negative $\Phi_{\star}$ towards high redshifts reported in previous studies. 
We find that our measurements are the most closely consistent with the measurements obtained with ALMA+SCUBA2 \citep{koprowski2017}, 
while we also find that our measurements are lower (higher) than the previous studies with single-dish telescopes \citep{magnelli2013, gruppioni2013, lwang2019, lim2020}.  
As discussed in Section \ref{sec:irlf}, 
the difference is likely explained by the combination of the partial coverage of $L_{\rm IR}$ and the coarse spatial resolutions in the previous single-dish surveys. 

Our measurements suggest a flattening in the correlation between the redshift and $L_{\star}$ at $z\sim$ 4. 
A similar flattening has been reported in the UV LF studies at $z\sim$ 2.5 \citep[e.g.,][]{bouwens2022}, which is thought to be explained by the maximum value of UV luminosity due to the increasing importance of dust extinction in the highest stellar mass and SFR sources. 
These results might indicate that there is also a maximum value in the IR luminosity, especially at such high redshifts due to the relatively short cosmic time scale for the required dust mass formation and evolution. 
The maximum IR luminosity might also be contributed by the decreasing trend of the dust continuum size towards high redshifts \citep[e.g.,][]{fujimoto2017, gomez2021}, where the Eddington limit would regulate the high surface density of IR luminosity. 

\subsection{Cosmic SFRD}
\label{sec:sfrd}

By integrating our IR LF measurements, we calculate the volume average of the integrated IR luminosity and thus investigate the obscured side of the cosmic SFR density (SFRD) $\psi_{\rm SFRD\,(IR)}$. 
We integrate the best-fit DPL functions down to our detection limit of $L_{\rm IR}=10^{10}\, L_{\odot}$. 
For conversion from $L_{\rm IR}$ to SFR, we use SFR = $\mathcal{K} L_{\rm IR}$, where $\mathcal{K}=4.5\times10^{-44} M_{\odot}$~year$^{-1}$~erg$^{-1}$~s, which is the same calibration as \cite{madau2014}, but we multiply a factor of 0.63 to apply the correction from the Salpeter IMF to the Chabrier IMF.

%%%%%%%%%%%%%%%%%%%%%%%
\begin{figure*}
\begin{center}
\includegraphics[trim=0cm 0cm 0cm 0cm, clip, angle=0,width=1.0\textwidth]{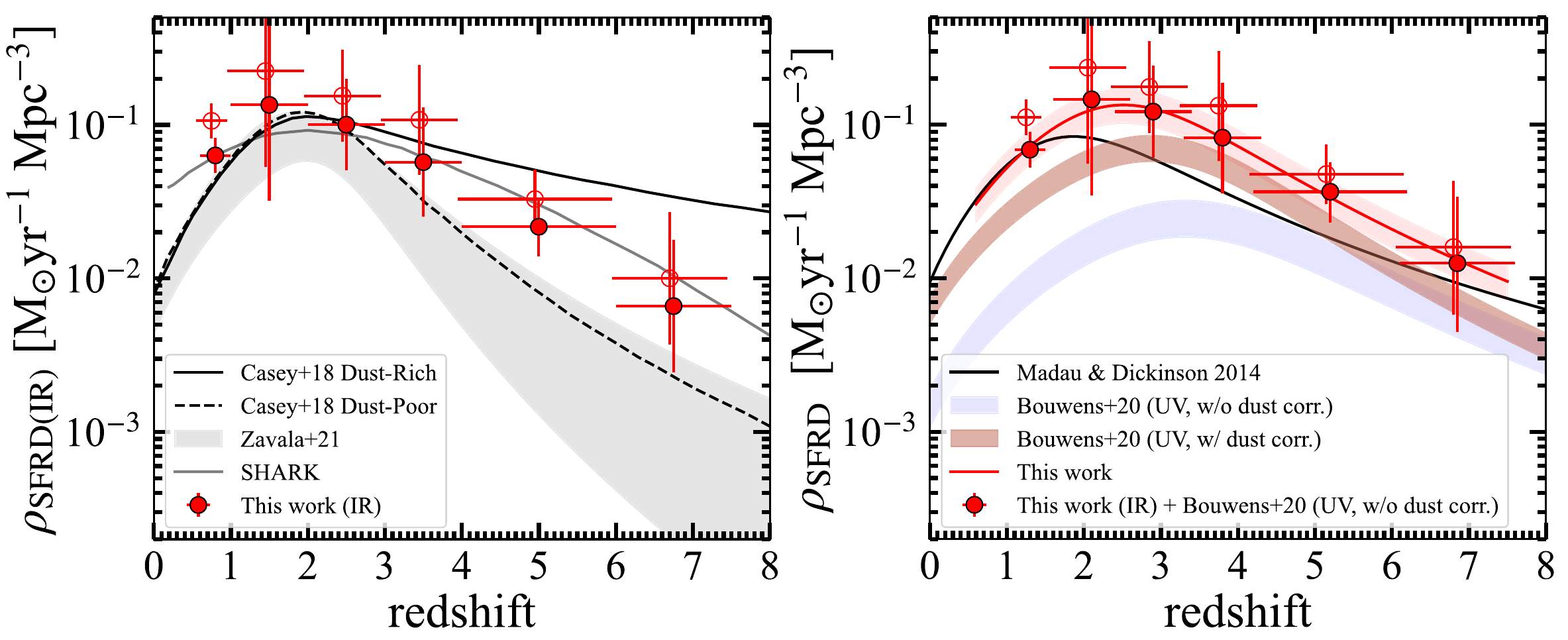}
 \caption{
{\it Left}: Cosmic IR SFRD measurements from ALCS IR LFs at $z=$0.6--7.5. 
The red circles are our measurements with a conversion of SFR [$M_{\odot}$~yr$^{-1}$] = $4.5\times10^{-44} L_{\rm IR}$ [erg~s$^{-1}$], which is the same calibration as \cite{madau2014}, but we multiply a factor of 0.63 to apply the correction from the Salpeter IMF to the Chabrier IMF. 
In both panel, the filled and open red circles indicate the integration range of IR LFs down to our detection limit of $L_{\rm IR}=10^{10}\,L_{\odot}$ and the potential turnover or flattening IR luminosity of $L_{\rm IR}=10^{8}\,L_{\odot}$ (see Section~\ref{sec:sfrd}). 
The black solid and dashed lines are two empirical models of ``Dust Rich'' and ``Dust Poor'' scenarios presented in \cite{casey2018b}, respectively. 
The grey shaded region denotes previous measurements of \cite{zavala2021} that perform a backward evolution model to constrain the evolution of the IR LFs to reproduce the observed number counts at 1.2~mm and 3~mm. 
The grey line is the prediction from the SHARK simulation \citep{lagos2018,lagos2020}, where the integrated $L_{\rm IR}$ range is matched with our measurements. 
{\it Right}: Cosmic total (=IR+UV) SFRD measurements. 
The black line shows the previous best estimate of the empirical model (Eq. \ref{eq:sfrd}) fitting to the SFRD$_{\rm total}$ and SFRD$_{\rm corr}$ measurements presented in \cite{madau2014}. 
The blue and brown shaded regions present the 1$\sigma$ range of the previous SFRD$_{\rm UV}$ estimates without and with the dust correction, respectively, drawn by fitting the empirical model (Eq. \ref{eq:sfrd}) to the measurements pres
The red filled circles represent our SFRD$_{\rm total}$ measurements estimated from our SFRD$_{\rm IR}$ and the previous SFRD$_{\rm UV}$ estimate without the dust correction \citep{bouwens2020}. 
The red line and shaded region are the best-fit empirical model (Eq. \ref{eq:sfrd}) for our SFRD$_{\rm total}$ measurements and its 1$\sigma$ range. 
For presentation purposes, we slightly shift the open red circles along the abscissa.
\label{fig:csfrd}}
\end{center}
\end{figure*}
%%%%%%%%%%%%%%%%%%%%%%% 

The left panel of figure \ref{fig:csfrd} shows the redshift evolution of $\rho_{\rm SFR\,(IR)}$ from our IR LF measurements. 
For comparison, we also present two possible scenarios of ``Dust-Rich'' and ``Dust-Poor'' in the universe presented in \cite{casey2018a} and the model prediction \citep{lagos2020}. 
We also show previous measurements presented in \cite{zavala2021} that perform a backward evolution model to constrain the evolution of the IR LFs to reproduce the observed number counts at 1.2~mm and 3~mm. 

We find that our measurements are overall consistent with the previous measurements and the prediction of the simulation within the errors. 
At $z>2$, our measurements fall between the two possible scenarios in \cite{casey2018a}. 
Although our measurements are generally higher than those of \cite{zavala2021}, this is probably because the fitting at the faint regime ($S_{\rm 1.2mm} < 0.1$ mJy) in their backward model relies on the ASPECS results that may be obtained from the underdense region (Section \ref{sec:number_counts}). 
Our measurements are well consistent with the prediction from the simulation of \cite{lagos2020}. 

In the right panel of Figure \ref{fig:csfrd}, we also show the evolution of the total (= IR+UV) SFRD $\psi_{\rm SFRD\, (IR+UV)}$ by combining our measurements and the un-obscured SFRD estimates without the dust extinction correction presented in \cite{bouwens2020}. 
The red line and shaded region denote the best-fit evolution and its 1$\sigma$ uncertainty range by using an empirical function form in \cite{madau2014} given by 
\begin{eqnarray}
\label{eq:sfrd}
\psi(z) = A \frac{(1+z)^{a}}{1+[(1+z)/B]^{b}} , 
\end{eqnarray}
where the evolution is propotional to $(1+z)^{a}$ at low redshift, and $(1+z)^{a-b}$ at high redshift, with $A$ and $B$ as coefficients. 
Because our measurements do not have the constrain at $z<0.6$, we calculate $\psi(z=0)$ with the best-fit parameters presented in \cite{madau2014} and include it in our fitting by assuming its error bar scale similar to our measurements. 
We obtain the best-fit parameters of $a=2.8\pm0.5$, $b=6.6\pm0.5$, $A=0.010\pm0.003$, and $B=3.3\pm0.3$ for our SFRD measurements.

Overall, we find that our measurements are generally consistent with the previous total SFRD measurements within the errors. 
At $z>2$, 
although we find that the central values in our measurements are higher than the previous measurements in both \cite{madau2014} and \cite{bouwens2020},  
the optical--NIR dark objects that have been recently reported even up to $z\simeq7$ \citep[e.g.,][]{fujimoto2016, franco2018, twang2019, yamaguchi2019, williams2019, casey2019, romano2020, fudamoto2021, gomez2021, talia2021, fujimoto2022, xiao2022, manning2022,barrufet2023a,guilietti2023,rodighiero2023} are not included in the optical--NIR surveys used in these previous measurements.  
Our blind survey for the mm sources includes the optical--NIR dark objects and thus provides a complete estimate of the total SFRD measurement, which is likely the cause of the slight excess in our estimate. 
Similarly at $z\lesssim2$, 
we also find that the central values in our measurements are higher than the previous studies. Although \cite{madau2014} include those measurements obtained from the IR surveys that unlikely miss the optical--NIR dark objects, those measurements are obtained with a flat faint-end slope assumption in the IR LFs that has not been well constrained in the previous {\it Herschel} studies at $z>0.3$. Therefore, the potential difference at $z\lesssim2$ is attributed to the faint-end slope in the IR LFs. 

Note that these cosmic SFRD estimates can vary with the integration range. 
In Section \ref{sec:cib}, we find that the faint-end slope in the 1.2-mm number counts may have a turnover or flattening ($S_{\rm turn}$) above $\sim0.2\mu$Jy not to make the integrated 1.2-mm flux density exceed the upper limit of the CIB.   
Given that $S_{\rm turn}=0.2\mu$Jy is $\sim2$-dex deeper than our detection limit in the direct counts of $S_{\rm 1.2mm}\sim 7\mu$Jy, 
we assume the turnover or flattening is also taking place in the IR LFs at $\sim2$-dex deeper than the detection limit of $L_{\rm IR}=10^{10}\,L_{\odot}$. 
We thus regard $L_{\rm IR}=10^{8}\,L_{\odot}$ as the turnover or flattening IR luminosity $L_{\rm IR, turn}$ and compute the integration of IR LFs down to $L_{\rm IR, turn}$, below which the integrated values are thought to be negligible. 
In Figure \ref{fig:csfrd}, the open circles present our SFRD measurements with $L_{\rm IR, turn}$. 
We find that the integrated values are not increased more than a factor of $\sim2$ in the entire redshift range, and at $z>4$, the increase is almost negligible. 
At $z>4$, 
the total cosmic SFRD measurements with $L_{\rm IR, turn}$ is estimated to be $161^{+25}_{-21}\%$ of the previous measurement of \cite{madau2014}. 
Our results indicate that the general understanding of the cosmic SFRD is unchanged by a factor of $\sim$2, while there may be an additional ($\approx 60$\%) SFRD component contributed by the faint-mm population, including NIR dark objects. 

\tcb{
In addition to the NIR-dark objects, recent ALMA observations have unveiled the presence of highly obscured star formation even in low-mass UV-selected galaxies at $z=4-7$ \citep[e.g.,][]{akins2022, mizener2024}. These results suggest that the previous CSFRD estimates, generally relying on UV-selected sources with dust correction based on the UV and continuum slope relation \citep[e.g.,][]{bouwens2014}, may be underestimated. 
Although the additional SFRD component could potentially introduce further tension with the cosmic stellar mass density (SMD) measurements at $0\leq z \leq3$ \citep[e.g.,][]{madau2014, leja2015}, these discrepancies may be explained by various causes such as overestimate in SFR, underestimate in stellar mass, uncertainty in the faint-end of UVLF and/or SMF, unknown IMF shapes at high-$z$, and the mass-loss factor of stellar mass within galaxy (see e.g., Section 5.3 in \citealt{madau2014}). 
Therefore, although actual contributions of NIR-dark galaxies to the CSFRD need to be further quantified by systematic ALMA follow-up observations for more statistics and/or more accurate measure of $L_{\rm IR}$ via multi-band observations, our findings of a potential $\sim$60\% excess in CSFRD measurements may be consistent with this recent discovery of NIR-dark objects and do not violate the cosmic SMD. However, we acknowledge that our results could be affected by the small number statistics, which is propagated to the uncertainty in our final CSFRD estimates. Therefore, we emphasize that our SFRD measurements are still consistent with previous measurements within the $1-2\sigma$ errors.
}

\section{Summary}
\label{sec:summary}
In this paper, we identify 180 ALMA continuum sources with a flux density of $\simeq$ 0.007–3.8~mJy in the 33 lensing clusters homogeneously observed in the large ALMA project of ALCS. 
The ALCS explores the faintest layer of mm wavelengths most comprehensively and efficiently with gravitational lensing support, resulting in a large, blind sample of faint mm sources out to $z\sim6$. 
This allows for the identification of unique sources, such as a strongly lensed and multiply imaged ($\mu\sim30$ and $\sim$160) galaxy at $z=6.07$, as well as NIR dark populations with no counterparts in deep \hst\ and even {\it Spitzer} images. 
Using this large sample, we derive the 1.2~mm number counts and discuss the contributions of these faint sources to the Cosmic Infrared Background light (CIB), including potential uncertainties in the faint-end shape. We also derive the infrared (IR) luminosity functions (LFs) and the total (= obscured and unobscured) cosmic star-formation rate density (SFRD) at $z=0.6$--7.5, in conjunction with the recent serendipitous detection of mm sources at $z>7$ in the literature. The main findings of this paper are summarized as follows:
\begin{enumerate}
\item 
The target clusters were selected from the best-studied 33 massive galaxy clusters drawn from \hst\ treasury programs of HFF, CLASH, and RELICS, where the deep \hst\ and {\it Spitzer} maps are available. The observations were carried out with two frequency setups in ALMA Band~6, spanning a total of 15~GHz width with the center at 261.25~GHz ($\sim$1.15~mm) to enlarge the blind identifications of the line emitters, in addition to the continuum sources. The total survey area above a relative sensitivity of 30\% in the mosaic map covers 133~arcmin$^{2}$ in the observed frame. 
After the lens correction at $z=6$, this is decreased down to 20~arcmin$^{2}$ and 4~arcmin$^{2}$ at detection limits of 0.1~mJy and 0.04~mJy, which are larger than those in ASPECS at the same detection limits by factors of $\sim$4.8 and $\sim$2.0, respectively. 
\item 
we produce natural-weighted and $uv$-tapered maps ($2\farcs0\times2\farcs0$) for each cluster not to miss strongly elongated mm sources due to the lensing effect. Based on the positive and negative source histogram as a function of peak SNR in the maps, we adopt thresholds of SNR $=$ 5.0 and 4.5 for the natural-weight and $uv$-tapered maps, respectively, to maintain a high purity $>0.99$. 
This yields 141 ALMA sources, referred to as the blind sample. 
From cross-matching with counterparts in the {\it Spitzer}/IRAC maps, we also identify additional 39 ALMA sources with SNR down to 4.0 in the natural-weighted maps. 
Among the blind sample, we find \tcb{18} ALMA sources with no \hst\ counterparts, some showing no counterparts even in {\it Spitzer}/IRAC ch1 and ch2 maps. 
\item 
After the lens correction and removing the dust emission from the cluster member galaxies, 
we obtain 146 ALCS sources which span in the intrinsic flux density range of $\simeq0.007$--3.8~mJy and $z\simeq0$--6 with the median values of 0.35~mJy and $z=2.03$ and are the most widely distributed in both parameter spaces among the sources identified in recent ALMA surveys. These median values are consistent with the general trend of the increasing median flux density at submm/mm wavelengths in the increasing median redshift. 
The median magnification factor in our ALCS sources is estimated to be 2.7. 
\item 
With careful corrections for the flux measurements and completeness via dedicated Monte Carlo (MC) simulations based on a constant FIR size assumption, we derive the 1.2~mm number counts by using the blind sample, exploring its faint end down to $\sim7\mu$Jy. 
We find that the number counts are well represented by both Schechter and double power law (DPL) functions and generally consistent with previous measurements in a wide flux range over a $\sim$3~dex scale. 
We estimate a faint-end slope of $\alpha=-2.05^{+0.12}_{-0.10}$ for our fiducial model.  These results suggest little flattening in the counts below $\sim0.1$~mJy as previously claimed by earlier ALMA studies. 
\item 
To assess the sensitivity of the completeness correction on the intrinsic source size, we derive 1.2~mm number counts by assuming a positive or negative correlation of FIR size with IR luminosity $L_{\rm IR}$ as reported in the literature and evaluate the impact of this on the form of the faint number counts.
We find that the faint-end slope varies depending on the FIR size assumptions, potentially changing the CIB resolved fraction by factors of $\sim$\,0.6--1.1. 
\item 
We also test the potential impact on the derived number counts of the varying precision of the lens model among the clusters by separately estimating the 1.2~mm counts for the HFF, CLASH, and RELICS clusters. The highest quality lens models, those for HFF,  yield the shallowest faint-end slope, $\alpha=-1.71^{+0.34}_{-0.29}$ (but with a larger cosmic variance uncertainty). This might indicate that the high magnification areas in the CLASH and RELICS clusters are underestimated due to their less detailed mass models with fewer multiple images, which decreases the granularity and small-scale structures included in the mass models. Combining all the clusters weighted by the quality of the lens models yields a slightly shallower value of $\alpha=-1.96^{+0.14}_{-0.13}$. We also note the estimate of the faint-end slope excluding the highly magnified sources ($\mu\gtrsim$3--10) also yields  $\alpha\sim-2.0$, similar to our fiducial model. Hence, despite the potential remaining uncertainties, 
all faint-end slope estimates agree within $\sim$1--2$\sigma$, suggesting a conservative $\alpha$ range of $-2.0^{+0.2}_{-0.3}$, regardless of the clusters used or the different FIR size assumptions. 
\item 
We integrate our fiducial Schechter function fit (which assumes a constant FIR size and no weights among clusters) to the 1.2~mm counts and obtain an integrated intensity of $20.7^{+8.5}_{-6.5}$~Jy~deg$^{-2}$ down to a flux limit of 7~$\mu$Jy. 
This integrated flux density corresponds to $\sim80$\% of the CIB measured by {\it COBE}/FIRAS assuming the model for Galactic foreground emission of \cite{odegard2019}. 
Extrapolating our fiducial Schechter function to $\lesssim7$~$\mu$Jy will exceed the CIB measurements unless there is a turnover or flattening in the counts at fainter flux densities or a shallower slope of $\alpha\gtrsim-1.8$ whose integration down to zero does not diverge. 
However, we note that there is also significant systematic uncertainty in the CIB constraint   due to the subtraction of the foreground emission.
\item 
We calculate $L_{\rm IR}$ of our ALCS sources by modeling the IR--mm SED with the {\it Spitzer}, {\it Herschel}, and ALMA photometry and also derive the IR luminosity functions (LFs) out to $z=7.5$, combining our ALCS sources with the faint-mm sources serendipitously detected in recent ALMA studies at $z>7$. 
We obtain the faint-end slope of $\simeq$0.9--1.0 in the Double Power Law (DPL) function, which is higher than previous IR LF measurements of $\simeq0.4$--0.6 at $z\lesssim0.3$, but close to reports in the UVLF studies of $\simeq1.5$--2.0. 
The difference may be explained by the relatively large luminosity range ($\sim2$--3~dex) covered by our IR LFs owing to the lensing support, compared to the previous studies, and/or by the different triggering mechanisms of dusty starbursts in the local universe due to the significant decrease of the gas fraction. 
Our ALCS IR LF results are generally consistent with the predictions from simulations within $\sim1$--2$\sigma$. 
\item 
We integrate the IR LFs and derive the redshift evolution of the obscured and total (= obscured + unobscured) star-formation rate density (SFRD). 
We find that our total cosmic SFRD measurement at $z=4-8$ is estimated to be  $161^{+25}_{-21}$\% of the previous measurements of \cite{madau2014}, where the measurements at $z>4$ relied only on optical--NIR surveys. 
This indicates that our current understanding of the cosmic SFRD might be underestimated by a factor of $\approx 1.6$ due to missing the SFRD component contributed by both the very faint mm populations and NIR-dark galaxies. 
\end{enumerate}

We thank the anonymous referee for the thorough review, insightful comments, and suggestions. 
We also thank Carlos G\'omez-Guijarro and Aswin Vijayan for sharing their measurements and the compilation of the literature. 
We are also grateful to Caitlin Casey for useful discussions on the evolution of the IR LFs, Steven Finkelstein, and Peter Behroozi for the useful comments on the cosmic star-formation rate and stellar mass density evolutions, and Soh Ikarashi for the helpful comments on the impact of the size assumptions to the faint-end of the 1.2-mm number counts. 
This paper makes use of the ALMA data: ADS/JAO. ALMA \#2018.1.00035.L, \#2019.1.00237.S, \#2019.1.00949.S, \#2021.1.01246.S, and \#2021.1.00407.S. 
ALMA is a partnership of the ESO (representing its member states), 
NSF (USA) and NINS (Japan), together with NRC (Canada), MOST and ASIAA (Taiwan), and KASI (Republic of Korea), 
in cooperation with the Republic of Chile. 
The Joint ALMA Observatory is operated by the ESO, AUI/NRAO, and NAOJ. 
This work is based on observations and archival data made with the {\it Spitzer Space Telescope}, which is operated by the Jet Propulsion
Laboratory, California Institute of Technology, under a contract with NASA, along with archival data from the NASA/ESA 
{\it Hubble Space Telescope}. This research also made use of the NASA/IPAC Infrared Science Archive (IRSA), 
which is operated by the Jet Propulsion Laboratory, California Institute of Technology, under contract with the National Aeronautics and Space Administration. 
This work was supported in part by World Premier International
Research Center Initiative (WPI Initiative), MEXT, Japan, and JSPS
KAKENHI Grant Numbers 22H01260, 20H05856,  20H00181, 17H06130, 22K21349, 20H00180, 21H04467, 20H01953, 22KK0231, and 23K20240. 
The Cosmic Dawn Center is funded by the Danish National Research Foundation under grant No. 140.
This work was supported by the joint research program of the Institute for Cosmic Ray
Research (ICRR), University of Tokyo. 
This project has received funding from the European Union’s Horizon 2020 research and innovation program under the Marie Sklodowska-Curie grant agreement No. 847523 `INTERACTIONS' and from NASA through the NASA Hubble Fellowship grant \#HST-HF2-51505.001-A awarded by the Space Telescope Science Institute, which is operated by the Association of Universities for Research in Astronomy, Incorporated, under NASA contract NAS5-26555.
S.F. acknowledges support from the European Research Council (ERC) Consolidator Grant funding scheme (project ConTExt, grant No. 648179). 
K.~Kohno acknowledges support from the NAOJ ALMA Scientific Research Grant Number 2017-06B and JSPS KAKENHI Grant Numbers JP17H06130, JP22H04939, and JP23K20035.
G.E.M. acknowledges the Villum Fonden research grants 13160 and 37440. 
I.R.S. acknowledges support from dSTFC (ST/T000244/1).
F.E.B. acknowledges support from ANID-Chile BASAL CATA FB210003, FONDECYT Regular 1200495 and 1190818, and Millennium Science Initiative Program--ICN12\_009. 
D.E. acknowledges support from a Beatriz Galindo senior fellowship (BG20/00224) from the Spanish Ministry of Science and Innovation, projects PID2020-114414GB-100 and PID2020-113689GB-I00 financed by MCIN/AEI/10.13039/501100011033, 
 project P20-00334  financed by the Junta de Andaluc\'{i}a, and project A-FQM-510-UGR20 of the FEDER/Junta de Andaluc\'{i}a-Consejer\'{i}a de Transformaci\'{o}n Econ\'{o}mica, Industria, Consejer\'a, Universidades.
 K.~Kirsten acknowledges support from the Knut and Alice Wallenberg Foundation (2017.0292 and 2019.0443) and the Swedish Research Council (2015-05580).

\software{{\sc casa} (v5.4.0, v5.6.1; \citealt{casa2022}), 
{\sc sourcextractor} (v2.5.0; \citealt{bertin1996}), 
{\sc eazy} \citep{brammer2008}, 
{\sc glafic} \citep{oguri2010}, {\sc lenstool} \citep{jullo2007}, {\sc ltm} \citep{zitrin2015}
}

\appendix

\section{Serendipitous and follow-up line detection}
\label{sec:appendix_line}

%%%%%%%%%%%%%%%%%%%%%%%
\begin{figure*}%[h]
\includegraphics[trim=0cm 0cm 0cm 0cm, clip, angle=0,width=1\textwidth]{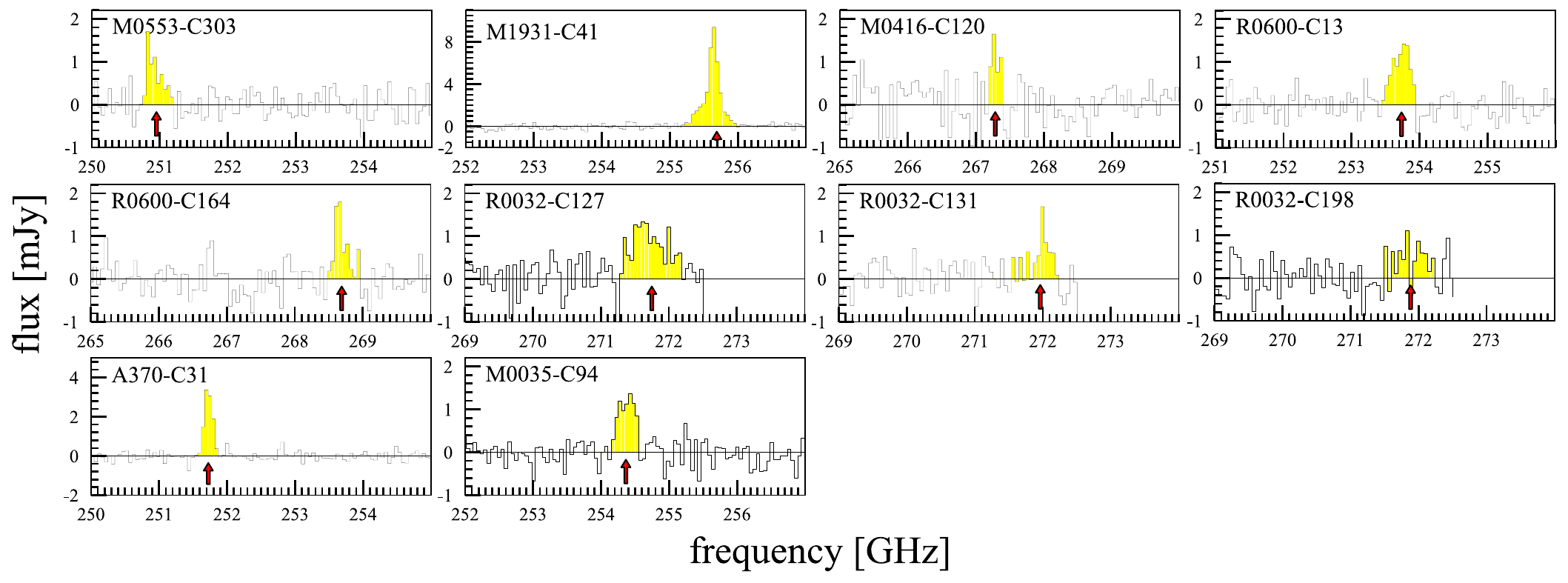}
 \caption{
ALCS Band~6 continuum-subtracted spectra for the 10 serendipitous line detection in the ALCS continuum sources. 
The S/N, central frequency, and line width are summarized in Table \ref{tab:line_prop}. 
\label{fig:app_line}}
\end{figure*}
%%%%%%%%%%%%%%%%%%%%%%% 

%%%%%%%%%%%%%%%%%%%%%%%
\begin{figure*}%[h]
\includegraphics[trim=0cm 0cm 0cm 0cm, clip, angle=0,width=1\textwidth]{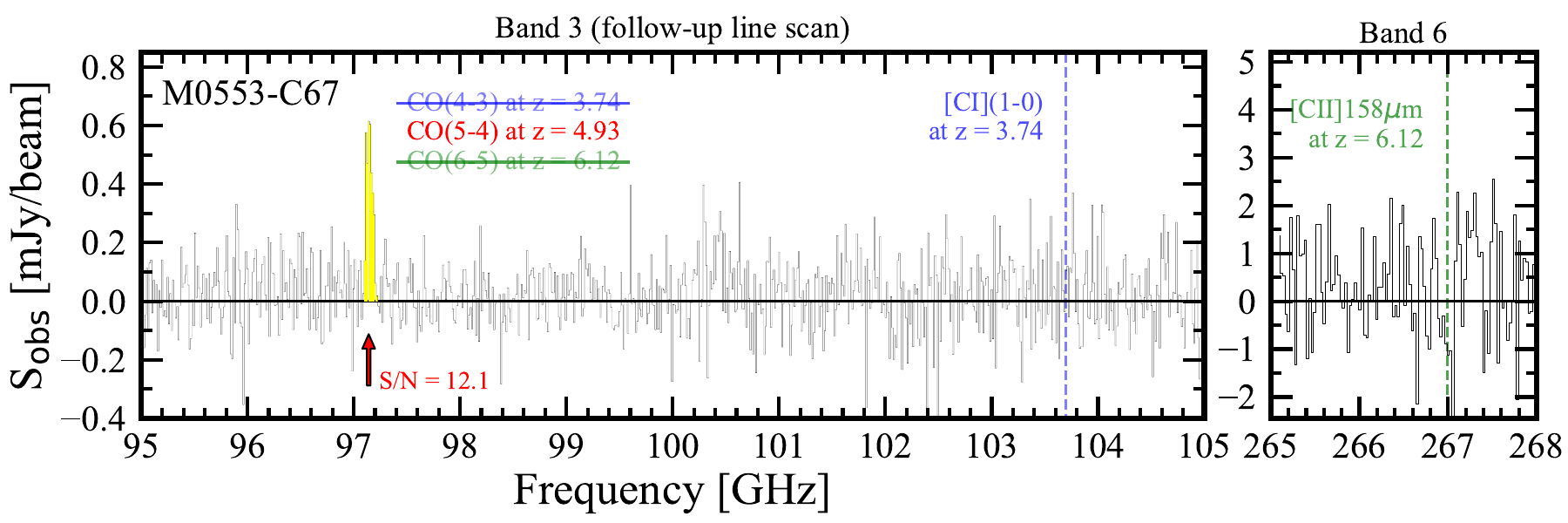}
 \caption{ 
Zoom-in spectra of the 27-GHz wide ($\sim$85--112~GHz) follow-up ALMA Band~3 line cube (left) and the ALCS 15-GHz wide Band~6 line cube (right) for a NIR-dark lensed galaxy of M0553-C67. 
The Band~3 follow-up line scan observations yield a robust single line detection (S/N=12.1) at $\sim97.1$~GHz. Based on the photometric redshift of $z_{\rm phot}\gtrsim4$ and the non-detection of neighboring lines, we regard the line as CO(5-4) at $z=4.93$ in this paper (see Appendix~\ref{sec:appendix_line}). The blue and green dashed vertical lines indicate the line frequency of [C{\sc \,i}](1-0) at $z=3.74$ and \cii158$\mu$m at $z=6.12$ in case the 97.1~GHz line corresponds to CO(4-3) and CO(6-5), respectively. 
\label{fig:follow-up}}
\end{figure*}
%%%%%%%%%%%%%%%%%%%%%%% 

We extract the ALMA Band 6 spectra with a $1\farcs0$-radius aperture for the 180 ALCS sources with the 15-GHz ALCS data cube and identify the serendipitous line emission from 13 ALCS continuum sources (Section \ref{sec:line}). 
Three lines are the CO emission line from the BCGs in the clusters, and the remaining ten lines are interpreted as \cii\ and CO emission lines from the sources behind the clusters.  
We summarize these 13 source properties in Table \ref{tab:line_prop}. 
The ALMA Band~6 line spectra for three sources are presented in Figure~\ref{fig:line_eg}, and 
Figure \ref{fig:app_line} presents the remaining 10 sources. 

Several ALMA CO line scan follow-up programs have been ongoing, and we additionally detect emission lines from 6 ALCS sources, which is summarized in Table~\ref{tab:line_prop2}. 
Figure~\ref{fig:follow-up} shows an ALMA Band~3 line spectrum for a NIR-dark lensed object of M0553-C67 that has been observed in a follow-up line scan program.  
We provide further details for each line emission as follows. 

{\flushleft \bf M1931-C41 (serendipitous)} -- 
The emission line is serendipitously identified in the ALCS data cube at 255.69~GHz with SNR = 46.0.
The source position corresponds to one of the BCGs in M1931, and the emission line is determined as CO(3-2) at $z=0.352$, which is in excellent agreement with the cluster redshift of M1931. 

{\flushleft \bf A370-C31 (serendipitous)} -- 
The emission line is serendipitously identified in the ALCS data cube at 251.77~GHz with SNR = 18.9 in the velocity integrated line flux density over 210 km~s$^{-1}$. 
The source position corresponds to one of the BCGs in A370, 
and the emission line is determined as CO(3-2) at $z=0.375$ which is in excellent agreement with the cluster redshift of A370. 

{\flushleft \bf M0035-C94 (serendipitous)} -- 
The emission line is serendipitously identified in the ALCS data cube at 254.43~GHz with SNR = 9.5 in the velocity integrated line flux density over 420 km~s$^{-1}$. 
The source position corresponds to one of the BCGs in M0035, and the emission line is determined as CO(3-2) at $z=0.359$, which is in excellent agreement with the cluster redshift of M0035. 

{\flushleft \bf R0600-C13 (serendipitous)} --
The emission line is serendipitously identified in the ALCS data cube at 253.77~GHz with SNR = 7.8 in the velocity integrated line flux density over 420 km~s$^{-1}$. 
The source falls outside of the HST/WFC3 images, but the template fitting with other multi-wavelength suggests $z_{\rm phot}=1.00^{+0.05}_{-0.05}$ (Appendix \ref{sec:app_znir}). An independent SED analysis based on the FIR photometry with {\sc magphys} suggests $z_{\rm phot}=1.23^{+0.40}_{-0.45}$.  
We thus regard this line as CO(5-4) at $z=1.271$, 
while this interpretation might be updated with a second-line detection in future follow-up observations. 

{\flushleft \bf R0600-C67 (follow-up)} --
Although the ALMA continuum is very brightly and securely detected ($\sim$2.3~mJy, SNR=62), M0600-C67 shows no counterparts in \hst/WFC3 and even IRAC maps (F160W $\gtrsim$ 27~mag, IRAC $\gtrsim$ 25~mag, 3$\sigma$; \citealt{coe2019}). 
Because of its uniquely NIR-faint nature even in the IRAC bands, despite the brightness in the ALMA Band~6, a follow-up ALMA Band~3+4 line scan observation was carried out (\#2021.1.00407.S, PI: F. Bauer). 
Only marginal line features are detected (SNR $=$ 4.6--5.0), where no convincing redshift solutions are found among the combinations of these marginal line features. 
We thus regard the marginal line feature in the Band3+4 spectrum with the highest SNR of 5.0 detected at 138.979~GHz as CO(7-6) at $z=4.80$ and use this redshift interpretation in our analysis throughout the paper. However, future follow-up observations might update this interpretation with a secure second-line detection.  
Further details of the follow-up line scan observation results will be presented in Bauer et al. (in prep.).

{\flushleft \bf R0949-C10 \& C19 (follow-up)} --
Similar to R0600-C13, R0949-C10 and R0949-C19 are characterized as very bright in ALMA Band~6 ($\sim$5.5~mJy \& $\sim$1.6~mJy, SNR=62 \& 13), but relatively very faint in NIR with no counterparts in \hst/WFC3 and even IRAC maps (F160W $\gtrsim$ 27~mag, IRAC $\gtrsim$ 25~mag, 3$\sigma$; \citealt{coe2019}). 
The upper limits of the flux ratio at 4.5~$\mu$m (IRAC ch2) and 1.2~mm (ALMA) indicate their source redshifts at $z>5.5$ and $z>4.5$ in R0949-C10 and R0949-C19, respectively, based on the composite SED of the high-$z$ dusty galaxies \citep{dudzeviciute2020}.   
A follow-up ALMA Band~3+4 line scan observation was conducted (\#2021.1.00407.S, PI: F. Bauer). 
The multiple lines are successfully detected in R0949-C10, determining the source redshift of $z=5.753$, which meets the above lower limit of the redshift. 
On the other hand, only marginal line features are detected in R0949-C19 (SNR $=$ 4.3--4.7), where no convincing redshift solutions are found among the combinations of these marginal line features. 
We thus regard the marginal line feature in the Band3+4 spectrum with the highest SNR of 4.7 detected at 88.266~GHz as CO(5-4) at $z=5.53$ and use this redshift interpretation for R0949-C19 in our analysis throughout the paper. However, future follow-up observations might update this interpretation with a secure second-line detection.  
Further details of the follow-up line scan observation results will be presented in Bauer et al. (in prep.).

{\flushleft \bf M0416-C120 (serendipitous)} -- 
The line is serendipitously identified in the ALCS data cube at 267.34~GHz with SNR = 5.3 in the velocity integrated line flux density over 240 km~s$^{-1}$. 
The source has a counterpart in the HST/F160W map and IRAC/ch1 and ch2, while it is not detected in any {\it Herschel} bands. 
The SED fitting with {\sc eazy} suggests $z_{\rm phot} = 9.62^{+0.07}_{-0.06}$ \citep{kokorev2022}. 
Although our ALCS 15-GHz data cube has the potential to detect bright FIR emission lines of \cii\ 158$\mu$m at $z=$ 5.97--6.17 and 6.38--6.60 as well as \oiii\ 88$\mu$m at $z=$ 11.45--11.80 and 12.18--12.57, the redshift solution falls outside of all of these redshift ranges. 
Moreover, the morphology in the HST/F160W map implies a large system (Figure \ref{fig:app_postage}) which deviates from typical galaxy sizes at $z>9$ (e.g., effective radius of $<0\farcs1$; e.g., \citealt{kawamata2018}) even with the lensing effect ($\mu=1.6$ at $z=9.6$) taken into account. 
We thus interpret that the Lyman-break and the 4000${\rm \AA}$-beak features are degenerated in the SED fitting, and $z_{\rm phot}\sim$2.2, instead of beyond 9. A public {\it HST}+IRAC catalog in the same field also shows a consistent estimate of $z_{\rm phot}=2.34 \pm 0. 20$ for this source \citep{shipley2018}. 
We thus regard this line as CO(8-7) at $2.448$, 
while this interpretation might be updated with a second-line detection in future follow-up observations.

{\flushleft \bf M0553-C67 (follow-up)}
Although the ALMA continuum is brightly and securely detected ($\sim$0.8~mJy, SNR=10), M0553-C67 shows no HST/WFC3 counterparts, but with a faint counterpart in IRAC ch1 and ch2 (F160W $\gtrsim$27~mag, IRAC $\sim$24~mag; \citealt{kokorev2022}). The template fitting with the IRAC and ALMA photometry (Section \ref{sec:redshift}) suggests its photometric redshift of $z_{\rm phot}=4.2\pm0.3$, while it could be higher because the IRAC photometry might be overestimated due to a nearby bright object.  
A follow-up $\sim$27-GHz wide ALMA Band~3 line scan observation was carried out (\#2019.1.00237.S, PI: S. Fujimoto), and we reduced the data in the same manner as the ALCS data cubes (Section \ref{sec:analysis}). A robust single line is detected at $97.140\pm0.003$~GHz with a line width of FWHM = $224\pm25$~km~s$^{-1}$ and SNR = 12.1 in the velocity-integrated map. 
In Figure~\ref{fig:follow-up}, we show the robust single-line detection from M0553-C67 in the follow-up Band~3 observations. 
Given the $z_{\rm phot}$ estimate and non-detection of the secondary lines in the $\sim$27-GHz wide Band~3 data cube, we interpret the line at 97.1~GHz as CO(4-3) at $z=3.74$, CO(5-4) at $z=4.93$, or CO(6-5) at $z=6.12$. 
In the case of CO(4-3) at $z=3.74$, \ci(1-0) should also be detected with SNR $>3$ in the same Band~3 data cube, given the typical line ratio of $\log$(\ci(1-0)/CO(4-3)) ranging in $[-0.6:0.5]$ among high-$z$ star-forming galaxies \citep{valentino2020b}.   
In the case of CO(6-5) at $z=6.12$, the \cii\ line should be detected with SNR $>5$ in the ALCS Band~6 data cube, given the typical ratio of $L_{\rm [CII]}/L_{\rm IR}$. 
Since we do not identify such secondary line features in the Band~3 and Band~6 data cubes (Figure \ref{fig:follow-up}), we regard the single line at 97.1~GHz as CO(5-4) and determine its redshift of $z=4.9323$. This interpretation might be updated with a second-line detection in future follow-up observations.

{\flushleft \bf R0600-C164 (serendipitous)} --
The line is serendipitously identified in the ALCS data cube at 268.68~GHz with SNR = 7.1 in the velocity integrated line flux density over 150 km~s$^{-1}$. 
The source has an HST counterpart with an arc shape elongated over $\sim6''$ scale, and the SNR is increased out to 9.1 with an optimized aperture along with the arc shape \citep{fujimoto2021}. 
The arc shows a clear Lyman-break feature at $z\sim6$ in the HST bands. 
The follow-up Gemini/GMOS spectroscopy detects the continuum break and confirms the source redshift at $z=6.19^{+0.06}_{-0.16}$ \citep{laporte2021}, 
and the line is determined as \cii\ 158~$\mu$m at $z=6.072$. 
The two multiple images of the arc are identified with the line detection at the same frequency and the same SED properties, while the dust continuum is not detected ($<4\sigma$) from these multiple images because of different magnifications among multiple images. 
Three independent lens models consistently suggest that an outskirt region of the galaxy behind the cluster is crossing the caustic line, which is strongly lensed into the arc shape in the image plane with a magnification factor of 163$^{+27}_{-13}$ (29$^{+4}_{-7}$) in the local (global) scale \citep{fujimoto2021}.
In this paper, we adopt the magnification factor of 29$^{+4}_{-7}$ in our analyses to estimate the intrinsic flux density of the host galaxy on the global scale, given the scopes in the number count and IRLF measurements. 

{\flushleft \bf M0553-C133/190/249 (serendipitous)} --
The lines are serendipitously and consistently identified in the ALCS data cube from nearby three ALCS sources at 269.12~GHz with SNR$\sim$15--30 in the velocity integrated line flux density over 660 km~s$^{-1}$ (Figure \ref{fig:line_eg}). 
Their HST counterparts are known to be a multiply imaged system spectroscopically confirmed at $z=1.14$ \citep{ebeling2017}, and thus these lines are determined as CO(5-4) at $z=1.14$. 
M0553-C113 falls in the FoV of follow-up ALMA Band 3 line scan observations for M0553-C67 (\#2019.1.00237.S, PI: S. Fujimoto), and we also detect the CO(2-1) line from M0553-C113 in the Band~3 data cube.

{\flushleft \bf M0553-C303 (serendipitous)} --
The line is serendipitously identified in the ALCS data cube at 250.95~GHz with SNR=9.5 in the velocity integrated line flux density over 270 km~s$^{-1}$. 
The source falls outside of the HST/WFC3 footprint, while it is detected in the IRAC ch1 and ch2 maps with the forced-aperture photometry (Appendix \ref{sec:app_znir}) and the {\it Herschel}/SPIRE maps with the debelended technique \citep{sun2022}. 
The template fitting with these IRAC, {\it Herschel}, and ALMA photometry suggests $z_{\rm phot}=0.65^{+0.05}_{-0.05}$ (Appendix \ref{sec:app_znir}), 
which is consistent with the independent FIR SED fitting results of $0.98^{+0.39}_{-0.39}$ \citep{sun2022} within the errors. 
We thus regard this line as CO(4-3) at $z=0.836$, 
while this interpretation might be updated with a second-line detection in future follow-up observations.
Note that the cluster redshift of M0553 is 0.430, and M0553-C303 is a lensed galaxy behind the cluster.

{\flushleft \bf R0032-C127/131/198 (serendipitous)} --
The lines are serendipitously consistently identified in the ALCS data cube at $\sim$272~GHz from nearby three ALCS sources  with SNR $\sim$ 5.0--9.2 in the velocity integrated line flux density over 800 km~s$^{-1}$. 
The counterparts are very faint, but identified in {\it HST}/WFC3 at their source positions, 
and secondary emission lines are securely detected at 101.97~GHz in a follow-up ALMA Band 3 \& 4 line scan program in this field (\#2021.1.01246.S, PI: K. Kohno). 
Although the central frequencies of the line features observed in the ALCS data cube are slightly different, this is likely explained either by the sensitivity or the differential magnification effect. Together with the similar NIR dark properties, we interpret these three sources as the multiply imaged system.  
Based on the frequencies of these two lines, we regard them as CO(8-7) in Band~6 and CO(3-2) in Band~3 at $z=$2.391. 
Our fiducial lens model also confirms the interpretation of the multiple images. 
Further details of the follow-up line scan observation results will be presented in Tsujita et al. (in prep.). 

{\flushleft \bf R0032-C208/281/304 (follow-up)} --
Although the ALMA continuum is brightly and securely detected ($>$1.0~mJy, SNR$\gtrsim$10), 
no \hst/WFC3 counterparts are identified at their source positions (F160W $\gtrsim$ 27~mag, 3$\sigma$; \citealt{coe2019}). 
The multiple lines are securely detected in follow-up ALMA Band 3 \& 4 line scan observations (\#2021.1.01246.S, PI: K. Kohno) exactly at the same frequencies at 86.80~GHz and 144.60~GHz. 
We conclude that these lines are CO(3-2) and CO(5-4) at $z=2.985$ and that these three sources are the multiply imaged system, which is consistent with the predictions from our fiducial lens model.
Further details of the follow-up line scan observation results will be presented in Tsujita et al. (in prep.). 

{\flushleft \bf M0417-C46/58/121 (follow-up)} --
Similar to the multiple images of R0032-C208/281/304, 
no \hst/WFC3 counterparts are identified at their source positions (F160W $\gtrsim$ 27~mag, 3$\sigma$; \citealt{coe2019}), 
although the ALMA continuum is brightly and securely detected ($>$2--4~mJy, SNR$>$20). 
A single line is robustly detected consistently at 99.11~GHz from all three sources in follow-up ALMA Band~3 line scan observations (\#2021.1.01246.S, PI: K. Kohno). The secondary line is also securely detected at 148.65~GHz from M0417-C46 in the Band~4 line scan observations in the same program, while the same observations were not carried out for the other two sources. 
Our fiducial lens model suggests that these three sources are multiple images, and given that the line at 99.1~GHz is detected exactly at the same frequency among these three sources, we conclude that these are multiple images. 
Based on the two emission lines, we conclude that their redshift at $z=3.652$. 
Further details of the follow-up line scan observation results will be presented in Tsujita et al. (in prep.). 

\setlength{\tabcolsep}{3pt}
\begin{table*}
\begin{center}
\caption{Serendipitous Line Detection from ALCS Continuum Sources in ALMA Band~6}
\label{tab:line_prop}
\vspace{-0.4cm}
\begin{tabular}{lcccccccc} \hline \hline
ID & S/N  & $\nu$ & FWHM & $I_{\rm line}$ & $f_{\rm cont}$ & Line & $z$ & Note \\ 
   &      & GHz   &  km~s$^{-1}$     & Jy~km~s$^{-1}$ & \%  & &  & \\
(1) & (2) &  (3)  &  (4)             & (5)            & (6) & (7)  & (8) & (9)\\ \hline
%\multicolumn{9}{c}{Serendipitious detection in ALCS 15-GHz data cube} \\ \hline
M1931-C41  &46.0  & $ 255.67 \pm 0.00$ & $ 224 \pm 11 $ & $ 2.05 \pm 0.14 $ & 1.7 & CO(3-2) & 0.352 & BCG\\
A370-C31   &18.9  & $ 251.75 \pm 0.00$ & $ 168 \pm 10 $ & $ 0.63 \pm 0.05 $ & 12.0 & CO(3-2) & 0.375 & BCG\\ 
M0035-C94  &9.5   & $ 254.40 \pm 0.02 $ & $ 332 \pm 50 $ & $ 0.48 \pm 0.1 $ & 10.9 & CO(3-2) & 0.359 & BCG \\
R0600-C164 & 9.1  &  $ 268.68 \pm 0.01 $ & $ 162 \pm 30 $ & $ 0.32 \pm 0.08 $ & 4.9 & \cii\  & 6.072 & {\it HST}-detected (F21,L21) \\
M0553-C133 & 15.2 & $269.07 \pm 0.02 $ & $ 557 \pm 62 $ & $ 1.53 \pm 0.22 $ & 3.2 & CO(5-4) & 1.142 & {\it HST}-detected, multiple images (E17) \\
M0553-C190 & 32.8 & $269.08 \pm 0.02 $ & $ 562 \pm 55 $ & $ 3.36 \pm 0.43 $ & 4.1 & CO(5-4) & 1.142 & {\it HST}-detected, multiple images (E17) \\
M0553-C249 &24.2  & $ 269.06 \pm 0.02 $ & $ 592 \pm 62 $ & $ 2.14 \pm 0.29 $ & 4.0 & CO(5-4) & 1.142 & {\it HST}-detected, multiple images (E17) \\
M0416-C120 & 5.4  & $ 267.30 \pm 0.02 $ & $ 157 \pm 55 $ & $ 0.25 \pm 0.12 $ & 6.7 & CO(8-7) & 2.448 & {\it HST}-detected, $z_{\rm phot}$ + single line  \\ 
R0600-C13  & 7.8  & $ 253.76 \pm 0.02 $ & $ 325 \pm 45 $ & $ 0.49 \pm 0.09 $ & 4.5 & CO(5-4) & 1.271 & {\it HST}-detected, $z_{\rm phot}$ + single line  \\
M0553-C303 & 6.2  & $ 250.92 \pm 0.02 $ & $ 260 \pm 56 $ & $ 0.34 \pm 0.1 $ & 3.4 & CO(5-4) & 0.836 & {\it HST}-detected, $z_{\rm phot}$ + single line  \\
R0032-C127 & 9.2   & $ 271.74 \pm 0.04 $ & $ 674 \pm 119 $ & $ 0.83 \pm 0.19 $ & 1.6 & CO(8-7) & 2.391 & {\it HST}-detected, two lines$\dagger$, multiple images  \\
R0032-C131 & 6.3   & $ 272.04 \pm 0.02 $ & $ 210 \pm 54 $ & $ 0.28 \pm 0.1 $ & 0.9 & CO(8-7) & 2.391 & {\it HST}-detected, two lines$\dagger$, multiple images  \\
R0032-C198 & 5.0   &  $ 271.93 \pm 0.08 $ & $ 551 \pm 219 $ & $ 0.33 \pm 0.17 $ & 2.1 & CO(8-7) & 2.391 & {\it HST}-detected, two lines$\dagger$, multiple images  \\ 
\hline \hline
\end{tabular}
\end{center}
\vspace{-0.4cm}
\tablecomments{
(1) ALCS continuum source ID. 
(2) SNR of the $1\farcs0$-diameter aperture photometry in the velocity-integrated map for the line emission. 
(3) Central frequency of the line emission.
(4) FWHM of the line width. 
(5) Velocity-integrated line flux.  
(6) Contribution of the line flux to the 1.2-mm continuum flux density estimate, which is corrected in Section \ref{sec:flux}. 
(7) Line classification. 
(8) Spectroscopic redshift from the line. 
(9) Reference: E17 \citep{ebeling2017}, F21 \citep{fujimoto2021}, and L21 \citep{laporte2021}.
The values in (3--5) are measured from a single Gaussian fitting in the spectrum. \\
$\dagger$ The second line is securely detected in follow-up Band 3 line scan observations  (Table~\ref{tab:line_prop2}) at a consistent frequency from these three sources.  
}
\end{table*}

\setlength{\tabcolsep}{3pt}
\begin{table*}
\begin{center}
\caption{Line Detection in Follow-up Line Scan in ALMA Band 3, 4 for NIR dark Galaxies in ALCS}
\label{tab:line_prop2}
\vspace{-0.4cm}
\begin{tabular}{lccccc} \hline \hline
ID & $z$ & Band & Line & Note & Ref.  \\ 
(1) & (2) &  (3) &  (4)  & (5) & (6)  \\ \hline
M0553-C67  & 4.932 & 3 & CO(5-4) &  No counterparts in {\it HST}/WFC3, $z_{\rm phot}$ + single line & App.~\ref{sec:appendix_line} \\
R0032-C127 & 2.391 & 3, 4 & CO(3-2), CO(4-3), \ci(1-0) & two lines, multiple images & T23 \\
R0032-C131 & 2.391 & 3, 4 &CO(3-2), CO(4-3), \ci(1-0)  & two lines, multiple images & T23  \\
R0032-C198 & 2.391 & 3, 4 &CO(3-2), CO(4-3), \ci(1-0)  & two lines, multiple images & T23  \\
R0032-C208 & 2.985 & 3, 4 &CO(3-2), CO(5-4)  & No counterparts in \hst/WFC3, two lines, multiple images & T23  \\
R0032-C281 & 2.985 & 3, 4 &CO(3-2), CO(5-4)  & No counterparts in \hst/WFC3, two lines, multiple images & T23   \\
R0032-C304 & 2.985 & 3, 4 &CO(3-2), CO(5-4)  & No counterparts in \hst/WFC3, two lines, multiple images & T23   \\ 
M0417-C46 & 3.652 & 3, 4  & CO(4-3), CO(6-5) & No counterparts in \hst/WFC3, two lines, multiple images & T23 \\
M0417-C58 & 3.652 & 3, 4 &CO(4-3), CO(6-5)  & No counterparts in \hst/WFC3, two lines, multiple images & T23 \\
M0417-C121 & 3.652 & 3, 4 &CO(4-3), CO(6-5) & No counterparts in \hst/WFC3, two lines, multiple images & T23 \\
R0600-C67  & (4.80) & 3, 4 & CO(7-6)$?$, CO(6-5)$?$ & No counterparts in \hst, IRAC , two tentative line + $z_{\rm phot}^{\dagger}$ & B23 \\
R0949-C10  & 5.753  & 3, 4 & CO(6-5), CO(5-4)       & No counterparts in \hst, IRAC, two lines  & B23 \\
R0949-C19  & (5.53) & 3, 4 & CO(5-4)$?$             & No counterparts in \hst, IRAC, single tentative line + $z_{\rm phot}^{\dagger}$ & B23 \\
\hline \hline
\end{tabular}
\end{center}
\vspace{-0.4cm}
\tablecomments{
(1) ALCS continuum source ID. 
(2) Redshift determined by the emission lines. 
(3) ALMA Band used in the follow-up line scan observations. 
(4) Lines detected in the follow-up observations. 
(4) Notes. 
(5) Reference: T23 (Tsujita et al. in prep.) and B23 (Bauer et al. in prep). \\ 
$\dagger$ The upper limit of the flux ratio at 4.5~$\mu$m (IRAC ch2) and 1.2~mm (ALMA) suggests the source redshift at $z>4.5$--5.0 based on the SED template of the high-$z$ dusty galaxy. We regard the line feature observed with the highest SNR ($\sim5$) in the Band 3+4 spectrum as the CO line, which satisfies the lowest-$z$ solution at $z>4.5$ (Appendix~\ref{sec:appendix_line}).  
}
\end{table*}

\section{$z_{\rm phot}$ estimate by template fitting}
\label{sec:app_znir}

In Section \ref{sec:redshift}, we identify 26 ALCS sources lacking the \hst\ counterparts 
due to their faintness in the \hst\ bands ($N=18$) and their sky positions outside of the \hst\ footprints ($N=8$), respectively. 
We obtain the forced-aperture photometry in IRAC ch1 and ch2 at the ALCS source position with a diameter of $3\farcs0$ on the residual IRAC maps after subtracting the models for the \hst-detected sources as well as the debelended {\it Herschel} photometry if detected \citep{sun2022}. 
There remain 4 ALCS sources that are not detected in any bands in IRAC and {\it Herschel} (ACT0102-C11, ACT0102-C251, R1347-C51, and M2129-C24). 
These 4 ALCS sources are all included in the primary catalog. 
With our purity cut for the primary catalog ($>$99\%) based on the positive and negative source distribution (Figure \ref{fig:pn_hist}), 
one of these 4 ALCS sources might be spurious. Still, the majority of them are the most likely real. 

Given the limited number of photometry data points, 
we simply perform the $\chi^{2}$ minimization with a fixed template to evaluate $P(z)$ for these 23 (=27-4) ALCS sources. 
We use the SED fitting code of {\sc stardust} \citep{kokorev2021} by implementing the SED template of Arp220 and a composite SED model from 707 dusty galaxies identified in the AS2UDS survey \citep[e.g.,][]{dudzeviciute2020}. 
We find that the AS2UDS template suggests a lower redshift solution than Arp220 in general. This is because the AS2UDS template is fainter in the rest-frame optical than Arp220 due to a lower typical stellar mass or much higher obscuration \citep{dudzeviciute2020}. 

%%%%%%%%%%%%%%%%%%%%%%%
\begin{figure*}
\begin{center}
\includegraphics[trim=0cm 0cm 0cm 0cm, clip, angle=0,width=0.85\textwidth]{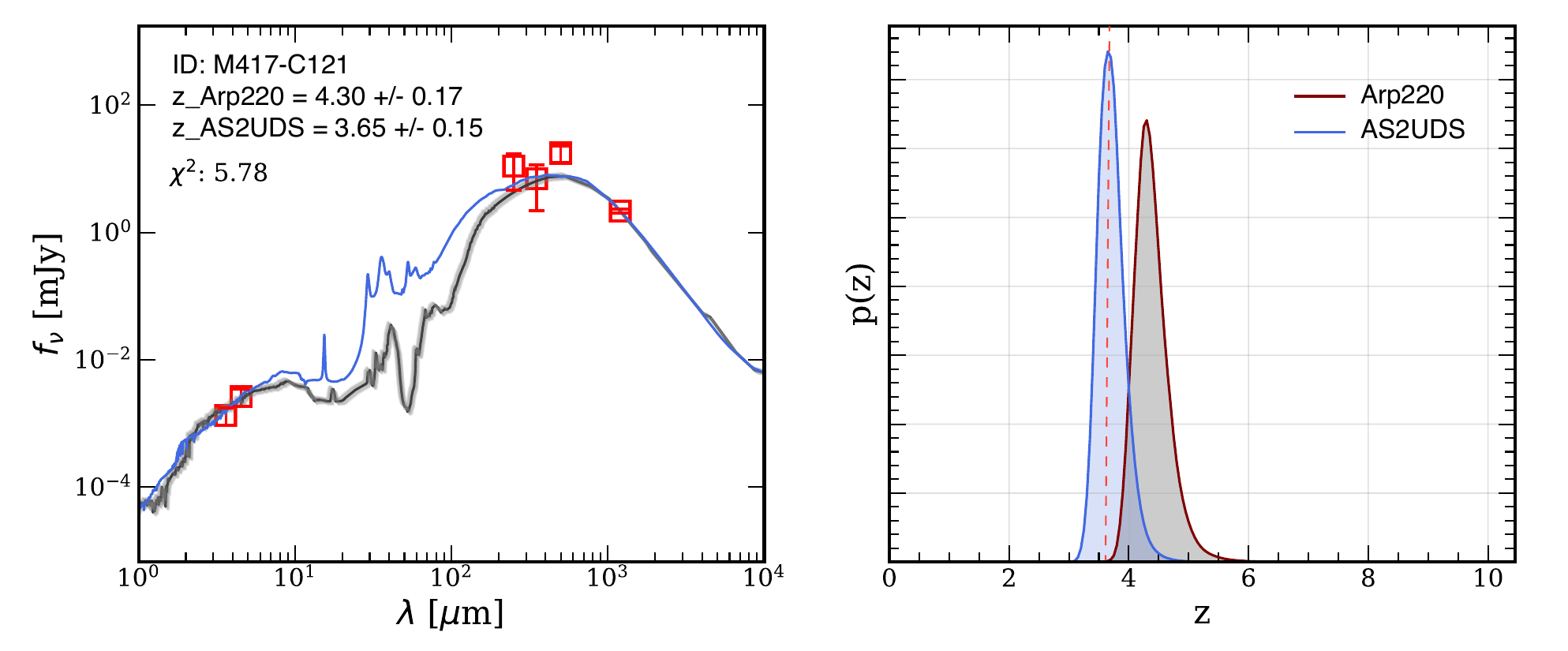}
 \caption{
Example of our template fitting for the ALCS sources with no \hst\ counterparts. 
{\it Left}: The best-fit SEDs for M0417-C121 based on the templates of Arp220 (black line) and the composite of 707 dusty galaxies in AS2UDS \citep{dudzeviciute2020}. 
{\it Right}: $P(z)$ from the $\chi^{2}$ minimazation. 
The red dashed line indicates $z_{\rm spec}=3.652$ of M0417-C121 determined from follow-up ALMA Band 3 \& 4 line scan observations with multiple line detection (Tsujita et al. in prep.).  
\label{fig:sed_temp}}
\end{center}
\end{figure*}
%%%%%%%%%%%%%%%%%%%%%%% 

Figure \ref{fig:sed_temp} shows an example of the NIR--mm photometry and the best-fit SED (left) and $P(z)$ (right) with the Arp220 and AS2UDS templates for M0417-C121.
M0417-C121 is robustly detected with ALMA (SNR$_{\rm nat}$=20.7) without any \hst\ counterparts, and follow-up ALMA Band 3 \& 4 line scan observations determine its redshift at $z_{\rm spec}=3.652$ with two secure CO line detection (Tsujita et al. in prep.). 
We find that these SED templates suggest the source redshift at $z\sim$ 3--4, where the peak of $P(z)$ with the AS2UDS template shows an excellent agreement with $z_{\rm spec}$ of M0417-C121. 
Given the intrinsically faint ($\sim$ less massive) and potentially high dust obscuration aspects without any \hst\ counterparts, 
the SEDs of these ALCS sources may be closer to the AS2UDS template.  
We thus use $P(z)$ obtained with the AS2UDS template for the 23 ALCS sources.

\section{Lens Models}
\label{sec:app_model_all}

We construct lens models using cluster member galaxies as well as multiple images behind clusters based on the photometric redshift, colors, and morphology of galaxies in the {\it HST} images with independent algorithms including {\sc glafic} \citep{oguri2010,kawamata2016, kawamata2018, okabe2020}, {\sc Lenstool} \citep{jullo2007, caminha2016, caminha2017, caminha2017b, caminha2019, richard2014, niemiec2020}, Light-Traces-Mass (Zitrin-LTM; \citealt{zitrin2013,zitrin2015}), and Pseudo-Isothermal Elliptical Mass Distribution plus elliptical Navarro–Frenk–White (Zitrin-dPIEeNFW; \citealt{zitrin2013,zitrin2015}). 
We use the lens model of {\sc glafic} as a fiducial model for our analyses throughout this paper, while other available models are also used to evaluate uncertainties in magnification factors (Appendix~\ref{sec:app_models}). 
Note that we do not use the median magnification estimate among different lens models, because the process of choosing the median estimate systematically reduces the effective survey area at high magnifications of $\mu\gtrsim10$ (Appendix~\ref{sec:app_models}). 
In Table~\ref{tab:model_all}, we summarize the number of the multiple images used for constructing the {\sc glafic} models and the r.m.s of positional differences between the observed and model-predicted multiple images. 

\setlength{\tabcolsep}{8pt}
\begin{table*}
\begin{center}
\caption{Summary of our fiducial and available lens models in the 33 ALCS clusters}
\label{tab:model_all}
\vspace{-0.4cm}
\begin{tabular}{lccccccccc} \hline \hline
Cluster	&	$z$	&	$N_{\rm src,all}$	&	$N_{\rm src,spec}$	&	$N_{\rm img}$	&	rms$_{\rm img}$	&	Ref.	&	Caminha	&	CATS	&	Zitrin	\\ 
(1) & (2) &  (3) &  (4)  & (5) & (6) & (7) & (8) & (9) & (10) \\ \hline
\multicolumn{10}{c}{HFF} \\ \hline
A2744	&	0.308	&	45	&	24	&	132	&	0.42	&	K18	&	N	&	Y	&	Y	\\
A370	&	0.375	&	49	&	19	&	135	&	0.50    &	K18		&	N	&	Y	&	Y	\\
M0416	&	0.396	&	75	&	34	&	202	&	0.44	&	K18	&	Y	&	Y	&	Y	\\
M1149	&	0.543	&	36	&	16	&	108	&	0.31	&	K16	&	N	&	Y	&	Y	\\
AS1063	&	0.348	&	53	&	19	&	141	&	0.38	&	K18   	&	Y	&	Y	&	Y	\\ \hline
\multicolumn{10}{c}{CLASH}   \\ \hline													
A209	&	0.206	&	3	&	0	&	7	&	0.52	&	Z15	&	N	&	Y	&	Y	\\
A383	&	0.187	&	8	&	6	&	23	&	0.41	&	Z15	&	N	&	Y	&	Y	\\
M0329	&	0.45	&	11	&	10	&	28	&	0.62	&	Ca19, R21	&	Y	&	Y	&	Y	\\
M0429	&	0.399	&	3	&	2	&	11	&	0.32	&	Ca19	&	Y	&	N	&	Y	\\
M1115	&	0.355	&	3	&	1	&	9	&	0.44	&	Ca19	&	Y	&	Y	&	Y	\\
M1206	&	0.439	&	37	&	37	&	109	&	0.47	&	Ca17, R21	&	Y	&	Y	&	Y	\\
M1311	&	0.494	&	3	&	1	&	8	&	0.57	&	Ca19	&	Y	&	N	&	Y	\\
R1347	&	0.451	&	40	&	38	&	129	&	0.65	&	U18, Ca19, R21	&	Y	&	Y	&	Y	\\
M1423	&	0.545	&	3	&	2	&	12	&	0.65	&	Z15	&	N	&	Y	&	Y	\\
M1931	&	0.352	&	7	&	7	&	19	&	0.39	&	Ca19	&	Y	&	Y	&	Y	\\
M2129	&	0.57	&	11	&	11	&	38	&	0.73	&	Ca19	&	Y	&	Y	&	Y	\\
R2129	&	0.234	&	7	&	7	&	22	&	0.43	&	Ca19   	&	Y	&	Y	&	Y	\\ \hline
\multicolumn{10}{c}{RELICS}  \\ \hline																		
R0032	&	0.396	&	12	&	2	&	36	&	0.49	&	A20	&	Y	&	Y	&	N	\\
M0035	&	0.352	&	5	&	0	&	13	&	0.24	&	\nodata	&	Y	&	N	&	N	\\
AC0102	&	0.87	&	10	&	0	&	28	&	0.52	&	Ce18	&	Y	&	N	&	N	\\
M0159	&	0.405	&	4	&	0	&	10	&	0.47	&	Ci18	&	Y	&	Y	&	N	\\
A295	&	0.3	&	6	&	4	&	18	&	0.22	&	Ci18	&	N	&	N	&	N	\\
M0257	&	0.505	&	4	&	0	&	12	&	0.36	&	Z11	&	N	&	Y	&	N	\\
P171	&	0.27	&	5	&	0	&	16	&	0.34	&	A18	&	N	&	N	&	Y	\\
A3192	&	0.425	&	5	&	2	&	16	&	0.51	&	H13	&	N	&	N	&	N	\\
M0417	&	0.443	&	21	&	8	&	57	&	0.45	&	M19 \& ALCS$^{\dagger}$	&	Y	&	Y	&	N	\\
M0553	&	0.43	&	10	&	2	&	30	&	0.73	&	E17	&	N	&	Y	&	N	\\
R0600	&	0.46	&	8	&	5	&	26	&	0.66	&	F21	&	Y	&	Y	&	N	\\
SM0723	&	0.39	&	4	&	0	&	13	&	0.27	&	\nodata	&	N	&	Y	&	N	\\
R0949	&	0.383	&	4	&	0	&	11	&	0.26	&	\nodata	&	N	&	Y	&	N	\\
A2163	&	0.203	&	4	&	3	&	15	&	0.3	&	Ce18, Re20	&	Y	&	N	&	N	\\
RJ2211	&	0.397	&	3	&	1	&	11	&	0.14	&	Ce18	&	N	&	N	&	N	\\
A2537	&	0.297	&	8	&	1	&	29	&	0.44	&	Ce18	&	N	&	Y	&	N	\\
\hline \hline
\end{tabular}
\end{center}
\vspace{-0.4cm}
\tablecomments{
(1) Cluster Name. 
(2) Cluster redshift. 
(3) Number of the multiple system identified in the cluster. 
(4) Number of multiple system spectroscopically confirmed identified in the cluster.  
(5) Number of multiple images identified in the image plane used for constructing our fiducial mass model with {\sc glafic} \citep{oguri2010}. 
(6) RMS of positional differences between observed and model-predicted multiple images, in units of arcseconds.  
(7) References for the multiple images: K18 \citep{kawamata2018}, Z15 \citep{zitrin2015}, Ca19 \citep{caminha2019}, R21 \citep{richard2021}, Ce18 \citep{cerny2018}, Ci18 \citep{cibirka2018}, Z11 \citep{zitrin2011}, M19 \citep{mahler2019}, E17 \citep{ebeling2017}, R20 \citep{rescigno2020}, U18 \citep{sueda2018}, H13 \citep{hsu2013}, F21 \citep{fujimoto2021}
(8--10) Availability (Y: Yes, No: No) of other mass models so far constructed by G. Caminha with {\sc Lenstool} \citep{caminha2016, caminha2017, caminha2017b, caminha2019}, CATS team with {\sc Lenstool} \citep{mahler2019, lagattuta2019, richard2010, richard2011, richard2021, jauzac2016, jauzac2021,  rexroth2018, mirka2017, livermore2012, repp2016, ebeling2017, newman2013b}, and A. Zitrin with LTM and dPIEeNFW algorithms \citep{zitrin2013, zitrin2015} in the ALCS 33 clusters. 
All model links will be available on the ALCS website. \footnote{
\url{http://www.ioa.s.u-tokyo.ac.jp/ALCS/}
} 
Interested readers may find more lens models on HFF\footnote{
\url{https://archive.stsci.edu/prepds/frontier/}
}, CLASH\footnote{
\url{https://www.stsci.edu/~postman/CLASH/}
}, and RELICS\footnote{
\url{https://relics.stsci.edu/}
} websites. \\
$\dagger$ The multiple images of the ALCS sources without \hst\ counterparts, M0417-C46/58/121, are also used for refining our fiducial lens model.  
}
\end{table*}

\section{Lens Models Uncertainty}
\label{sec:app_models}

%%%%%%%%%%%%%%%%%%%%%%%
\begin{figure*}
\begin{center}
\includegraphics[trim=0cm 0cm 0cm 0cm, clip, angle=0,width=1.0\textwidth]{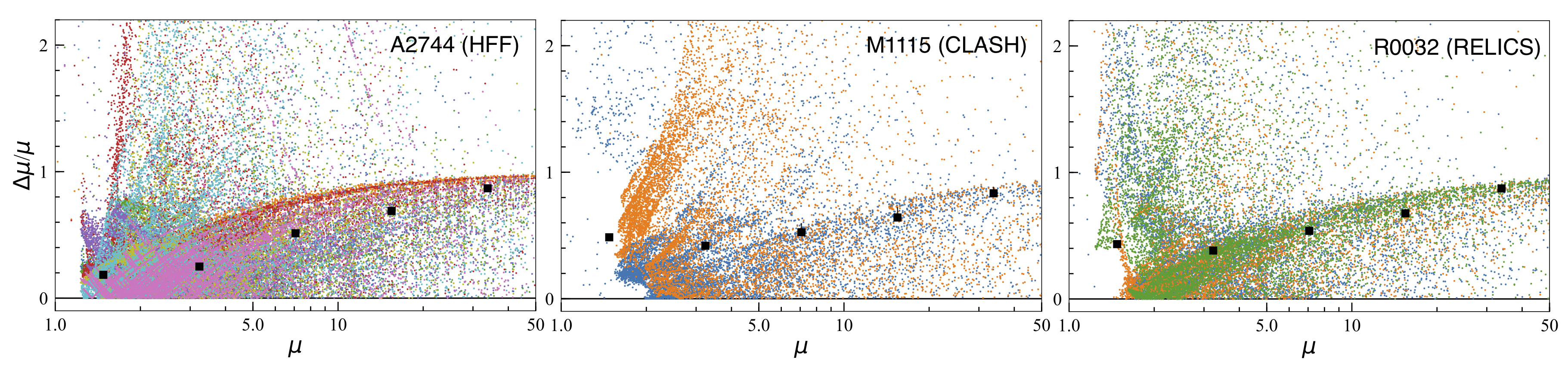}
 \caption{
Difference of the magnification factor at random positions among different lens models. 
We evaluate the difference with $\Delta \mu$/$\mu$ and $\Delta \mu=|\mu-\mu_{\rm other}|$, where $\mu$ and $\mu_{\rm other}$ are the magnification factors with our fiducial and other lens models, respectively. 
The different color plots indicate different lens models for $z=2$, where there are 8, 3, and 4 publically available lens models in A2744 (HFF, left), M1115 (CLASH, middle), and R0032 (RELICS, right), respectively, including our fiducial lens model constructed with {\sc glafic} \citep{oguri2010}. 
The black squares denote the median of $\Delta \mu$/$\mu$ in each $\mu$ bin. 
\label{fig:app_model}}
\end{center}
\end{figure*}
%%%%%%%%%%%%%%%%%%%%%%% 

We compare the magnification factors between our fiducial and other lens models at random pixel positions of the maps and evaluate the systematic uncertainty from the choice of the lens model. 
To evaluate the systematic uncertainty, we calculate $\Delta\mu/\mu$ ($\equiv|\mu_{\rm other} - \mu|/\mu$). 
The accuracy of the lens model generally depends on the richness of the multi-wavelength data in the cluster field for identifying the multiple images and the cluster member galaxies. 
Thus we separately calculate $\Delta\mu/\mu$ for HFF, CLASH, and RELICS. 
The clusters of A2744, M1115, and R0032 are used for the calculation of $\Delta\mu/\mu$ as the representatives of HFF, CLASH, and RELICS, respectively, and we fix the redshift at $z=2$ for the lens model. 

%%%%%%%%%%%%%%%%%%%%%%%
\begin{figure}[t]
\begin{center}
\includegraphics[trim=0cm 0cm 0cm 0cm, clip, angle=0,width=0.5\textwidth]{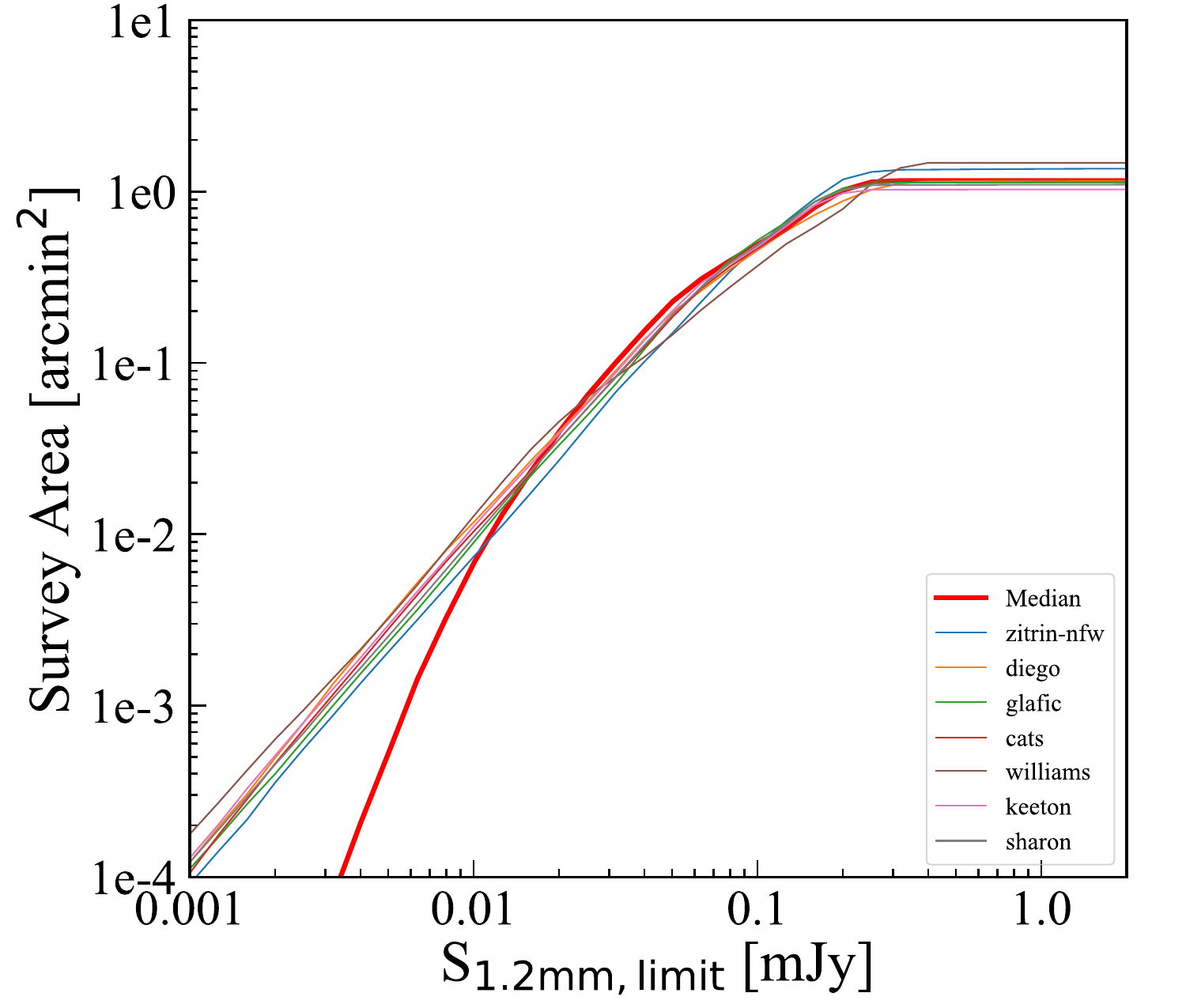}
 \caption{
Comparison of the effective survey area among 7 different lens models publically available in AS1063. 
The red line is drawn from a median magnification map created from these lens models. 
We find that the median magnification map underestimates the effective survey, especially at $0.01$~mJy which corresponds to the area with $\mu\gtrsim10$. 
This is because the highly magnified regions are smeared out by sampling the median values due to slight differences in the critical curve positions among different lens models. 
\label{fig:app_area}}
\end{center}
\end{figure}
%%%%%%%%%%%%%%%%%%%%%%% 

In Figure \ref{fig:app_model}, we show $\Delta\mu/\mu$ at random positions in the cluster fields. 
The median values as a function of $\mu$ are also plotted with black squares. 
We find an increasing trend in $\Delta\mu/\mu$ as a function of $\mu$. 
We also find that $\Delta\mu/\mu$ is the smallest in HFF at $\mu \leq 5$, while there is no large difference in $\Delta\mu/\mu$ among HFF, CLASH, and RELICS at $\mu > 5$. 
Based on these results, we adopt 20\% (40\%), 50\%, 60\%, and 80\% for the systematic uncertainty of the different lens models at $\mu\leq5$, $5<\mu\leq10$, $10<\mu\leq30$, and $\mu >30$ in HFF (CLASH \& RELICS). 

We also evaluate the difference in the effective survey area among different lens models. 
In Figure \ref{fig:app_area}, we show the survey area after correcting the lensing effect with 7 different lens models in AS1063. 
The redshift is fixed again at $z=2$ for all these lens models. 
We find that the survey area is unchanged by more than $\sim$10\% in the wide magnification ranges. This indicates that the overall shape of the magnification distribution is not significantly changed among the different lens models, although the small variations should remain. 
On the other hand, however, if we create the median magnification map from these 7 different lens models, the survey area is underestimated at high magnifications ($\mu\gtrsim10$). 
This is because such highly magnified regions are smeared out even by the small variation of the critical curve positions among the different models.

\begin{figure}
\begin{center}
\includegraphics[trim=0cm 0cm 0cm 0cm, clip, angle=0,width=0.5\textwidth]{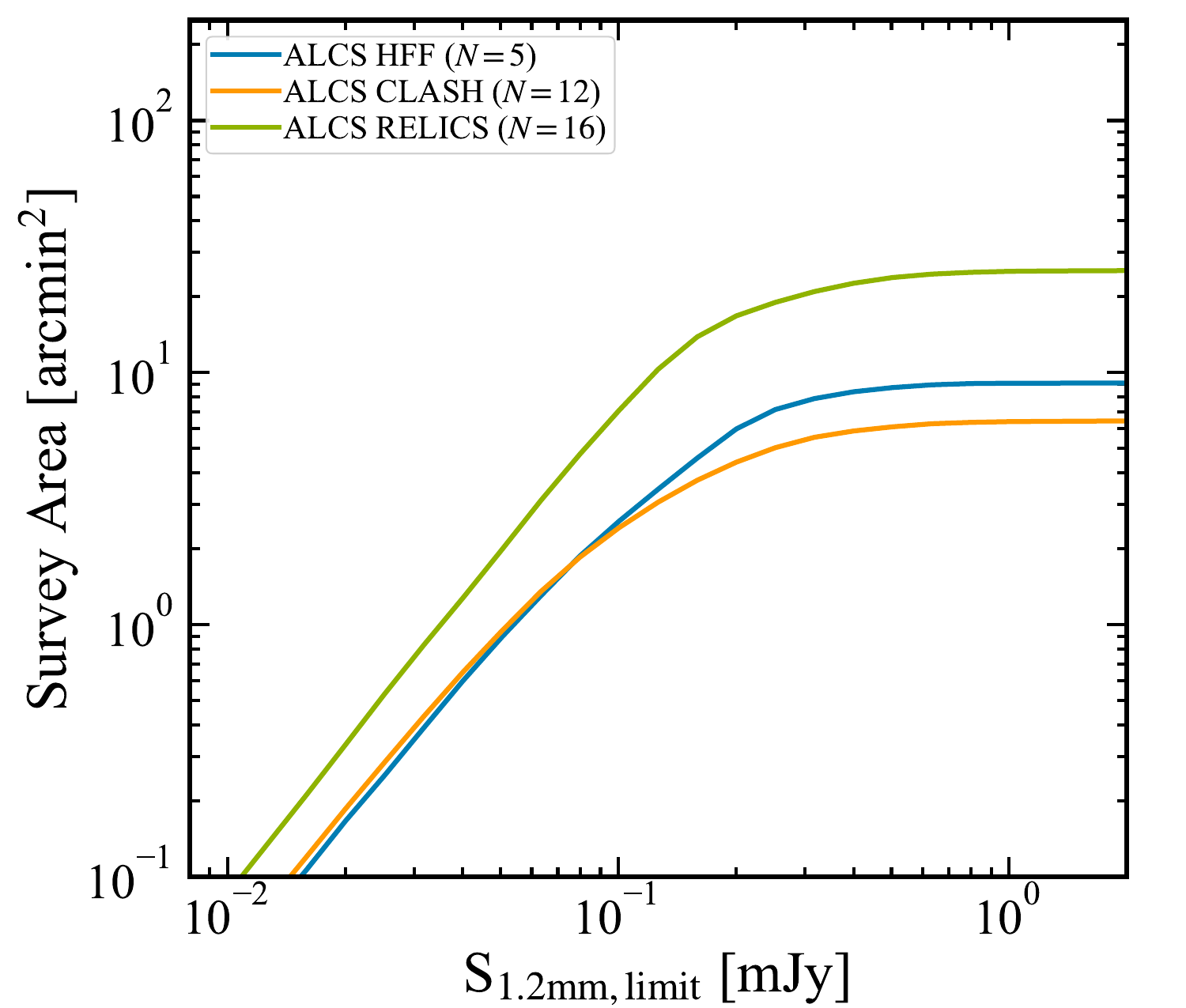}
 \caption{
Comparison of the effective survey area among the HFF, CLASH, and RELICS clusters based on our fiducial lens model at $z=2$. 
\label{fig:diff_clusters}}
\end{center}
\end{figure}

Figure~\ref{fig:diff_clusters} shows the comparison of the effective survey area among the HFF, CLASH, and RELICS. As discussed in Section~\ref{sec:caveats}, the less detailed models with fewer multiple images in CLASH and RELICS than HFF may lead to underestimating the survey area at high magnification ($\mu\gtrsim3$) by factors of $\sim1.4$--1.8 \citep{jauzac2015}. 
The sum of the survey areas from CLASH and RELICS at the high magnification regime contribute to almost constantly $\sim$80\% of the total survey area, suggesting that our survey area estimate can be increased by factors of $\sim$1.1--1.5 in future deep follow-up observations in the CLASH and RELICS clusters. 
On the other hand, the faint-end estimates in the number counts and LFs may not be changed much, because the increase of the high magnification area also increases the sources with high magnification estimates. 

%%%%%%%%%%%%%%%%%%%%%%%
%%%%%%%%%%%%%%%%%%%%%%% 

\section{Predictions from the EAGLE simulation}
\label{sec:app_eagle}

For EAGLE, we retrieved the publicly available spectral energy distributions of all galaxies from $z=10$ to $z=0$ (see \citealt{camps2018} for details of how these fluxes are computed). We compute the total IR luminosity using all four (rest-frame) IRAC bands available in the database and the conversion between those fluxes and the total IR luminosity of \cite{sanders1996}. From those luminosities, we construct the IR luminosity functions at all redshifts. To compute the number counts at 1.2mm, we retrieve the observer-frame ALMA band 6 fluxes and use all the snapshots of the simulation to construct a lightcone. The lightcone is constructed as follows: for each redshift, we compute the implied projected sky area of a 100 Mpc$^{2}$ assuming the EAGLE cosmology \citep{planck2014xi} and the number of galaxies per unit area at each flux bin contributed by that redshift $N(z,S)$. We then integrate under the $N(z,S)$ vs $z$ curve to obtain $N(S)$. We note that the fluxes available in the EAGLE database are computed only for galaxies that have a number of particles representing the galaxy’s body of dust $>250$. The latter mostly removes quenched galaxies that are more prevalent at low redshift. Hence, we do not think that affects the 1.2mm number counts in the flux range investigated here.

\section{Footprints of 33 target clusters in ALCS}
\label{sec:app_fields}

\begin{figure*}
\begin{center}
\includegraphics[trim=0cm 0cm 0cm 0cm, clip, angle=0,width=1.0\textwidth]{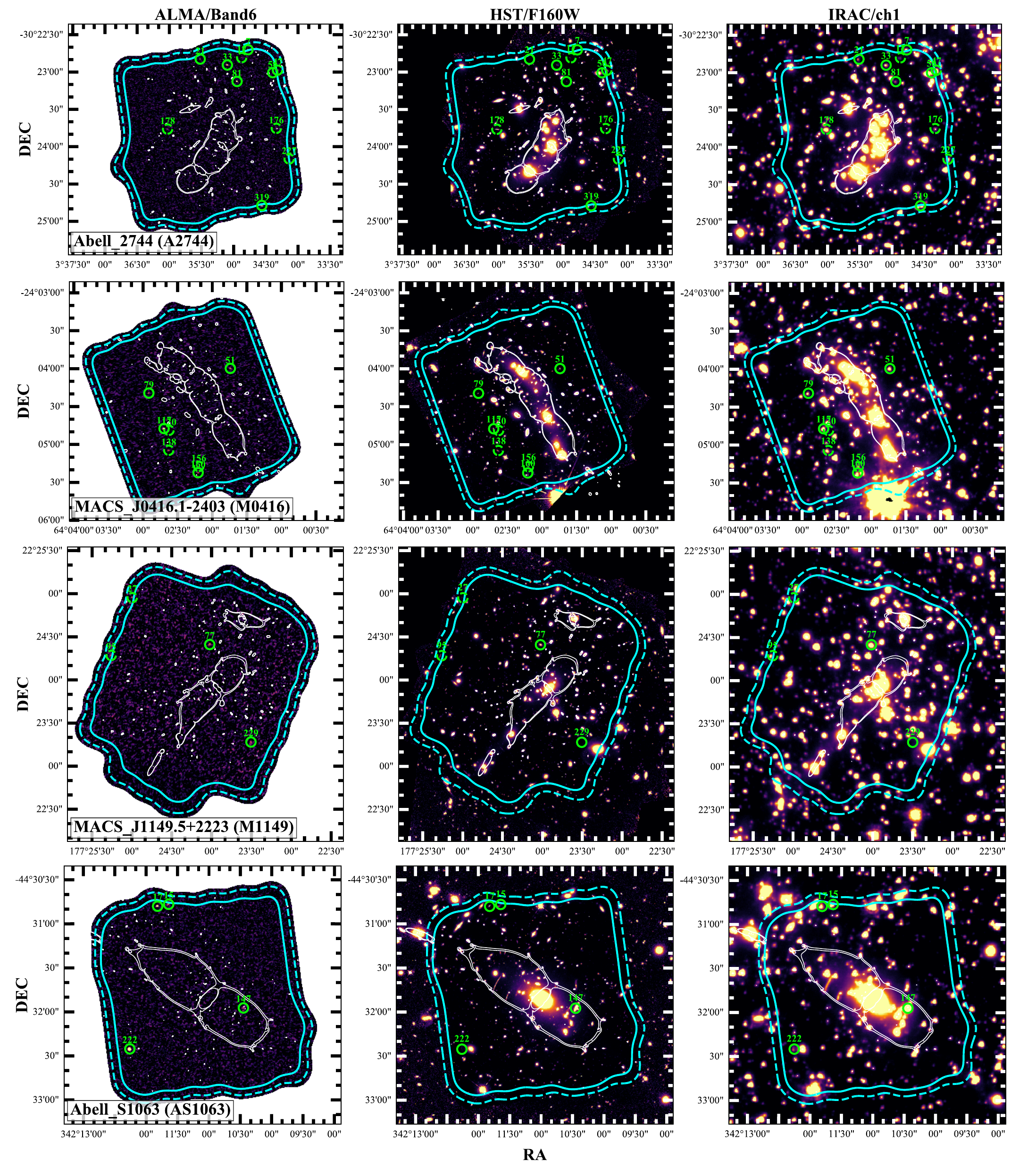}
 \caption{
 Same as Figure~\ref{fig:fig1}, but for the remaining 32 lensing clusters and the sky positions of the ALCS sources. 
%Overview of the ALCS 33 lensing clusters and the sky positions of the 180 ALCS sources. 
ALMA Band~6, HST/F160W, and IRAC/ch1 maps are presented from left to right. 
The green solid and dashed circles show the ALCS sources in the primary and secondary samples, respectively. 
The white lines denote the magnification curve of $\mu=200$ at $z=2$ with our fiducial lens model. 
The solid and dashed cyan lines indicate the PB sensitivity at 50\% and 30\%, respectively. 
\label{fig:app_cluster}}
\end{center}
\end{figure*}

In Figure \ref{fig:app_cluster}, we present the footpritns of ALMA Band~6, HST/F160W, and IRAC/ch1 for the 33 target lensing clusters in the same format as Figure \ref{fig:fig1}. 

\section{Full list of ALCS 180 sources}
\label{sec:postage}

\setlength{\tabcolsep}{2pt}
\startlongtable
\begin{deluxetable*}{lllcccccccc}
\tablecaption{ALCS Continuum Source Catalog}
\tablehead{
\colhead{ID} & \colhead{R.A.} & \colhead{Dec}   & \colhead{SNR} & \colhead{PB} & \colhead{$S_{\rm nat}$} & \colhead{$S_{\rm tap}$}  &  \colhead{$S_{\rm ap}$} & \colhead{$S_{\rm imf}$} & \colhead{flag} \\
  \colhead{}  & \colhead{(deg)}   & \colhead{(deg)}  &  \colhead{}    &  \colhead{}    & \colhead{(mJy/beam)}&  \colhead{(mJy/beam)} &   \colhead{(mJy)}    &  \colhead{(mJy)}     &  \colhead{}  \\
\colhead{(1)} & \multicolumn{2}{c}{(2)} & \colhead{(3)} & \colhead{(4)} & \colhead{(5)} & \colhead{(6)} & \colhead{(7)} & \colhead{(8)} & \colhead{(9)}}
\startdata
\multicolumn{10}{c}{Blind Catalog ($N=141$)} \\   \hline
AC0102-C11 & 15.7477863 &$ -49.2904001 $& 5.4 (3.1) & 0.96 & 0.39 $\pm$ 0.07 & 0.34 $\pm$ 0.11 & 0.28 $\pm$ 0.12 & 0.35 $\pm$ 0.1 &$ 0 $\\
AC0102-C22 & 15.7626821 &$ -49.2864566 $& 7.9 (7.4) & 0.79 & 0.71 $\pm$ 0.09 & 0.97 $\pm$ 0.13 & 1.23 $\pm$ 0.14 & 0.96 $\pm$ 0.19 &$ 2 $\\
AC0102-C50 & 15.7475095 &$ -49.2821233 $& 5.8 (5.6) & 0.99 & 0.41 $\pm$ 0.07 & 0.58 $\pm$ 0.1 & 0.66 $\pm$ 0.11 & 0.67 $\pm$ 0.16 &$ 2 $\\
AC0102-C52 & 15.7561196 &$ -49.2823856 $& 5.8 (3.2) & 0.99 & 0.41 $\pm$ 0.07 & 0.33 $\pm$ 0.11 & 0.04 $\pm$ 0.11 & 0.4 $\pm$ 0.12 &$ 0 $\\
AC0102-C118 & 15.7128798 &$ -49.2607919 $& 34.3 (26.5) & 0.92 & 2.63 $\pm$ 0.08 & 2.99 $\pm$ 0.11 & 3.29 $\pm$ 0.12 & 2.93 $\pm$ 0.14 &$ 2 $\\
AC0102-C160 & 15.7508527 &$ -49.2677044 $& 4.7 (4.9) & 0.98 & 0.34 $\pm$ 0.07 & 0.52 $\pm$ 0.11 & 0.72 $\pm$ 0.12 & 0.87 $\pm$ 0.23 &$ 2 $\\
AC0102-C215 & 15.7288441 &$ -49.2540865 $& 42.4 (32.7) & 0.99 & 3.0 $\pm$ 0.07 & 3.4 $\pm$ 0.1 & 3.83 $\pm$ 0.11 & 3.6 $\pm$ 0.18 &$ 2 $\\
\enddata
\tablecomments{
(1) Source ID, including the prefix of the cluster name + ``AC'' (ALCS continuum). 
(2) Source coordinate of the continuum peak in the natural-weighted map.
(3) Signal-to-noise ratio of the peak pixel in the natural-weighted map before primary beam correction. 
(4) Primary beam sensitivity in the mosaic map. 
(5) Source flux density at 1.2 mm measured by the peak pixel count in the natural-weighted map.  
(6) Source flux density at 1.2 mm measured by the peak pixel count in the $uv$-tapered ($2''\times2''$) map. 
(7) Source flux density at 1.2 mm measured by a $2.0''$-radius aperture in the natural-weighted map.  
(8) Source flux density at 1.2 mm measured by 2D elliptical Gaussian fitting with CASA {\sc imfit}.  
(9) Flag of the {\sc imfit} output (0: point source, 1: mergry resovled, 2: fully resolved, $-$1: fitting error, $-2$: near the edge of the map). 
}
\label{tab:alcs_catalog_full}
\end{deluxetable*}

\startlongtable
\begin{deluxetable*}{lcccccccc}
\tablecaption{Physical Properties of ALCS Continuum Sources}
\tablehead{
\colhead{ID}  & \colhead{$z_{\rm spec}$} & \colhead{$z_{\rm phot}$} & \colhead{$S_{\rm obs}$} & \colhead{$L_{\rm IR}$} & \colhead{$\mu$}  & \colhead{$z$ type} & \colhead{Ref.} & \colhead{Note} \\
\colhead{}    &       \colhead{}          &  \colhead{}               &  \colhead{(mJy)}           & \colhead{($L_{\odot}$)}              & \colhead{}        & \colhead{}         & \colhead{}     & \colhead{}  \\
\colhead{(1)} & \colhead{(2)}           & \colhead{(3)}                 & \colhead{(4)}                & \colhead{(5)}     & \colhead{(6)}      & \colhead{(7)}  & \colhead{(8)} & \colhead{(9)}}
\startdata
\multicolumn{9}{c}{Blind Catalog \tcb{($N=$125)$^{\dagger}$}} \\   \hline
AC0102-C11 &    \nodata           & $ (3.0\pm2.0) $ 		   & $ 0.3\pm 0.10 $ & $ 12.00\pm0.19 $ & 1.6$_{-0.6}^{+0.7}$ & 5 &  \nodata   & No counterparts in {\it HST}, IRAC, {\it Herschel}\\
AC0102-C22 &    \nodata           & $ 3.45^{+0.17}_{-0.17}$   & $ 1.11\pm0.13 $ & $ 12.53\pm0.19 $ & 2.1$_{-0.8}^{+0.8}$ & 4 & App.\ref{sec:app_znir}  & No counterparts in {\it HST} \\
AC0102-C50 &    \nodata           & $ 2.38^{+0.10}_{-0.11}$   & $ 0.60\pm0.10 $ & $ 12.85\pm0.07 $ & 2.4$_{-1.0}^{+1.0}$ & 3 & K22 &  \\
AC0102-C52 &    \nodata           & $ 2.02^{+0.35}_{-0.16}$   & $ 0.30\pm0.09 $ & $ 12.36\pm0.11 $ & 2.3$_{-0.9}^{+1.0}$ & 3 & K22 &  \\
AC0102-C160 &    \nodata           & $ 1.99^{+0.18}_{-0.14}$ & $ 0.65\pm0.10 $ & $ 12.61\pm0.13 $ & 2.3$_{-0.9}^{+0.9}$ & 3 & K22 &  \\
AC0102-C224 & 4.320                 &  \nodata                          & $ 3.19\pm0.15 $ & $ 13.07\pm0.14 $ & 5.1$_{-3.6}^{+3.6}$ & 1  & C21 & multiple images (AC0102-C118/215/224)  \\
AC0102-C241 &    \nodata           & $ 2.82^{+0.05}_{-0.67}$ & $ 0.97\pm0.23 $ & $ 12.70\pm0.14 $ & 2.0$_{-0.8}^{+0.8}$ & 3 & K22 &  \\
\enddata  
\tablecomments{
(1) Source ID. 
(2) Spectroscopic redshift (Section \ref{sec:redshift}). 
(3) Photometric redshift (Section \ref{sec:redshift}).  
(4) Observed flux density, including the primary beam correction and systematic offsets through the MC simulations (Section \ref{sec:flux}), 
\tcb{but without the lens correction.}
The errors in the flux measurement and corrections are propagated.  
%where the uncertainties are propagated. 
For the multiple-image system, we show an observed flux density calculated by multiplying the average values of the intrinsic flux densities and magnifications. 
(5) Magnification factor based on our fiducial lens model, where the error is evaluated from the propagation of the redshift uncertainty and the systematic uncertainty from different lens models (Section \ref{sec:lens}). 
(6) Redshift category.  
1: $z_{\rm spec}$,  
2: $z_{\rm phot}$ + multiple images used in the construction of the lens model, 
3: $z_{\rm phot}$ with {\sc eazy}, 
4: $z_{\rm phot}$ with the composite SED from 707 dusty galaxies in AS2UDS for the sources without {\it HST} counterparts/data (Appendix \ref{sec:app_znir}), 
5: $z=3.0\pm2.0$ assumed for four sources with no counterparts in all bands other than ALMA). 
(7): Reference: 
B13 \citep{biviano2013}, 
C21 \citep{caputi2021}, 
Ca19 \citep{caminha2019}, 
Ce18 \citep{cerny2018}, 
D17 \citep{mirka2017}, 
E09 \citep{ebeling2009}, 
E17 \citep{ebeling2017}, 
F21 \citep{fujimoto2021}, 
G12 \citep{gomez2012}, 
K22 \citep{kokorev2022}, 
L21 \citep{laporte2021}
R21 \citep{richard2021}, 
MA22 \citep{munoz-arancibia2022}, 
S04 \citep{sand2004}, 
S14 \citep{schmidt2014}, 
S99 \citep{soucail1999}, 
T15 \citep{treu2015}, 
W15 \citep{wang2015}, and 
W12 \citep{wold2012}. \\
\tcb{$\dagger$ This is after correcting the lensing effect for the multiply imaged sources, resulting in a slightly lower number than that in Table~\ref{tab:alcs_catalog_full}. }
} \label{tab:alcs_catalog2_full}
\end{deluxetable*}

In Table \ref{tab:alcs_catalog_full} and \ref{tab:alcs_catalog2_full}, we summarize the observed flux density, source redshift, and the magnification factor for the ALCS 180 sources. 

In Figure \ref{fig:app_postage}, we also summarize the image cutouts of ALMA/Band6, HST/F814W, HST/F160W, and IRAC/ch1 for the ALCS 180 sources. 

%%%%%%%%%%%%%%%%%%%%%%%

\begin{figure*}
\begin{center}
\includegraphics[trim=0cm 0cm 0cm 0cm, clip, angle=0,width=1.0\textwidth]{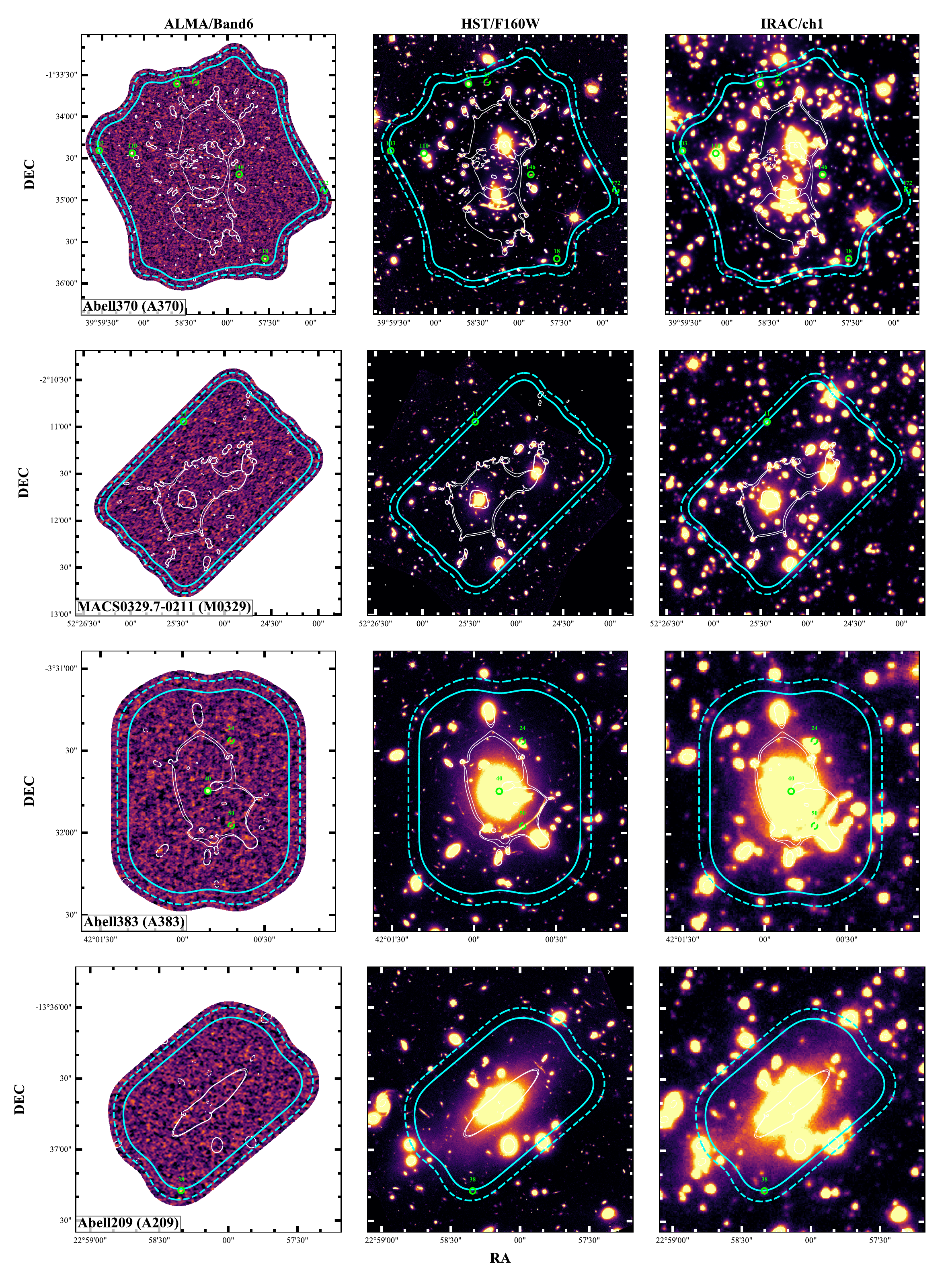}
{\bf Figure \ref{fig:app_cluster}} (continued)
\end{center}
\end{figure*}

\begin{figure*}
\begin{center}
\includegraphics[trim=0cm 0cm 0cm 0cm, clip, angle=0,width=1.0\textwidth]{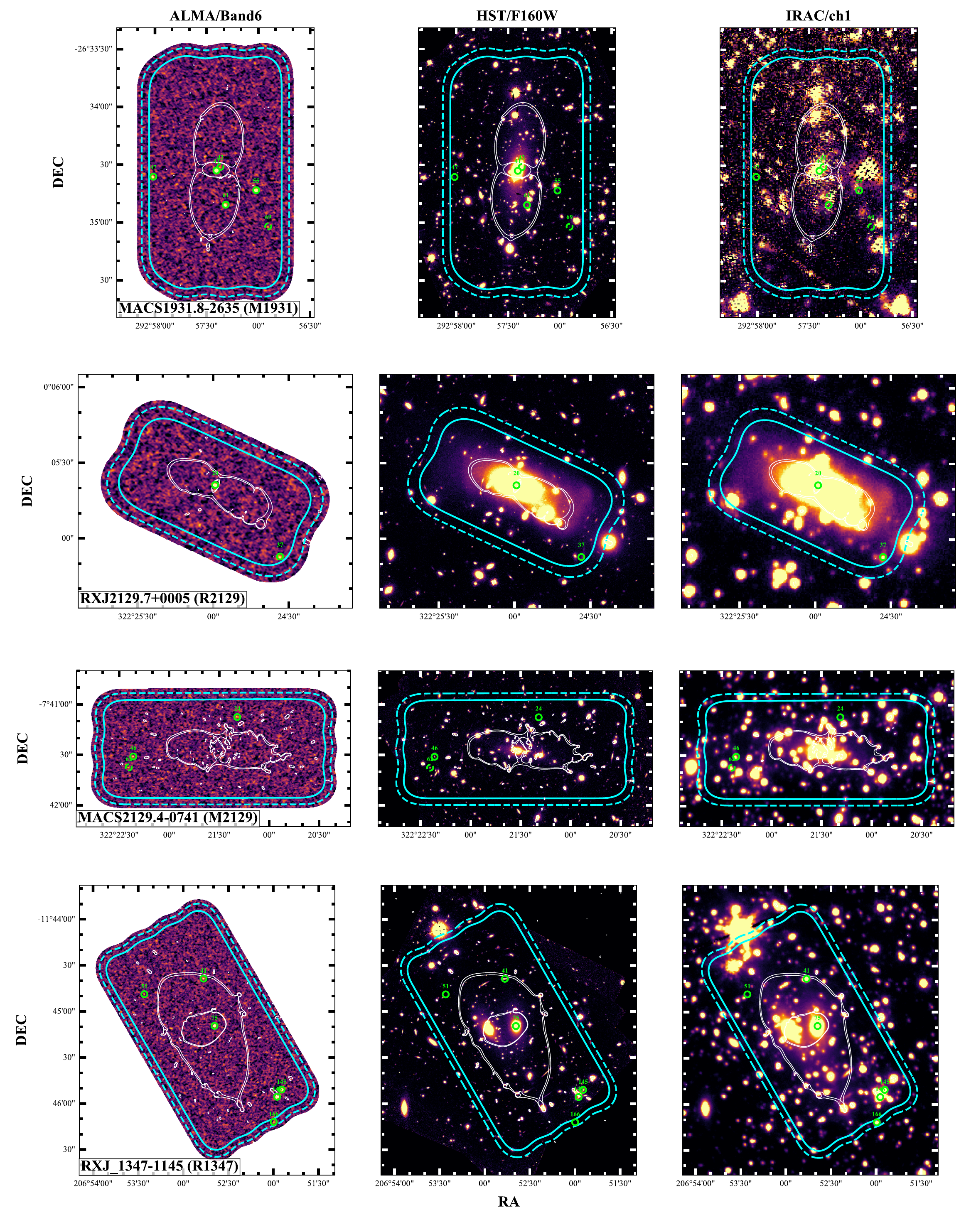}
{\bf Figure \ref{fig:app_cluster}} (continued)
\end{center}
\end{figure*}

\begin{figure*}
\begin{center}
\includegraphics[trim=0cm 0cm 0cm 0cm, clip, angle=0,width=1.0\textwidth]{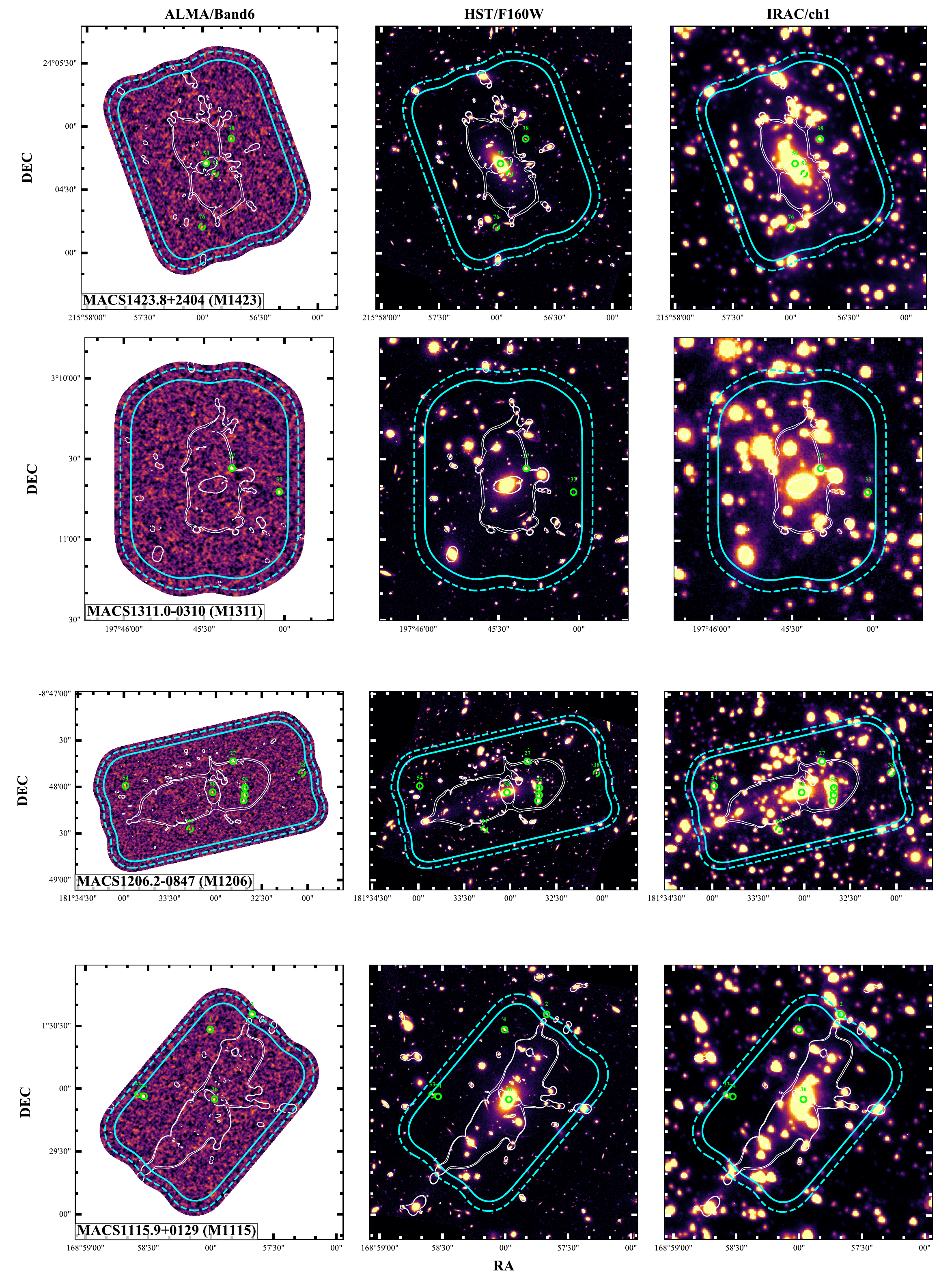}
{\bf Figure \ref{fig:app_cluster}} (continued)
\end{center}
\end{figure*}

\begin{figure*}
\begin{center}
\includegraphics[trim=0cm 0cm 0cm 0cm, clip, angle=0,width=1.0\textwidth]{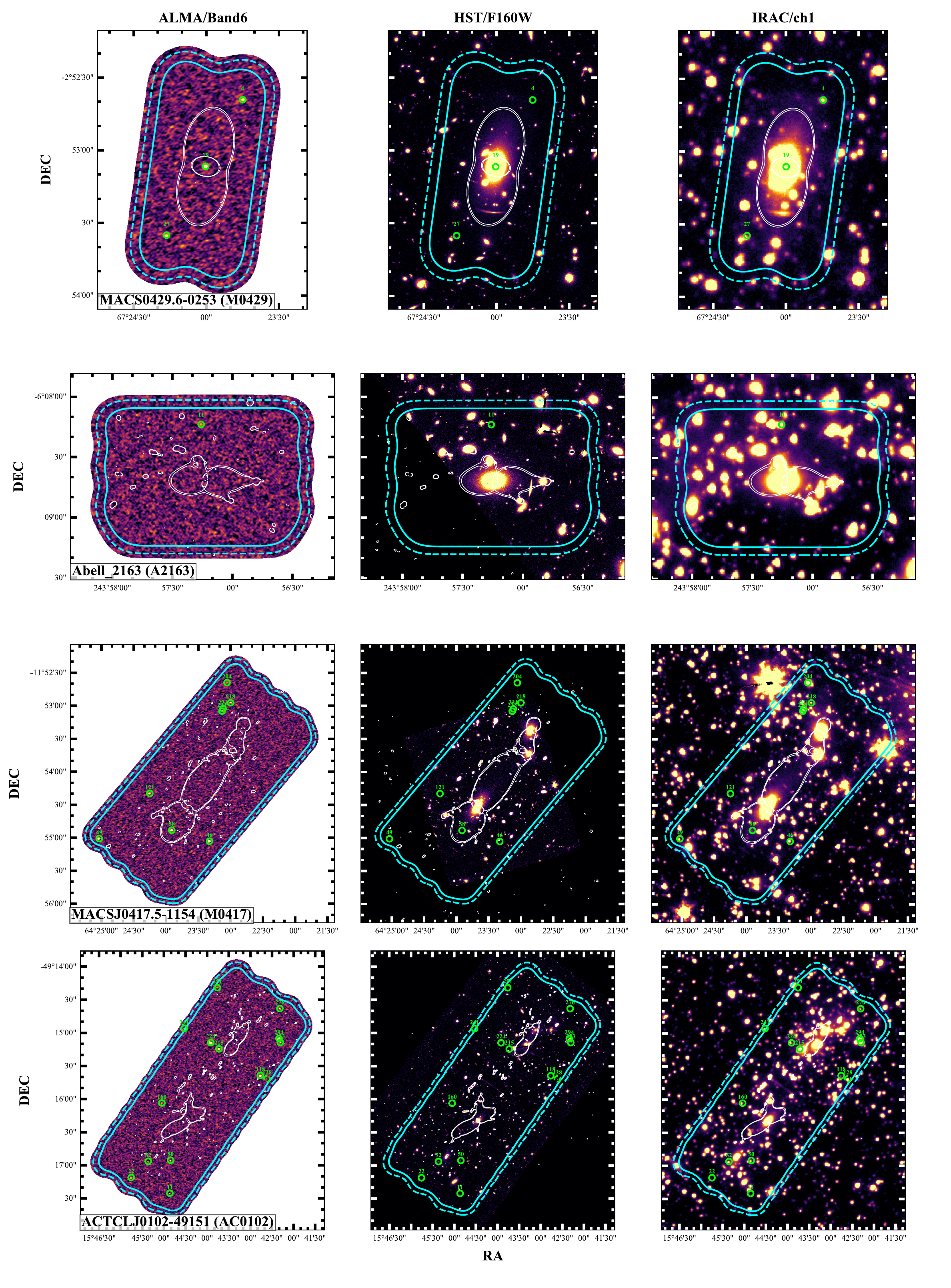}
{\bf Figure \ref{fig:app_cluster}} (continued)
\end{center}
\end{figure*}

\begin{figure*}
\begin{center}
\includegraphics[trim=0cm 0cm 0cm 0cm, clip, angle=0,width=1.0\textwidth]{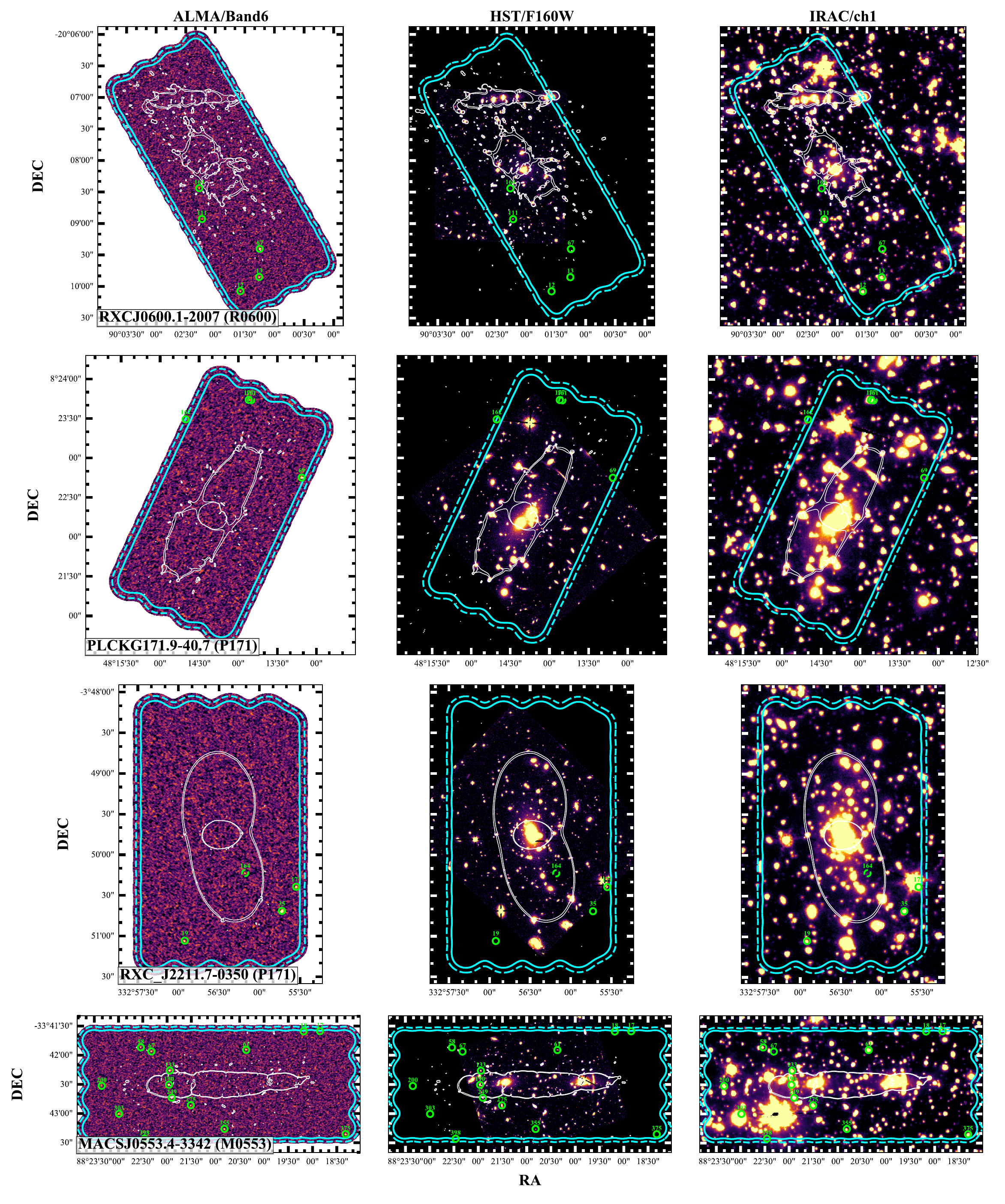}
{\bf Figure \ref{fig:app_cluster}} (continued)
\end{center}
\end{figure*}

\begin{figure*}
\begin{center}
\includegraphics[trim=0cm 0cm 0cm 0cm, clip, angle=0,width=1.0\textwidth]{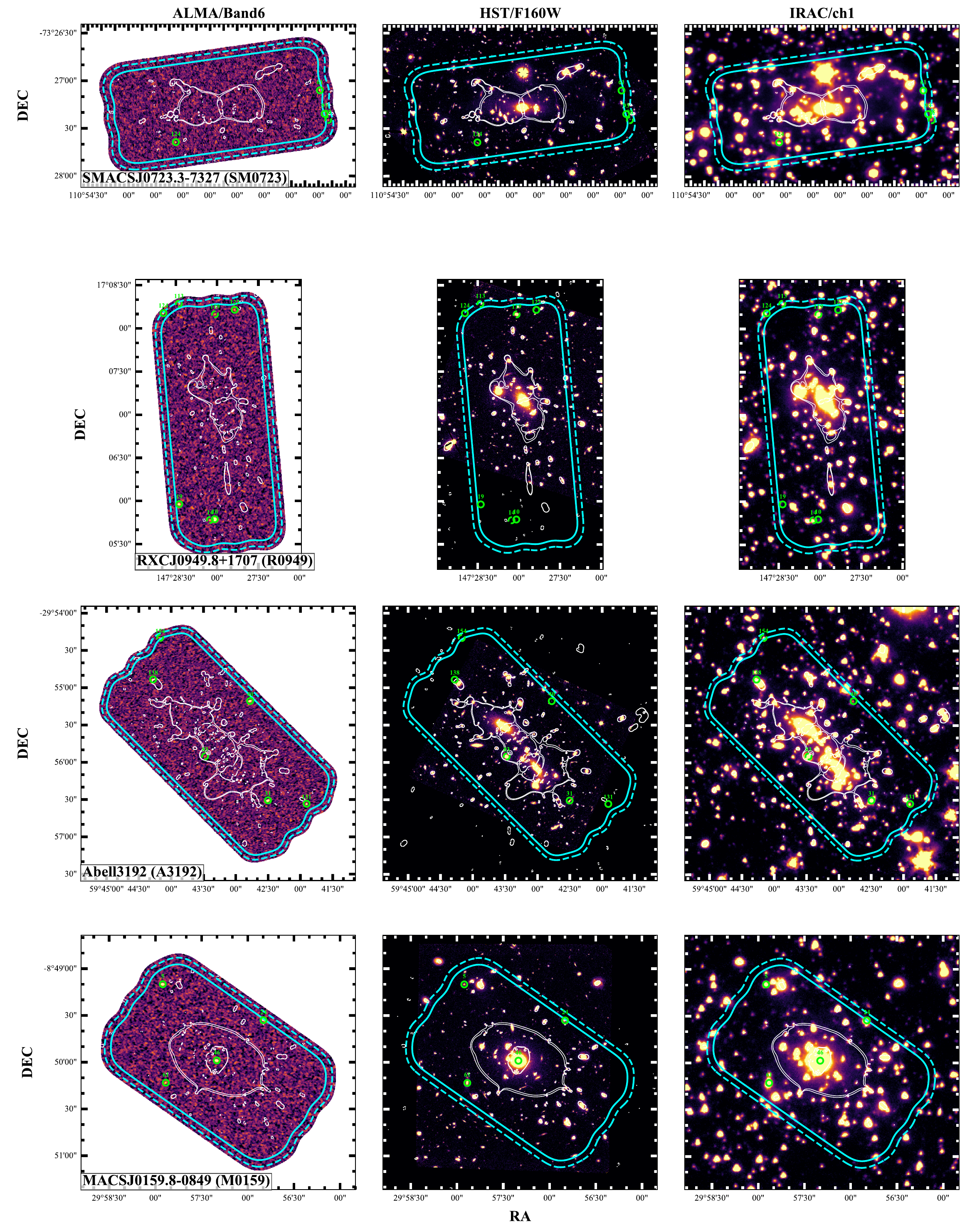}
{\bf Figure \ref{fig:app_cluster}} (continued)
\end{center}
\end{figure*}

\begin{figure*}
\begin{center}
\includegraphics[trim=0cm 0cm 0cm 0cm, clip, angle=0,width=1.0\textwidth]{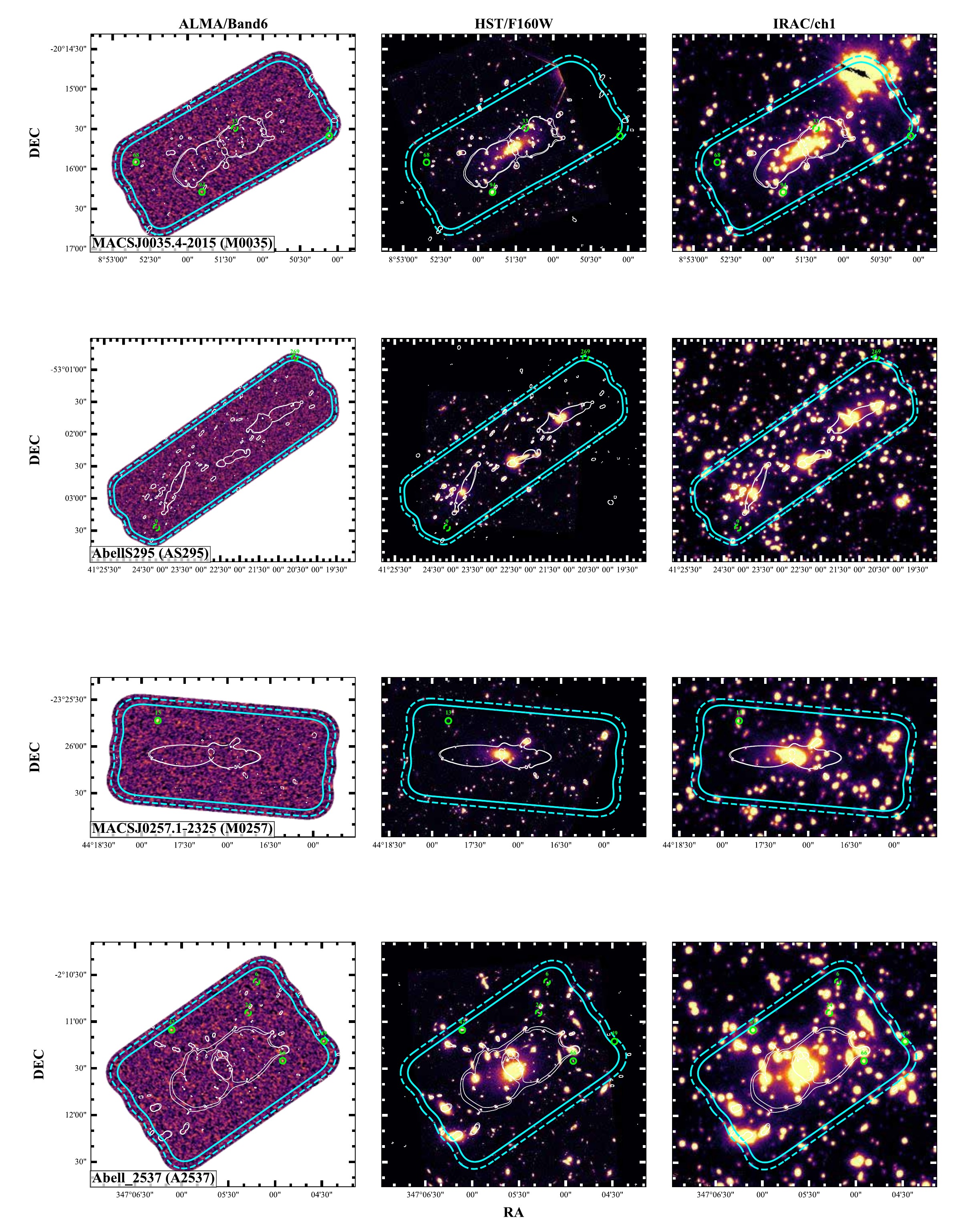}
{\bf Figure \ref{fig:app_cluster}} (continued)
\end{center}
\end{figure*}

%%%%%%%%%%%%%%%%%%%%%% 

%%%%%%%%%%%%%%%%%%%%%%%
\begin{figure*}
\begin{center}
\includegraphics[trim=0cm 0cm 0cm 0cm, clip, angle=0,width=1.0\textwidth]{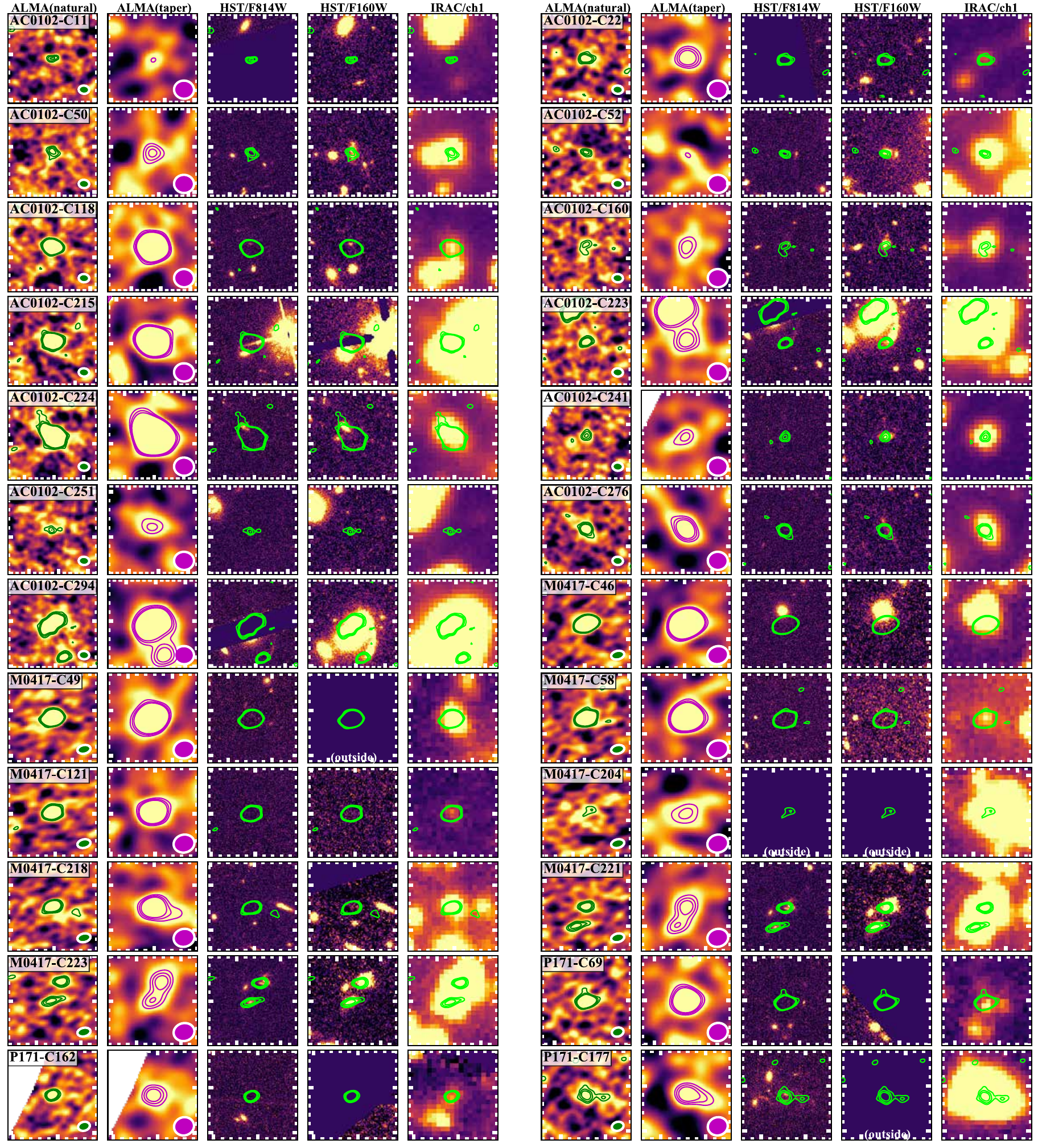}
 \caption{
$10''\times10''$ image cutouts -- ALMA 1.2-mm (natural-weighted), ALMA 1.2-mm ($uv$-tapered, HST/F814W, HST/F160W, and IRAC/ch1 from left to right -- for 141 ALCS sources in the blind catalog. 
The contours indicate the intensity at 3$\sigma$, 4$\sigma$, and 5$\sigma$ levels. 
\label{fig:app_postage}}
\end{center}
\end{figure*}

\begin{figure*}
\begin{center}
\includegraphics[trim=0cm 0cm 0cm 0cm, clip, angle=0,width=1.0\textwidth]{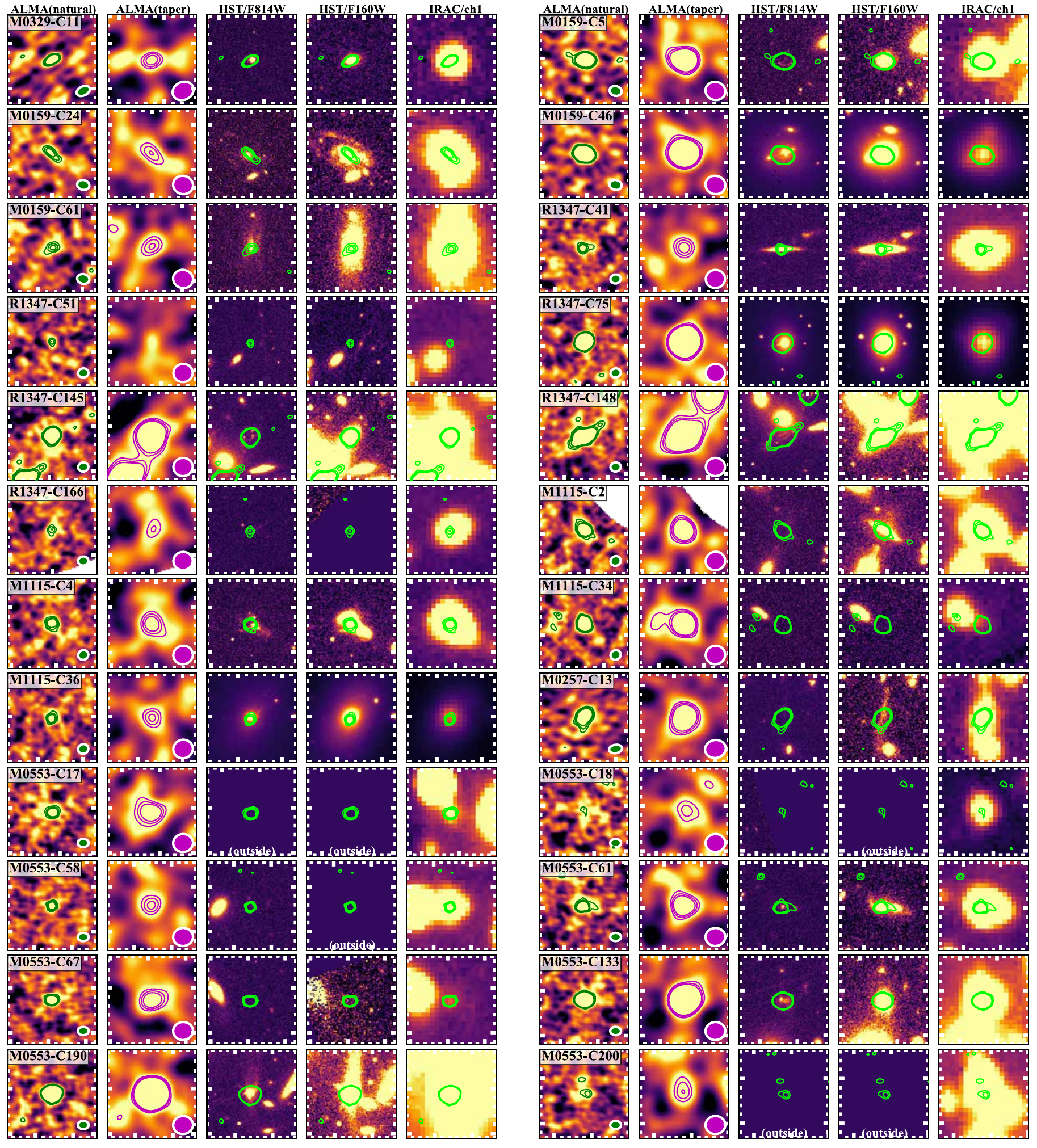}
{\bf Figure \ref{fig:app_postage}} (continued)
\end{center}
\end{figure*}

\begin{figure*}
\begin{center}
\includegraphics[trim=0cm 0cm 0cm 0cm, clip, angle=0,width=1.0\textwidth]{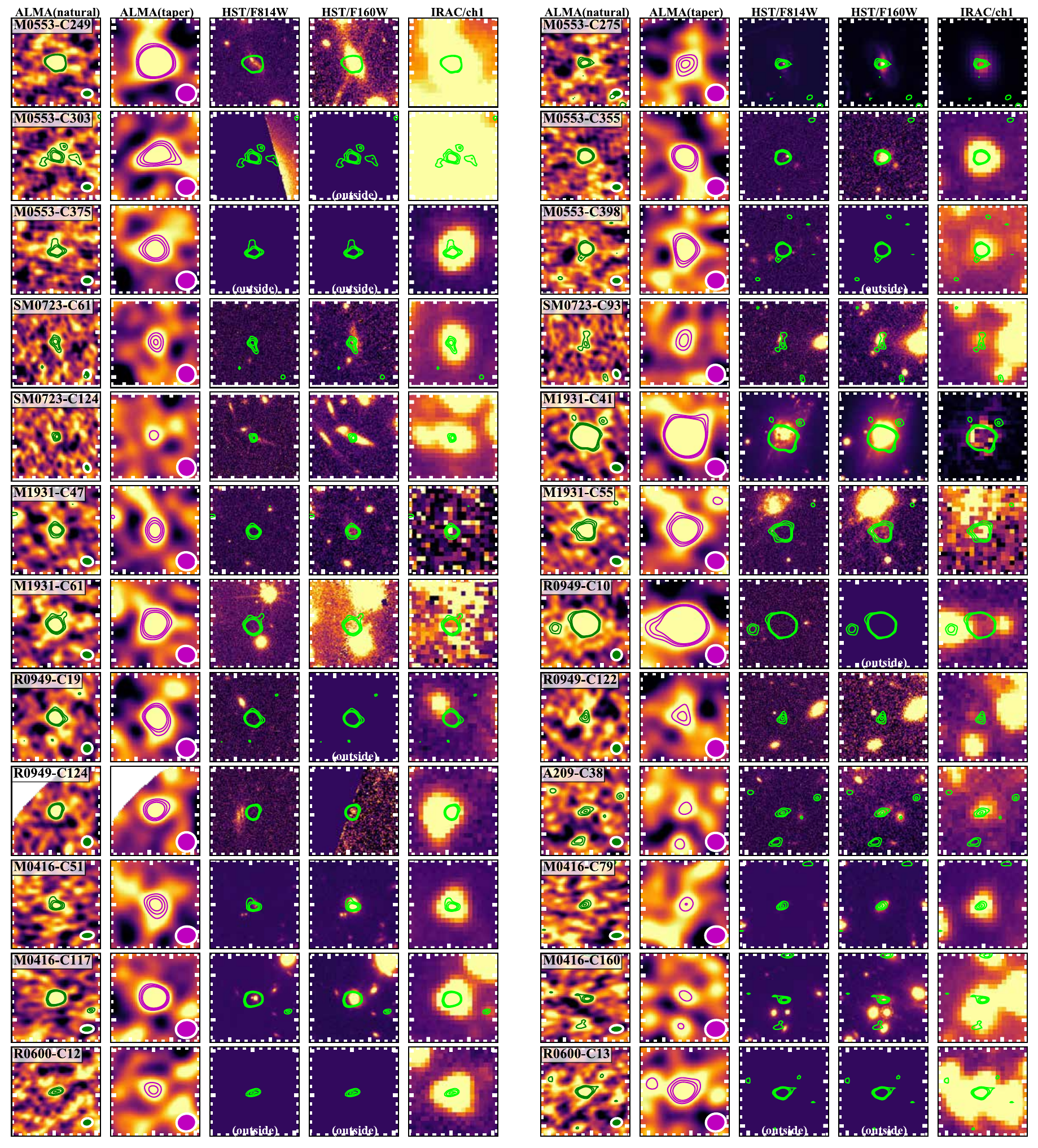}
{\bf Figure \ref{fig:app_postage}} (continued)
\end{center}
\end{figure*}

\begin{figure*}
\begin{center}
\includegraphics[trim=0cm 0cm 0cm 0cm, clip, angle=0,width=1.0\textwidth]{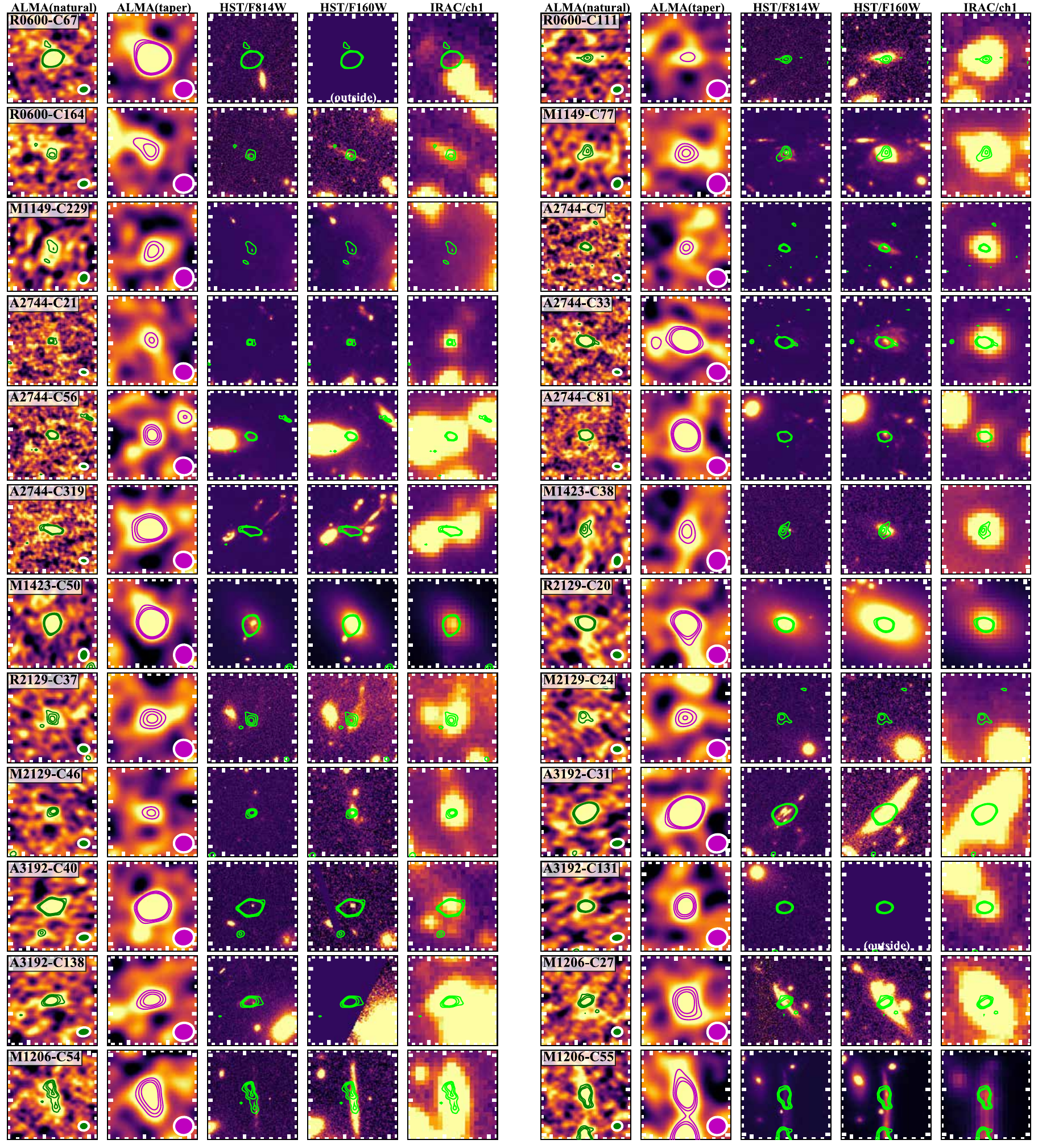}
{\bf Figure \ref{fig:app_postage}} (continued)
\end{center}
\end{figure*}

\begin{figure*}
\begin{center}
\includegraphics[trim=0cm 0cm 0cm 0cm, clip, angle=0,width=1.0\textwidth]{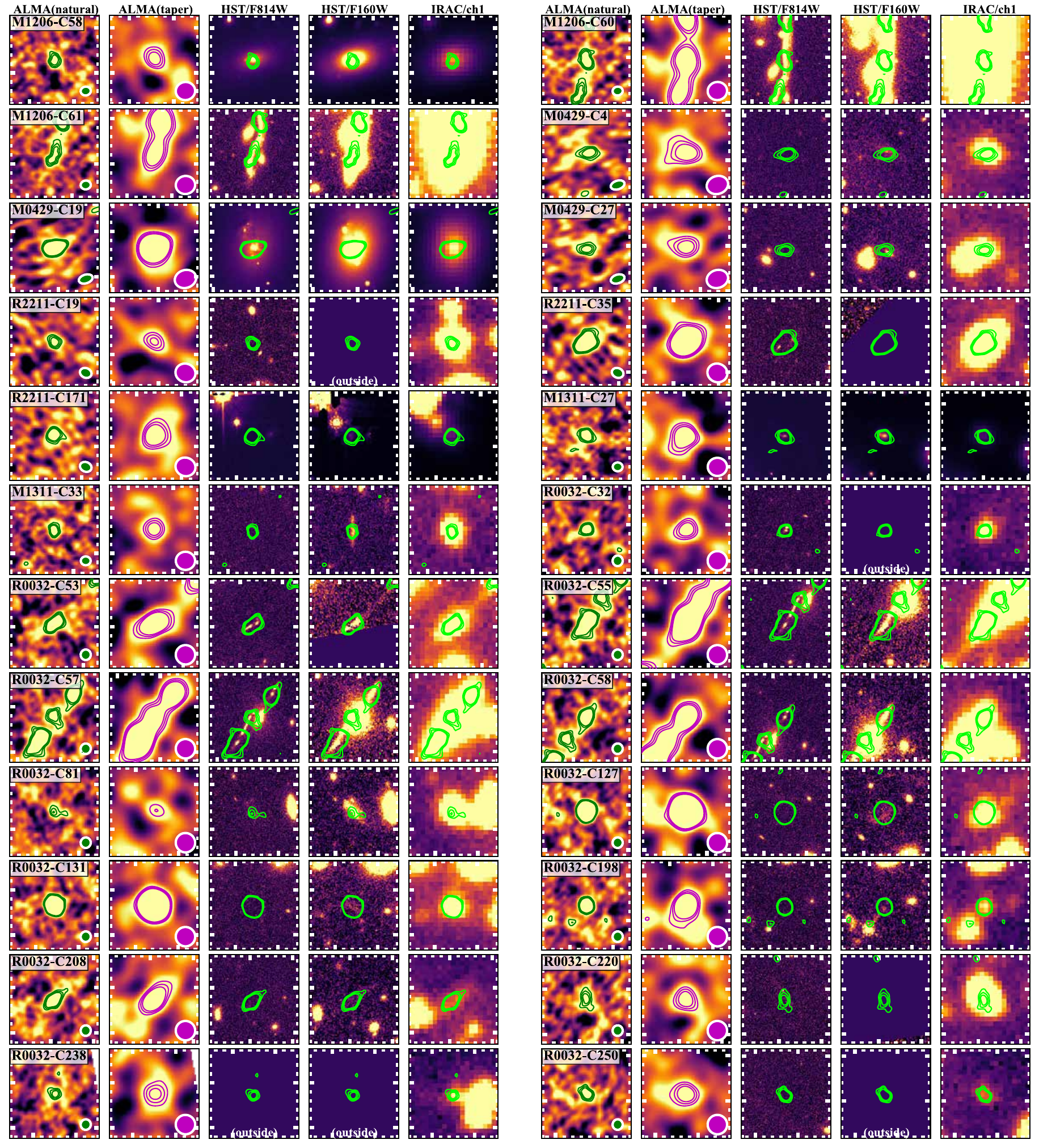}
{\bf Figure \ref{fig:app_postage}} (continued)
\end{center}
\end{figure*}

\begin{figure*}
\begin{center}
\includegraphics[trim=0cm 0cm 0cm 0cm, clip, angle=0,width=1.0\textwidth]{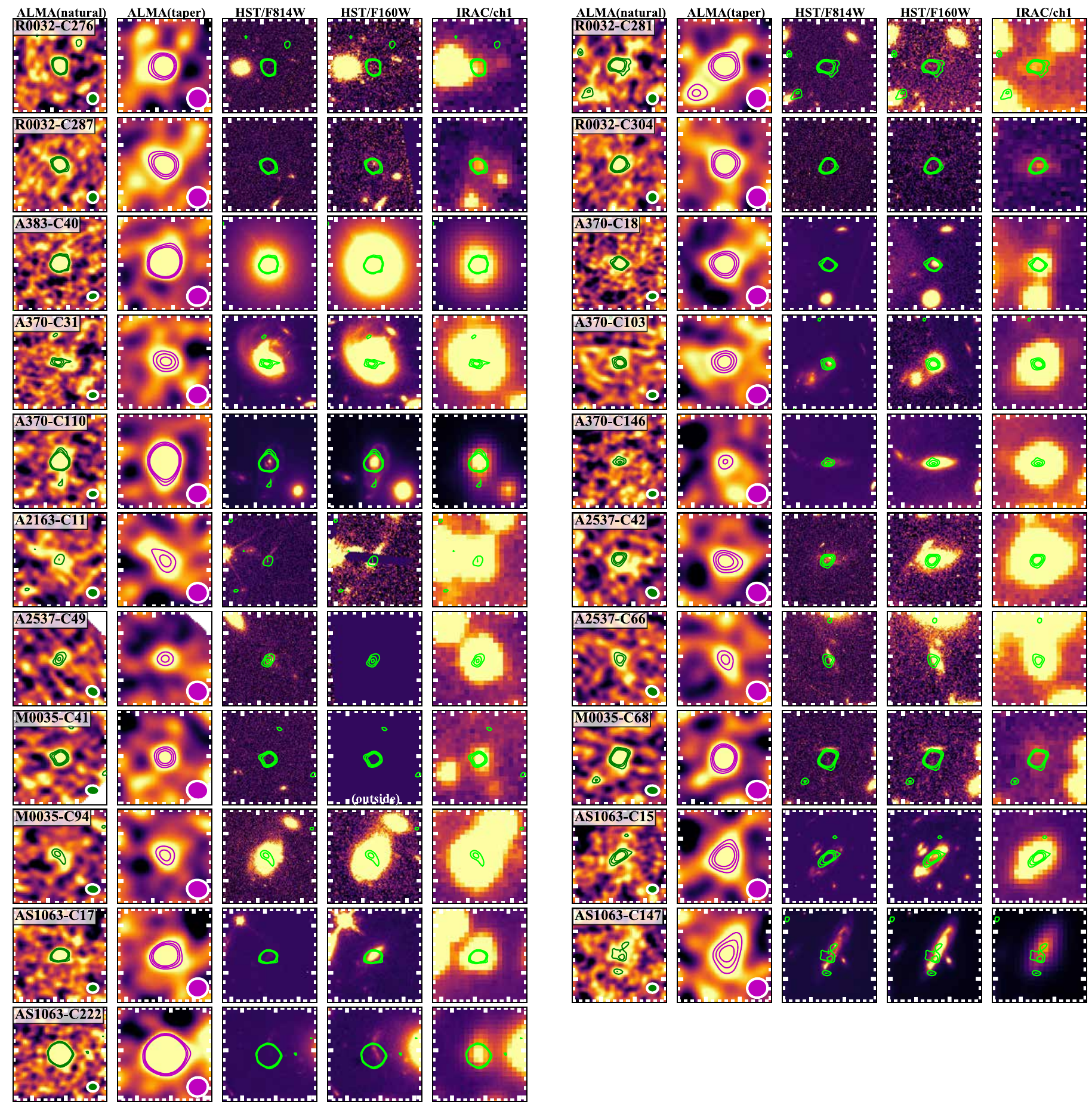}
{\bf Figure \ref{fig:app_postage}} (continued)
\end{center}
\end{figure*}

\begin{figure*}
\begin{center}
\includegraphics[trim=0cm 0cm 0cm 0cm, clip, angle=0,width=1.0\textwidth]{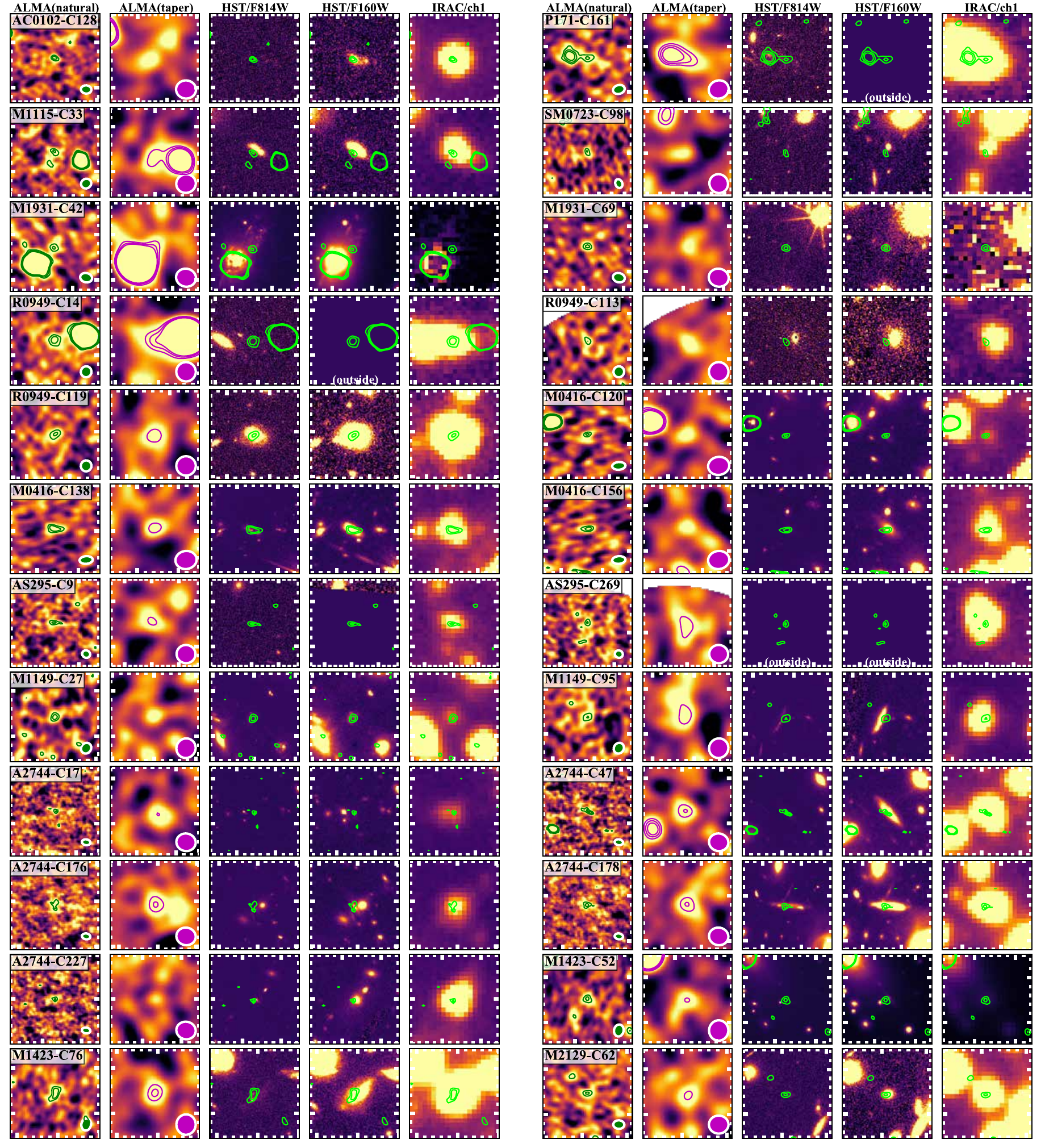}
 \caption{
Same as Figure \ref{fig:app_postage}, but for 39 ALCS sources in the prior catalog. 
\label{fig:app_postage2}}
\end{center}
\end{figure*}

\begin{figure*}
\begin{center}
\includegraphics[trim=0cm 0cm 0cm 0cm, clip, angle=0,width=1.0\textwidth]{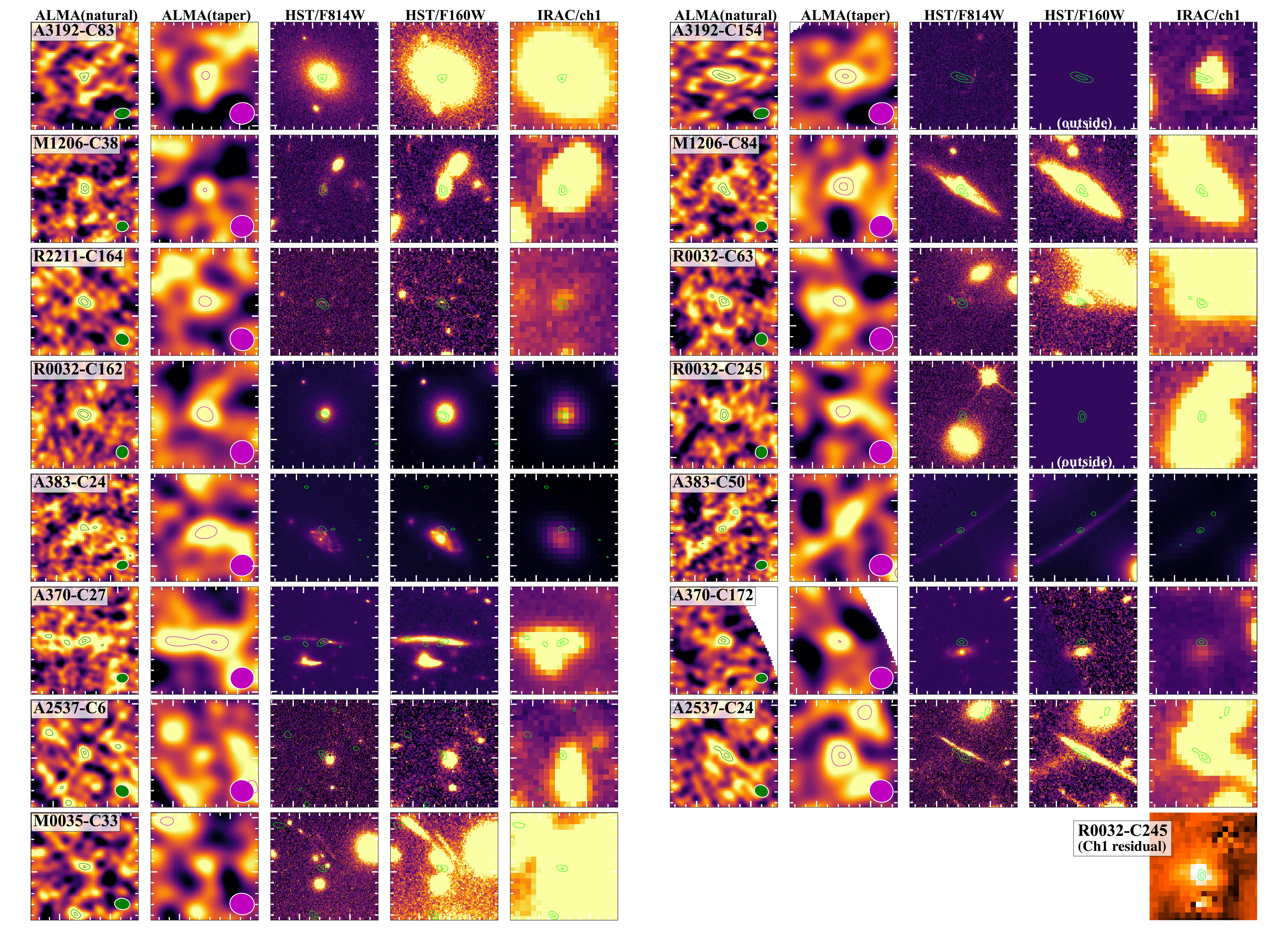}
{\bf Figure \ref{fig:app_postage2}} (continued)
\end{center}
\end{figure*}

%%%%%%%%%%%%%%%%%%%%%% 

\bibliographystyle{apj}
\bibliography{apj-jour,reference}

\end{document}